\documentclass{mypinchcrn} 
\pdfoutput=1

\usepackage{graphicx}
\usepackage{braket,slashed,mathrsfs}
\usepackage{bm,amsmath,amssymb}

\definecolor{refkey}{rgb}{1.0,0.0,0.0}
\definecolor{labelkey}{rgb}{0,0.5,0.0}

\usepackage{tikz}
\usetikzlibrary{decorations.pathmorphing}
\usetikzlibrary{automata,positioning}

\usepackage{cancel}
\usepackage[normalem]{ulem}

\newcommand\vev[1]{\left \langle #1 \right \rangle}
\newcommand\abs[1]{\left | #1 \right |}
\newcommand{\MSbar}{\ensuremath{\overline{\text{MS}}}}
\newcommand{\lqcd}{\ensuremath{\Lambda_{\text{QCD}}}}
\newcommand{\lchi}{\ensuremath{\Lambda_\chi}}
\newcommand\nn{\nonumber \\ }
\newcommand\rd{\text{d}}
\newcommand\eUV{\epsilon_\text{UV}}
\newcommand\eIR{\epsilon_\text{IR}}
\def\LL{\mathscr{L}}
\def\HH{\mathscr{H}}
\def\d{\mathsf{d}} 
\def\opdim{\mathscr{D}}

\renewcommand{\O}{\mathcal{O}}

\title{
\centerline{\vspace{2cm}\hspace{0.7cm}\hbox{\includegraphics[width=18cm]{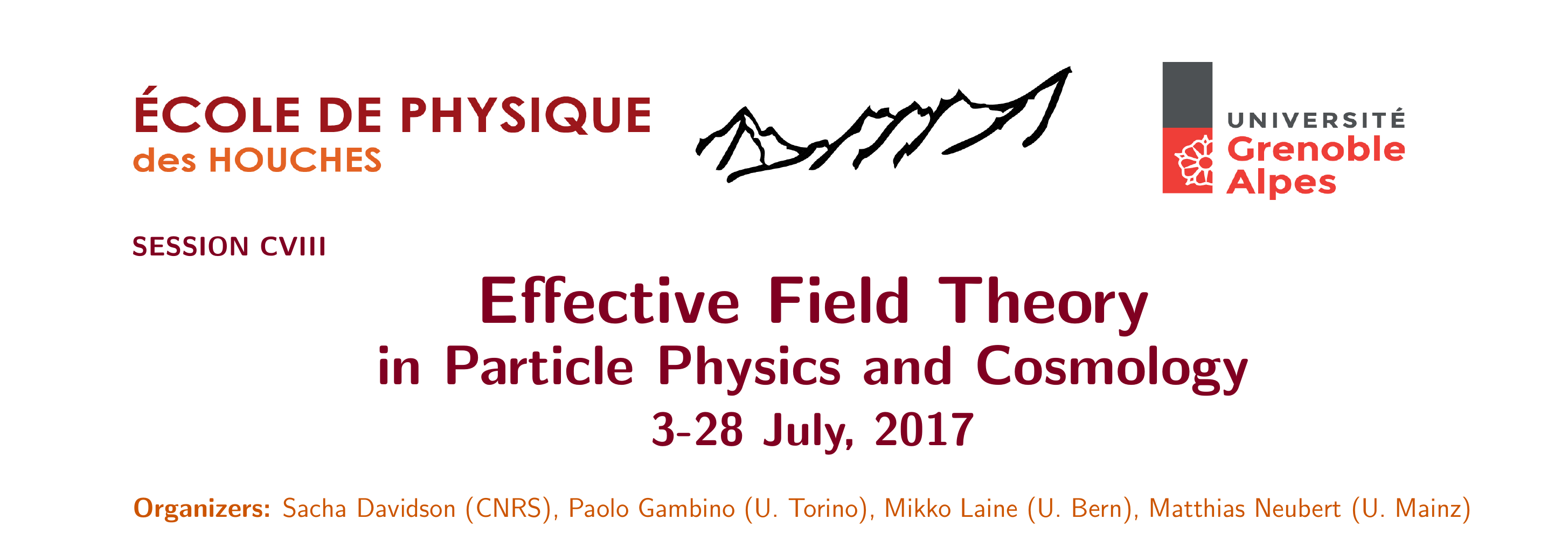}}} 
 Introduction to Effective Field Theories }

\author{Aneesh V. Manohar}
\affiliation{Department of Physics 0319,
  University of California at San Diego, \\ 9500 Gilman Drive, La Jolla, CA 92093, USA
}

\begin{document}

\maketitle

\tableofcontents

\maintext

\chapter{Introduction}

This is an introductory set of lectures on the basic ideas and methods of effective field theories (EFTs). Other lectures at the school will go into more details about the most commonly used effective theories in high energy physics and cosmology. Professor Neubert's lectures~\cite{Neubert:dp}, delivered concurrently with mine, provide an excellent introduction to renormalization in quantum field theory (QFT), the renormalization group equation, operator mixing, and composite operators, and this knowledge will be assumed in my lectures. I also have some 20 year old lecture notes from the Schladming school~\cite{Manohar:1996cq} which should be read in conjunction with these lectures. Additional references are~\cite{Georgi:1985kw,Kaplan:1995uv,Rothstein:2003mp,Stewart:fv}.
The Les Houches school and these lecture notes focus on aspects of EFTs as used in high energy physics and cosmology which are relevant for  making contact with experimental observations.

The intuitive idea behind effective theories is that you can calculate without knowing the exact theory. Engineers are able to design and build bridges without any knowledge of strong interactions or quantum gravity. The main inputs in the design are Newton's laws of mechanics and gravitation, the theory of elasticity, and fluid flow. The engineering design depends on parameters measured on macroscopic scales of order meters, such as the elastic modulus of steel. Short distance properties of Nature, such as the existence of weak interactions, or the mass of the Higgs boson are not needed.

In some sense, the ideas of EFT are ``obvious.'' However, implementing them in a mathematically consistent way in an interacting quantum field theory is not so obvious. These lectures provide pedagogical examples of how one actually implements EFT ideas in particle physics calculations of experimentally relevant quantities. Additional details on specific EFT applications are given in other lectures in this volume.

An EFT is a quantum theory in its own right, and like any other QFT, it comes with a regularization and renormalization scheme necessary to obtain finite matrix elements. One can compute $S$-matrix elements in an EFT from the EFT Lagrangian, with no additional external input, in the same way that one can compute in QED starting from the QED Lagrangian. In many cases, an EFT is the low-energy limit of a more fundamental theory (which might itself be an EFT), often called the ``full theory.'' 

Effective field theories allow you to compute an experimentally measurable quantity with some \emph{finite} error. Formally, an EFT has a small expansion parameter $\delta$, known as the power counting parameter. Calculations are done in an expansion to some order $n$ in $\delta$, so that the error is of order $\delta^{n+1}$. Determining the order in $\delta$ of a given diagram is done using what is referred to as a power counting formula.

A key aspect of EFTs is that one has a systematic expansion, with a well-defined procedure to compute higher order corrections in $\delta$. Thus one can compute to arbitrarily high order in $\delta$, and make the theoretical error as small as desired, by choosing $n$ sufficiently large. Such calculations might be extremely difficult in practice because higher order diagrams are hard to compute, but they are possible in principle. This is very different from modeling, e.g.\ the non-relativistic quark model provides a good description of hadron properties at the 25\% level. However, it is \emph{not} the first term in a systematic expansion, and it is not possible to systematically improve the results.

In many examples, there are multiple expansion parameters $\delta_1$, $\delta_2$, etc.\ For example, in heavy quark effective theory (HQET)~\cite{Isgur:1989vq,Isgur:1989ed,Manohar:2000dt,Shifman:1987rj}, $b$ decay rates have an expansion in $\delta_1=\lqcd/m_b$ and $\delta_2=m_b/M_W$. In such cases, one has to determine which terms $\delta_1^{n_1} \delta_2^{n_2}$ must be retained to reach the desired accuracy goal. Usually, but not always, the expansion parameter is the ratio of a low-energy scale such as the external momentum $p$, or particle mass $m$, and a short-distance scale usually denoted by $\Lambda$, $\delta = p/\Lambda$. In many examples, one also has a perturbative expansion in a small coupling constant such as $\alpha_s(m_b)$ for HQET.

EFT calculations to order $\delta^n$ depend on a finite number of Lagrangian parameters $N_n$. The number of parameters $N_n$ generally increases as $n$ increases. One gets parameter-free predictions in an EFT by calculating more experimentally measured quantities than $N_n$. For example, HQET computations to order $\lqcd^2/m_b^2$ depend on two parameters $\lambda_1$ and $\lambda_2$ of order $\lqcd^2$. There are many experimental quantities that can be computed to this order, such as the meson masses, form factors, and decay spectra~\cite{Manohar:2000dt}. Two pieces of data are used to fix $\lambda_1$ and $\lambda_2$, and then one has parameter-free predictions for all other quantities.

 EFTs can be used even when the dynamics is non-perturbative. The most famous example of this type is chiral perturbation theory ($\chi$PT), which has an expansion in $p/\lchi$, where $\lchi \sim 1$\,GeV is the chiral symmetry breaking scale. Systematic computations in powers of $p/\lchi$ are in excellent agreement with experiment~\cite{Gasser:1983yg,Pich:1995bw,Pich:cs,Weinberg:1978kz}.
 
The key ingredient used in formulating EFTs  is locality, which leads to a separation of scales, i.e.\ factorization of the field theory amplitudes into short-distance Lagrangian coefficients and long-distance matrix elements. The short-distance coefficients are universal, and  independent of the long-distance matrix elements computed~\cite{Wilson:1969zs}. The experimentally measured quantities $\O_i$ are then given as the product of these short-distance coefficients $C$ and long-distance matrix elements. Often, there are multiple coefficients and matrix elements, so that $\O_i = \sum_i C_{ij} M_j$. Sometimes, as in deep-inelastic scattering, $C$ and $M$ depend on a variable $x$ instead of an index $i$, and the sum becomes a convolution
\begin{align}
\label{1.1}
\O &= \int_0^1 \frac{\rd x}{x} C(x) M(x)\,.
\end{align}
The short distance coefficient $C(x)$ in this case is called the hard-scattering cross section, and can be computed in QCD perturbation theory. The long-distance matrix elements are the parton distribution functions, which are determined from experiment. The hard-scattering cross-section is universal, but the parton distribution functions depend on the hadronic target. 

 EFTs allow one to organize calculations in an efficient way, and to estimate quantities using the power counting formula in combination with locality and gauge invariance. The tree-level application of EFTs is straightforward; it is simply a series expansion of the scattering amplitude in a small parameter. The true power lies in being able to compute radiative corrections. It is worth repeating that EFTs are full-fledged quantum theories, and one can compute measurable quantities such as $S$-matrix elements \emph{without any reference or input from a underlying UV theory. } The 1933 Fermi theory of weak interactions~\cite{Fermi:1933jpa} was used long before the Standard Model was invented, or anyone knew about electroweak gauge bosons. Pion-nucleon scattering lengths \cite{Tomozawa:1966jm,Weinberg:1966kf} and $\pi-\pi$ scattering lengths~\cite{Weinberg:1966kf} were computed in 1966, without any knowledge of QCD, quarks or gluons.

\bigskip

Here are some warm-up exercises which will be useful later.
\begin{exercisebn}

Show that for a \emph{connected} graph, $V-I+L=1$, where $V$ is the number of vertices, $I$ is the number of internal lines, and $L$ is the number of loops. What is the formula if the graph has $n$ connected components?

\end{exercisebn}
\begin{exercisenn}

Work out the transformation of fermion bilinears $\overline \psi(\mathbf{x},t)\, \Gamma\, \chi(\mathbf{x},t)$ under $C$, $P$, $T$, where $\Gamma=P_L, P_R,\gamma^\mu P_L, \gamma^\mu P_R ,\sigma^{\mu \nu} P_L, \sigma^{\mu \nu}P_R$. Use your results to find the transformations under $CP$, $CT$, $PT$ and $CPT$.

\end{exercisenn}
\begin{exercisenn}\label{ex:nfierz}

Show that for $SU(N)$,
\begin{align}\label{sun}
[T^A]^\alpha_{\ \beta}\, [T^A]^{\lambda}_{\ \sigma} &= \frac12 \delta^\alpha_\sigma\, \delta^\lambda_\beta - \frac{1}{2N} \delta^\alpha_\beta \, \delta^\lambda_\sigma,
\end{align}
where  the $SU(N)$ generators are normalized to $\text{Tr}\, T^A T^B=\delta^{AB}/2$. From this, show that
\begin{align}\
\delta^\alpha_{\ \beta}\, \delta^{\lambda}_{\ \sigma} &= \frac{1}{N} \delta^\alpha_\sigma\, \delta^\lambda_\beta + 2 [T^A]^\alpha_{\ \sigma} \, [T^A]^\lambda_{\ \beta}, \nn
[T^A]^\alpha_{\ \beta}\, [T^A]^{\lambda}_{\ \sigma} &= \frac{N^2-1}{2N^2} \delta^\alpha_\sigma\, \delta^\lambda_\beta - \frac{1}{N} [T^A]^\alpha_{\ \sigma}\, [T^A]^\lambda_{\ \beta}.
\end{align}

\end{exercisenn}
\begin{exercisenb}\label{ex:spinfierz}

Spinor Fierz identities are relations of the form
\begin{align*}
(\overline A\, \Gamma_1\, B)(\overline C\, \Gamma_2\, D) = \sum_{ij} c_{ij} (\overline C\, \Gamma_i\, B)(\overline A\, \Gamma_j\, D)
\end{align*}
where $A,B,C,D$ are fermion fields, and $c_{ij}$ are numbers. They are much simpler if written in terms of chiral fields using $\Gamma_i=P_L, P_R,\gamma^\mu P_L, \gamma^\mu P_R ,\sigma^{\mu \nu} P_L, \sigma^{\mu \nu}P_R$, rather than Dirac fields.  Work out the Fierz relations for
\begin{align*}
& (\overline A P_L B)(\overline C  P_L D), && 
(\overline A \gamma^\mu P_L B)(\overline C \gamma_\mu P_L  D), &&
(\overline A \sigma^{\mu \nu} P_L   B)(\overline C \sigma_{\mu \nu} P_L  D), \nn
& (\overline A P_L B)(\overline C  P_R D),&&
(\overline A \gamma^\mu P_L B)(\overline C \gamma_\mu P_R  D), &&
(\overline A \sigma^{\mu \nu} P_L   B)(\overline C \sigma_{\mu \nu} P_R  D).
\end{align*}
Do not forget the Fermi minus sign. The $P_R \otimes P_R$ identities are obtained from the $P_L \otimes P_L$ identities by using $L \leftrightarrow R$. 

\end{exercisenb}

\chapter{Examples}

In this section, we discuss some qualitative examples of EFTs illustrating the use of power counting, symmetries such as gauge invariance, and dimensional analysis. Some of the examples are covered in detail in other lectures at this school.

\section{Hydrogen Atom}

A simple example that should be familiar to everyone is the computation of the hydrogen atom energy levels, as done in a quantum mechanics class. The Hamiltonian for an electron of mass $m_e$ interacting via a Coulomb potential with a proton treated as an infinitely heavy point particle is
\begin{align}
\label{1.2}
\HH &= \frac{\mathbf{p}^2}{2m_e} - \frac{\alpha}{r}\,.
\end{align}
The binding energies, electromagnetic transition rates, etc.\ are computed from eqn~(\ref{1.2}). The fact that the proton is made up of quarks, weak interactions, neutrino masses, etc.\ are irrelevant, and we do not need any detailed input from QED or QCD. The only property of the proton we need is that its charge is $+1$; this can be measured at long distances from the Coulomb field.

Corrections to eqn~(\ref{1.2}) can be included in a systematic way. Proton recoil is included by replacing $m_e$ by the reduced mass $\mu=m_e m_p/(m_e+m_p)$, which gives corrections of order $m_e/m_p$. At this point, we have included one  strong-interaction parameter, the mass $m_p$ of the proton, which can be determined from experiments done at low energies, i.e.\ at energies much below $\lqcd$.

The hydrogen fine structure is calculated by including  higher order (relativistic) corrections to the Hamiltonian, and gives corrections of relative order $\alpha^2$. The hydrogen hyperfine structure (the famous 21\,cm line) requires including the spin-spin interaction between the proton and electron, which depends on their magnetic moments. The proton magnetic moment $\mu_p=2.793 \,e \hbar/(2m_pc)$ is the second strong interaction parameter which now enters the calculation, and can be measured in low-energy NMR experiments. The electron magnetic moment is given by its Dirac value $-e \hbar/(2 m_e c)$.

Even more accurate calculations require additional non-perturbative parameters, as well as QED corrections. For example, the proton charge radius $r_p$, $g-2$ for the electron, and QED radiative corrections for the Lamb shift all enter to obtain the accuracy required to compare with precision experiments. 

For calculations with an accuracy of $10^{-13}$\,eV $\sim 50$\,Hz, it is necessary to include the weak interactions. The weak interactions give a very small shift in the energy levels, and are a tiny correction to the energies. But they are the leading contribution to atomic parity violation effects. The reason is that the strong and electromagnetic interactions conserve parity. Thus the relative size of various higher-order contributions depends on the quantity being computed---there is no universal rule that can be unthinkingly followed in all examples. Even in the simple hydrogen atom example, we have multiple expansion parameters $m_e/m_p$, $\alpha$, and $m_p/M_W$.

\section{Multipole Expansion in Electrostatics}\label{sec:mult}

\begin{figure}
\centering
\includegraphics[height=6cm]{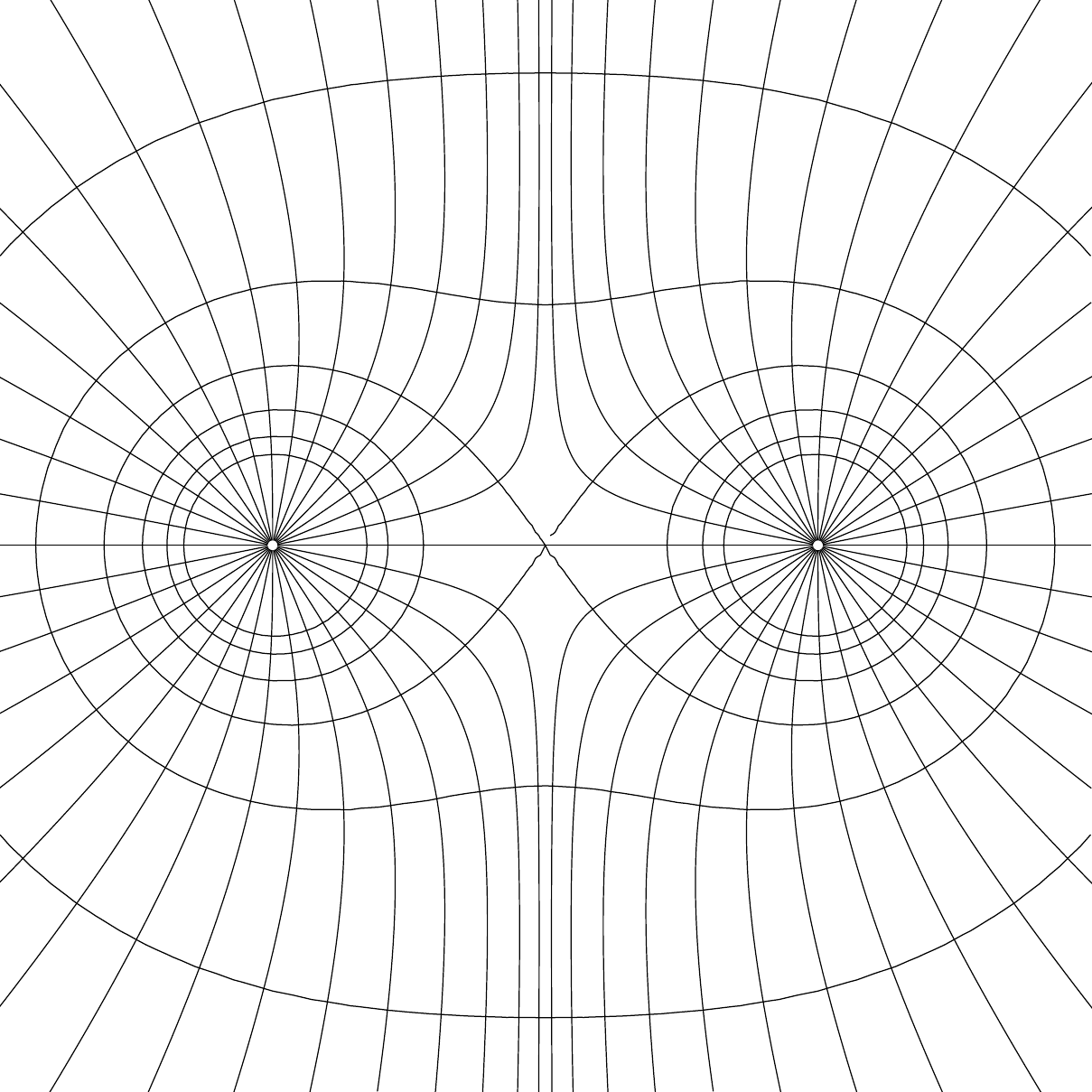}\hspace{0.2cm}
\includegraphics[height=6cm]{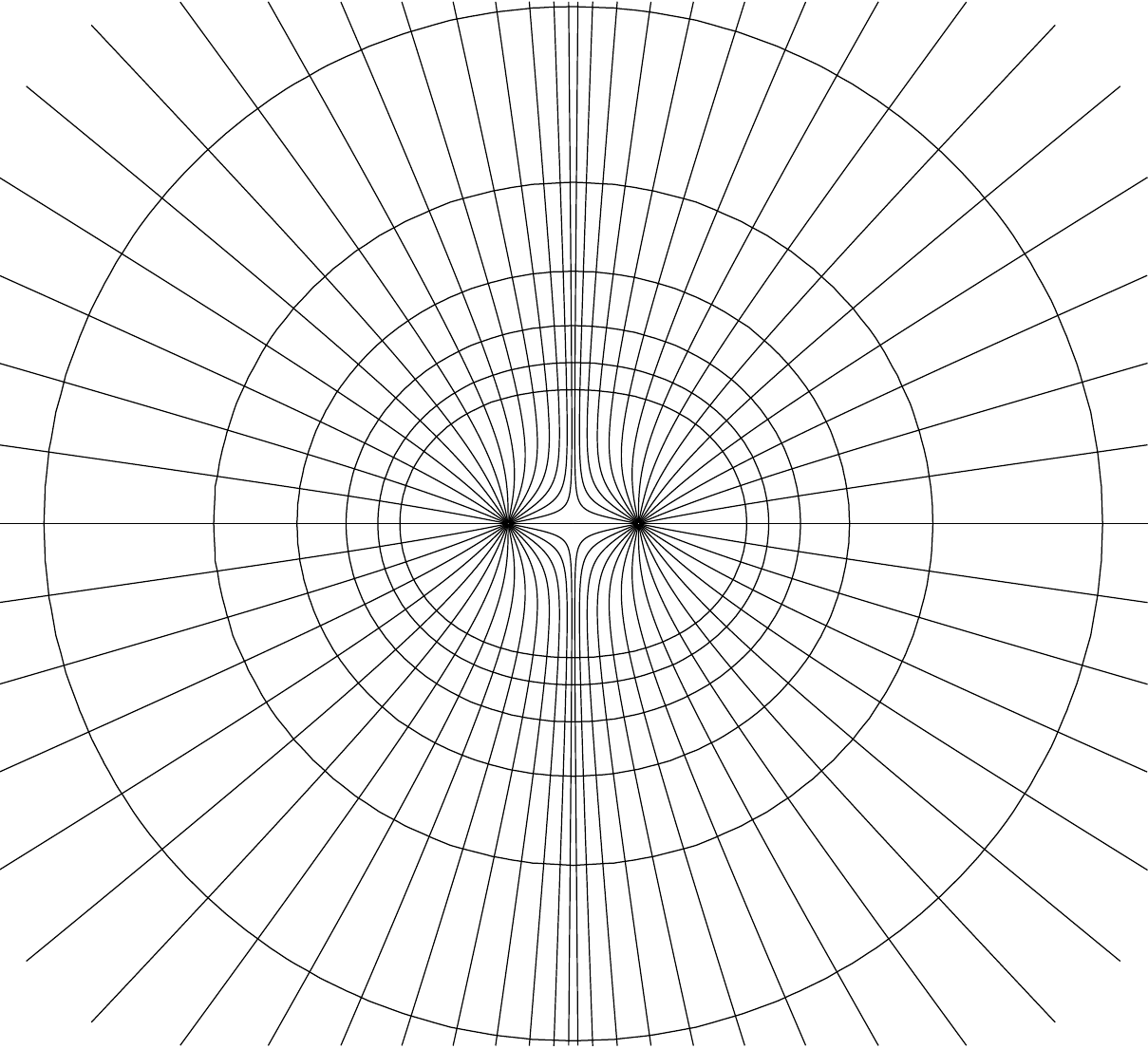}
\caption{\label{fig:multipole} The electric field and potential lines for two point charges of the  same sign. The right figure is given by zooming out the left figure.}
\end{figure}

A second familiar example is the multipole expansion from electrostatics,
\begin{align} \label{1.3}
V(\mathbf{r}) &= \frac{1}{r} \sum_{l,m} b_{lm} \frac{1}{r^l}  Y_{lm}(\Omega)\,,
\end{align}
which will illustrate a number of useful points.  A  sample charge configuration with its electric field and equipotential lines is shown in Fig.~\ref{fig:multipole}. 

While the discussion below is in the context of the electrostatics example, it holds equally well for other EFT examples. If the typical spacing between charges in Fig.~\ref{fig:multipole} is of order $a$, eqn~(\ref{1.3}) can be written as
\begin{align}\label{1.3a}
V(\mathbf{r}) &=  \frac{1}{r} \sum_{l,m} c_{lm}  \left( \frac{a}{r} \right)^l Y_{lm}(\Omega) \,,
& b_{lm} &\equiv c_{lm}a^l\,,
\end{align}
using dimensionless coefficients $c_{lm}$.
\begin{itemize}
\item As written, eqn~(\ref{1.3a}) has two scales $r$ and $a$, with $r \gg a$. $r$ is the long-distance, or infrared (IR) scale, and $a$ is the short-distance or ultraviolet (UV) scale. The small expansion parameter is the ratio of the IR and UV scales $\delta = a/r$. The expansion is  useful if the two scales are widely separated, so that $\delta \ll 1$. We often work in momentum space, so that the IR scale is $p \sim 1/r$,  the UV scale is $\Lambda \sim 1/a$, and  $\delta = p/\Lambda$.
\item A far away (low-energy) observer measures the potential $V(r)$ as a function of $r$ and $\Omega=(\theta,\phi)$. By Fourier analysis, the observer can determine the short distance coefficients $b_{lm}=c_{lm}a^l \sim c_{lm}/\Lambda^l$. These coefficients are dimensionful, and suppressed by inverse powers of $\Lambda$ as $l$ increases. 
\item More accurate values of the potential are given by including more multipoles. The terms in eqn~(\ref{1.3},\ref{1.3a}) get smaller as $l$ increases. A finite experimental resolution implies that $c_{lm}$ can only be experimentally determined up to a finite maximum value $l_\text{max}$ that depends on the resolution. More accurate experiments probe larger $l_\text{max}$.
\item One can factor out powers of $a$, as shown in eqn~(\ref{1.3a}), and use $c_{lm}$ instead of $b_{lm}$. Then $c_{lm}$ are order unity. This is dimensional analysis. There is no precise definition of $a$, and any other choice for $a$ of the same order of magnitude works equally well. $a$ is given from observations by measuring $b_{lm}$ for large values of $r$, and inferring $a$ by letting $b_{lm} = c_{lm} a^l$, and seeing if some choice of $a$ makes all the $c_{lm}$ of order unity.
\item Some $c_{lm}$ can vanish, or be anomalously small due to an (approximate) symmetry of the underlying charge distribution. For example, cubic symmetry implies $c_{lm}=0$ unless $l \equiv 0$ (mod 2) and $m \equiv 0$ (mod 4). Measurements of $b_{lm}$ provide information about the short-distance structure of the charge distribution, and possible underlying symmetries. 
\item More accurate measurements require higher order terms in the $l$ expansion. There are only a finite number, $(l_\text{max}+1)^2$, parameters including all terms up to order $l_\text{max}$.
\item We can use the $l$ expansion without knowing the underlying short-distance scale $a$, as can be seen from the first form eqn~(\ref{1.3}). The parameters $b_{lm}$ are determined from the variation of $V(r)$ w.r.t.\ the IR scale $r$. Using $b_{lm}=c_{lm}a^l $ gives us an estimate of the size of the charge distribution. We can determine the short-distance scale $a$ by accurate measurements at the long-distance scale $r \gg a$, or by less sophisticated measurements at shorter distances $r$ comparable to $a$. 
\end{itemize}

The above analysis also applies to  searches for BSM (beyond Standard Model) physics. Experiments are searching for new interactions at short distances $a \sim 1/\Lambda$, where $\Lambda$ is larger than the electroweak scale $v \sim 246$\,GeV. Two ways of determining the new physics scale are by making more precise measurements at low-energies, as is being done in $B$ physics experiments, or by making measurements at even higher energies, as at the LHC.

Subtleties can arise even in the simple electrostatic problem. 
\begin{figure}
\centering
\includegraphics[height=7cm]{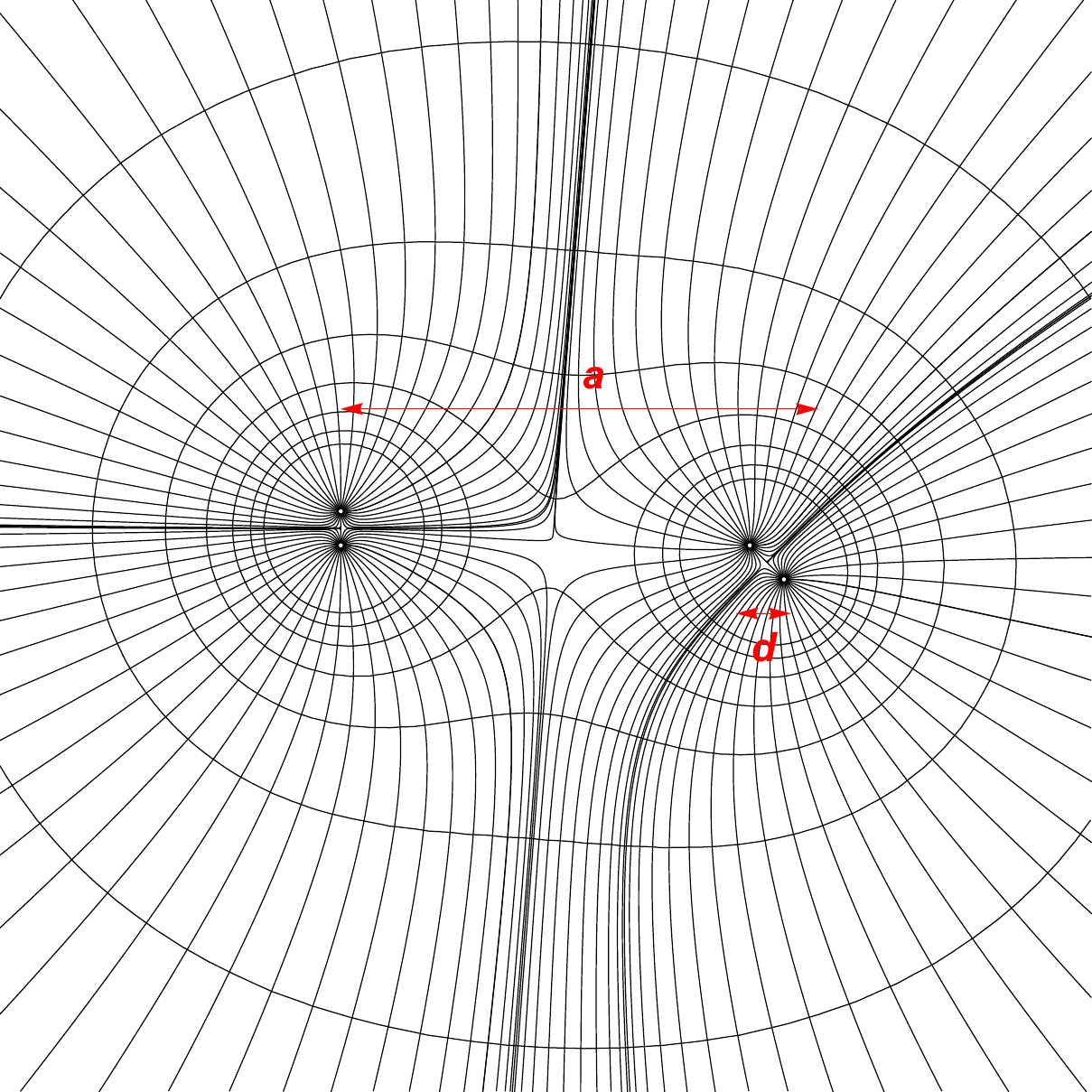}
\caption{\label{fig:multipole2} A charge distribution with two intrinsic scales: $d$, the size of each clump, and $a$, the distance between clumps.}
\end{figure}
Consider the charge configuration shown in Fig.~\ref{fig:multipole2}, which is an example of a multiscale problem. The system has two intrinsic scales, a shorter scale $d$ characterizing the individual charge clumps, and a longer scale $a$ characterizing the separation between  clumps. Measurements at large values of $r$ determine the scale $a$. Very accurate measurements of $c_{lm}$ can determine the shorter distance scale $d$. Discovering $d$ requires noticing patterns in the values of $c_{lm}$. It is much easier to determine $d$ if one knows ahead of time that there is a short distance scale $d$ that must be extracted from the data. $d$ can  be easily determined by making measurements at shorter distances (higher energies) $d \ll r \ll a$, i.e.\ if one is allowed to measure the electrostatic potential between the two clumps of charges.

Multiscale problems are common in EFT applications. The Standard Model EFT (SMEFT) is an EFT used to characterize BSM physics. The theory has a scale $\Lambda$, of order a few TeV, which is the expected scale of BSM physics in the electroweak sector, as well as higher scales  $\Lambda_{\slashed{L}}$ and $\Lambda_{\slashed{B}}$ at which lepton and baryon number are broken. $\chi$PT has the scales $m_\pi\sim 140$\,MeV, $m_K\sim 500$\, MeV and the chiral symmetry breaking scale $\lchi\sim 1$\,GeV. HQET has the scales $m_b$, $m_c$ and $\lqcd$. EFT methods allow us to separate scales in a multi-scale problem, and organize the calculation in a systematic way.

\section{Fermi Theory of Weak Interactions}

The Fermi theory of weak interactions~\cite{Fermi:1933jpa} is an EFT for weak interactions at energies below the $W$ and $Z$ masses. It is a low-energy EFT constructed from the SM. The EFT power counting parameter is $\delta=p/M_W$, where $p$ is of order the momenta of particles in the weak decay. For example, in $\mu$ decay, $p$ is of order the muon mass. In hadronic weak decays, $p$ can be of order the hadron (or quark) masses, or of order $\lqcd$. The theory also has the usual perturbative expansions in  $\alpha_s/(4\pi)$ and $\alpha/(4\pi)$.  Historically, Fermi's theory was used for weak decay calculations even when the scales $M_W$ and $M_Z$ were not known. We will construct the Fermi interaction in Sec.~\ref{sec:fermi}.

\section{HQET/NRQCD}

Heavy quark effective theory (HQET) and non-relativistic QCD (NRQCD~\cite{Caswell:1985ui}) describe the low-energy dynamics of hadrons containing a heavy quark. The theories are applied to hadrons containing $b$ and $c$ quarks. In HQET, the expansion parameter is $\lqcd/m_Q$, where $m_Q=m_b,m_c$ is the mass of the heavy quark. The theory also has an expansion in powers of $\alpha_s(m_Q)/(4\pi)$. The matching from QCD to HQET can be done in perturbation theory, since $\alpha_s(m_Q)/(4\pi)$ is small, $\alpha_s(m_b) \sim 0.22$, $\alpha_s(m_b)/(4\pi) \sim0.02$.  Calculations in HQET contain non-perturbative corrections, which can be included in a systematic way in an expansion in $\lqcd/m_Q$.

NRQCD is similar to HQET, but treats $Q \overline Q$ bound states such as the $\Upsilon$ meson. The heavy quarks move non-relativistically, and the expansion parameter is the velocity $v$ of the heavy quarks, which is of order $v  \sim \alpha_s(m_Q)$.

HQET and NRQCD are covered in Professor T.~Mannel's lectures at this school~\cite{Mannel:fy}.

\section{Chiral Perturbation Theory}

Chiral perturbation theory describes the interactions of pions and nucleons at low momentum transfer. The theory was developed in the 1960's, and the method closest to the modern way of calculating was developed by Weinberg. $\chi$PT describes the low-energy dynamics of QCD. In this example, the full theory is known, but it is not possible to analytically compute the matching onto the EFT, since the matching is non-perturbative. Recent progress has been made in computing the matching numerically~\cite{Aoki:2016frl}. The two theories, QCD and $\chi$PT, are not written in terms of the same fields. The QCD Lagrangian has quark and gluon fields, whereas $\chi$PT has meson and baryon fields. The parameters of the chiral Lagrangian are usually fit to experiment.

Note that computations in $\chi$PT, such as Weinberg's calculation of $\pi\pi$ scattering, were done using $\chi$PT \emph{before} QCD was even invented. This example shows rather clearly that one can compute in an EFT without knowing the UV origin of the EFT. 

The expansion parameter of $\chi$PT is $p/\lchi$, where $\lchi \sim 1$\,GeV is referred to as the scale of chiral symmetry breaking. $\chi$PT can be applied to baryons even though baryon masses are comparable to $\lchi$. The reason is that baryon number is conserved, and so baryons can be treated as heavy particles analogous to heavy quarks in HQET as long as the momentum transfer is smaller than $\lchi$. There is an interesting relation between the large-$N_c$ expansion of QCD and baryon chiral perturbation theory~\cite{Jenkins:1998wy,Manohar:1998xv}.

$\chi$PT is covered in Professor A.~Pich's lectures at this school~\cite{Pich:cs}.

\section{SCET}

Soft-collinear effective theory (SCET~\cite{Bauer:2000ew,Bauer:2000yr,Bauer:2001ct,Bauer:2001yt}) describes energetic QCD processes where the final states have small invariant mass compared to the center-of-mass energy of the collision, such as in jet production in high-energy $pp$ collisions. The underlying theory is once again QCD. The expansion parameters of SCET are $\lqcd/Q$, $M_J/Q$ and $\alpha_s(Q)/(4\pi)$, where $Q$ is the center-of-mass energy of the hard-scattering process, and $M_J$ is the invariant mass of the jet. SCET was originally developed for the decay of $B$ mesons to light particles, such as $B \to X_s \gamma$ and $B \to \pi \pi$.

SCET is covered in T.~Becher's lectures at this school~\cite{Becher:zt}.

\section{SMEFT}

SMEFT is the EFT constructed out of SM fields, and is used to analyze deviations from the SM, and search for BSM physics. The higher dimension operators in SMEFT are generated at a new physics scale $\Lambda$, which is not known. Nevertheless, one can still perform systematic computations in SMEFT, as should be clear from the multipole expansion example in Sec.~\ref{sec:mult}. SMEFT is discussed in Sec.~\ref{sec:smeft}. 

\section{Reasons for using an EFT}

There are many reasons for using an EFT, which are summarized here. The points are treated in more detail later in these lectures, and also in the other lectures at this school.

\begin{itemize}
\item {\bf Every theory is an effective field theory.} For example, QED, the first relativistic quantum field theory developed, is an approximation to the SM. It is an EFT obtained from the SM by integrating out all particles other than the photon and electron.
\item {\bf EFTs simplify the computation by dealing with only one scale at a time:} For example the $B$ meson decay rate depends on $M_W$, $m_b$ and $\lqcd$, and one can get horribly complicated functions of the ratios of these scales. In an EFT,  we deal with only one scale at a time, so there are no functions, only constants. This is done by using a series of theories, $\text{SM} \to \text{Fermi Theory} \to \text{HQET}$.
\item {\bf EFTs make symmetries manifest:} QCD has a spontaneously broken chiral symmetry, which is manifest in the chiral Lagrangian. Heavy quarks have an Isgur-Wise~\cite{Isgur:1989vq} spin-flavor symmetry under which $b \uparrow,\ b \downarrow, c \uparrow, c \downarrow$ transform as a four-dimensional representation of $SU(4)$. This symmetry is manifest in the HQET Lagrangian~\cite{Georgi:1990um}, which makes it easy to derive the consequences of this symmetry.  Symmetries such as spin-flavor symmetry are only true for certain limits of QCD, and so are hidden in the QCD Lagrangian. 
\item  {\bf EFTs include only the relevant interactions:} EFTs have an explicit power counting estimate for the size of various interactions. Thus one can only include the relevant terms in the EFT Lagrangian needed to obtain the required accuracy of the final result.
\item {\bf Sum logs of the ratios of scales:} This allows one to use renormalization-group improved perturbation theory, which is more accurate, and has a larger range of validity than fixed order perturbation theory. For example, the semileptonic $B$ decay rate depends on powers
\begin{align}
\label{1.20}
\left( \frac{\alpha_s}{4\pi} \ln \frac{M_W}{m_b} \right)^n\,.
\end{align}
Even though $\alpha_s/(4\pi)$ is small, it is multiplied by a large log, and fixed order perturbation theory  can break down. RG improved perturbation theory sums the corrections in eqn~(\ref{1.20}), so that the perturbation expansion is in powers of $\alpha_s/(4\pi)$, without a multiplicative log. The resummation of logs is even more important in SCET, where there are two powers of a log for each $\alpha_s$, the so-called Sudakov double logarithms.

The leading-log corrections are not small. For example, the strong interaction coupling changes by a factor of two between $M_Z$ and $m_b$,
\begin{align*}
\alpha_s(M_Z) &\sim 0.118, &
\alpha_s(m_b) &\sim 0.22.
\end{align*}

While summing logs might seem like a technical point, it is one of the main reasons why EFTs (or equivalent methods such as factorization formul\ae\ in QCD) are used in practice. In QCD collider processes, resummed cross sections can be dramatically different from fixed order ones.
\item {\bf Sum IR logs by converting them to UV logs:} This is related to the previous point. UV logs are summed by the renormalization group equations, since they are related to anomalous dimensions and renormalization counterterms. There is no such summation method for IR logs. However, IR logs in the full theory can be converted to UV logs in the EFT, which can then be summed by integrating the renormalization group equations in the EFT (see Sec.~\ref{sec:5.8}). QCD leads to a number of different effective theories, HQET, NRQCD, SCET and $\chi$PT. Each one is designed to treat a particular IR regime, and sum the corresponding IR logs.
\item {\bf Non-perturbative effects can be included in a systematic way:} In HQET, powers of $\lqcd$   are included through the matrix elements of higher dimension operators, giving the $(\lqcd/m_b)^n$ expansion.
\item {\bf Efficient method to characterize new physics:} EFTs provide an efficient way to characterize new physics, in terms of coefficients of higher dimension operators. This method includes the constraints of locality, gauge invariance and Lorentz invariance. All new physics theories can be treated in a unified framework using a few operator coefficients.

\end{itemize}

\chapter{The EFT Lagrangian}

\section{Degrees of Freedom}

To write down an EFT Lagrangian, we first need to determine the dynamical degrees of freedom, and thus the field content of the Lagrangian. In cases where the EFT is a weakly coupled low-energy version of a UV theory, this is simple---just retain the light fields. However, in many cases, identifying the degrees of freedom in an EFT can be non-trivial.

NRQCD describes $Q \bar Q$ bound states, and is an EFT which follows from QCD. One formulation of NRQCD has multiple gluon modes, soft and ultrasoft gluons, which describe different momentum regions contributing to the $Q \bar Q$ interaction. SCET describes the interactions of energetic particles, and is applicable to processes such as jet production by $q \bar q \to q \bar q$ interactions. It has collinear gluon fields for each energetic particle direction, as well as ultrasoft gluon fields.

A famous example which shows that there is no unique ``correct'' choice of fields to use in an interacting quantum field theory is the sine-Gordon -- Thirring model duality in $1+1$ dimensions~\cite{Coleman:1974bu}. The sine-Gordon model is a bosonic theory of a real scalar field with Lagrangian
\begin{align}
\LL &= \frac12 \partial_\mu \phi\, \partial^\mu \phi + \frac{\alpha}{\beta^2} \cos \beta\phi,
\end{align}
and the Thirring model is a fermionic theory of a Dirac fermion with Lagrangian
\begin{align}
\LL &= \bar \psi \left( i \slashed{\partial} - m \right) \psi -\frac12
g \left(\bar \psi \gamma^\mu \psi \right)^2.
\end{align}
Coleman showed that the two theories were \emph{identical};  they map into each other with the couplings related by
\begin{align}\label{13.3}
\frac{\beta^2 }{ 4 \pi} &= \frac{1 }{ 1 + g/\pi}.
\end{align}
The fermion in the Thirring model is the soliton of the sine-Gordon model, and the boson of the sine-Gordon model is a fermion-antifermion bound state of the Thirring model. The duality exchanges strongly and weakly coupled theories. This example also shows that one cannot distinguish between elementary and composite fields in an interacting QFT.

\section{Renormalization}

A quick summary of renormalization in QCD is presented here, to define the notation and  procedure we will use for EFTs.  A detailed discussion is given in Neubert's lectures~\cite{Neubert:dp}.

QCD is a quantum field theory with Lagrangian
\begin{align}
\LL =-\frac14 F^A_{\mu \nu} F^{A\mu \nu} + \sum_{r=1}^{N_F} \left[ \overline \psi_r i \slashed{D} \psi_r - m_r \overline \psi_r \psi_r \right]
+ \frac{\theta g^2}{32 \pi^2} F^A_{\mu \nu} \widetilde F^{A\mu \nu}\,,
\label{3}
\end{align}
where $N_F$ is the number of flavors. The covariant derivative is $D_\mu = \partial_\mu + i g A_\mu$, and the $SU(3)$ gauge field is a matrix $A_\mu=T^A A^A_\mu$, where the generators are normalized to $\text{Tr}\, T^A T^B=\delta^{AB}/2$. Experimental limits on the neutron electric dipole moment give $\theta \lesssim 10^{-10}$, and we will neglect it here.

The basic observables in a QFT are $S$-matrix elements---on-shell scattering amplitudes for particles with physical polarizations. Green functions of $\psi$ and $A_\mu$, which are the correlation functions of products of fields, are gauge dependent and not experimental observables. The QCD Lagrangian eqn~(\ref{3}) is written in terms of fields, but \emph{fields are not particles}. The relation between $S$-matrix elements of particles and Green functions for fields is through the LSZ reduction formula~\cite{Lehmann:1954rq} explained in Sec.~\ref{sec:LSZ}. One can use \emph{any} field $\phi(x)$ to compute the $S$-matrix element involving a particle state $\ket{p}$ as long as
\begin{align}
\braket{p | \phi(x) | 0} \not=0\,,
\label{4}
\end{align}
i.e.\ the field can create a one-particle state from the vacuum.

Radiative corrections in QCD are infinite, and we need a regularization and renormalization scheme to compute finite $S$-matrix elements. The regularization and renormalization procedure is part of the \emph{definition} of the theory. The standard method used in modern field theory computations is to use dimensional regularization and the \MSbar\ subtraction scheme. We will use dimensional regularization in $d=4-2\epsilon$ dimensions. A brief summary of the procedure is given here.

The QCD Lagrangian for a single flavor in the $CP$-conserving limit (so that the $\theta$ term is omitted) that gives finite $S$-matrix elements is
\begin{subequations}
\begin{align}
\LL &=-\frac14 F^A_{0\mu \nu} F_0^{A\mu \nu} + \overline \psi_0 i (\slashed{\partial} + i g_0 \slashed{A}_0) \psi_0 - m_0 \overline \psi_0 \psi_0 
\label{5a} \\
&=-\frac14 Z_A F^A_{\mu \nu} F^{A\mu \nu} + Z_\psi \overline \psi i (\slashed{\partial} + i g \mu^\epsilon Z_g Z_A^{1/2}\slashed{A}) \psi - m Z_m Z_\psi \overline \psi \psi 
\label{5}
\end{align}
\end{subequations}
where $\psi_0$, $A_0$, $g_0$ and $m_0$ are the bare fields and parameters, which are related to the renormalized fields and parameters
$\psi$, $A$, $g$ and $m$ by
\begin{align}
\psi_0 &= Z_\psi^{1/2} \psi, &
A_{0\mu} &= Z_A^{1/2} A_\mu, &
g_0 &= Z_g g \mu^\epsilon, &
m_0 &= Z_m m.
\label{6}
\end{align}
The renormalization factors $Z_a$ have an expansion in inverse powers of $\epsilon$,
\begin{align}
Z_a &= 1 + \sum_{k=1}^\infty \frac{Z^{(k)}_a}{\epsilon^k}, & a&=\psi,A,g,m,
\label{7}
\end{align}
with coefficients which have an expansion in powers of $\alpha_s=g^2/(4\pi)$,
\begin{align}
Z^{(k)}_a &=  \sum_{r=1}^\infty Z^{(k,r)}_a \left( \frac{\alpha_s}{4\pi}\right)^r\,.
\label{8}
\end{align}
The renormalized parameters $g$ and $m$ are finite, and depend on $\mu$. The renormalization factors $Z_a$ are chosen to give finite $S$-matrix elements.

Separating out the $1$ from $Z_a$, the Lagrangian eqn~(\ref{5}) can be written as
\begin{align}
\LL
&=-\frac14  F^A_{\mu \nu} F^{A\mu \nu} +  \overline \psi i (\slashed{\partial} + i g \mu^\epsilon \slashed{A}) \psi - m \overline \psi \psi 
+ \text{c.t.}
\label{9}
\end{align}
where $\text{c.t.}$ denotes the renormalization counterterms which are pure poles in $1/\epsilon$,
\begin{align}
\LL_\text{c.t.}
&=-\frac14 \left(Z_A-1\right) F^A_{\mu \nu} F^{A\mu \nu} + \left(Z_\psi-1\right) \overline \psi i \slashed{\partial}  \psi 
+\left(Z_\psi Z_g Z_A^{1/2} -1\right)   \overline \psi i g\mu^\epsilon \slashed{A} \psi  \nn
& - \left(Z_\psi Z_m -1\right)  m \overline \psi \psi \,.
\label{10}
\end{align}

The Lagrangian eqn~(\ref{5a}) contains 2 bare parameters, $g_0$ and $m_0$. The Lagrangian eqn~(\ref{5}) contains two renormalized parameters $g(\mu)$, $m(\mu)$ and the renormalization scale $\mu$. As discussed in Neubert's lectures~\cite{Neubert:dp}, the renormalization group equation, which follows from the condition that the theory is $\mu$-independent, implies that there are only two free parameters, for example $g(\mu_0)$ and $m(\mu_0)$ at some chosen reference scale $\mu_0$. The renormalization group equations determine how $m$ and $g$ must vary with $\mu$ to keep the observables the same. We will see later how the freedom to vary $\mu$ allows us to sum logarithms of the ratio of scales. The variation of renormalized parameters with $\mu$ is sometimes referred to as the renormalization group flow.

The bare parameters in the starting Lagrangian eqn~(\ref{5a}) are infinite. The infinities cancel with those in loop graphs, so that $S$-matrix elements computed are finite. Alternatively, one starts with the Lagrangian split up into the renormalized Lagrangian with finite parameters plus counterterms, as in eqn~(\ref{9}). The infinite parts of loop graphs computed from the renormalized Lagrangian are cancelled by the counterterm contributions, to give finite $S$-matrix elements. The two methods are equivalent, and give the usual renormalization procedure in the \MSbar\ scheme. Usually, one computes in perturbation theory in the coupling $g$, and determines the renormalization factors $Z_a$ order by order in $g$ to ensure finiteness of the $S$-matrix.

\begin{exercise}

Compute the mass renormalization factor $Z_m$ in QCD at one loop. Use this to determine the one-loop mass anomalous dimension $\gamma_m$,
\begin{align}
\mu \frac{\rd m}{\rd \mu} &= \gamma_m m,
\end{align}
by differentiating $m_0=Z_m m$, and noting that $m_0$ is $\mu$-independent.

\end{exercise}

\section{Determining the couplings}\label{sec:3.3}

How do we determine the parameters in the Lagrangian?  The bare Lagrangian parameters are infinite, and cannot be measured directly. The renormalized Lagrangian parameters are finite. However, in general, they are scheme dependent, and also not directly measurable.  In QCD, the \MSbar\ quark mass $m_b(\mu)$ is not a measurable quantity. Often, people refer to the quark pole mass $m_b^\text{pole}$ defined by the location of the pole in the quark propagator in perturbation theory.  It is related to the \MSbar\ mass by
\begin{align}
\label{polemass}
m_b^\text{pole} &= m_b(m_b) \left[ 1 + \frac{4 \alpha_s(m_b)}{3 \pi} + \ldots \right]\,.
\end{align}
$m_b^\text{pole}$ is independent of $\mu$, and hence is renormalization-group invariant.
Nevertheless, $m_b^\text{pole}$ is not measurable---quarks are confined, and there is no pole in gauge-invariant correlation functions at $m_b^\text{pole}$. Instead one determines the $B$ meson mass $m_B$ experimentally. The quark mass $m_b(\mu)$ or $m_b^\text{pole}$ is fixed by adjusting it till it reproduces the measured meson mass. To actually do this requires a difficult non-perturbative calculation, since $m_b^\text{pole}$ and $m_B$ differ by order $\lqcd$ effects. In practice, one uses observables which are easier to compute theoretically, such as the electron energy spectrum in inclusive $B$ decays, or the the  $e^+ e^- \to b \overline b$ cross section near threshold, to determine the quark mass. Similarly,  the gauge coupling $g(\mu)$ is not an observable, and must be determined indirectly.

\begin{exercise}

Verify the one-loop relation between the \MSbar\ and pole masses, eqn~(\ref{polemass}).

\end{exercise}

Even in QED, the Lagrangian parameters are not direct observables.  QED has two Lagrangian parameters, and two experimental inputs are used to fix these parameters. One can measure the electron mass $m_e^{\text{obs}}$ (which is the pole mass, since electrons are not confined), and the electrostatic potential at large distances, $-\alpha_\text{QED}/r$. These two measurements fix the values of the Lagrangian parameters $m_e(\mu)$ and $e(\mu)$. All other observables, such as positronium energy levels, the Bhabha scattering cross section, etc.\ are then determined, since they are functions of $m_e(\mu)$ and $e(\mu)$.

The number of Lagrangian parameters $N_\LL$ tells you how many inputs are needed to completely fix the predictions of the theory. In general, one computes a set of observables $\left\{O_i\right\}, i=1,\ldots,N_\LL$ in terms of the Lagrangian parameters. $N_\LL$ observables are used to fix the parameters, and the remaining $N_O-N_\LL$ observables are predictions of the theory:
\begin{align}
\label{1.12}
\underbrace{O_1,\ldots,O_{N_\LL} }_{\text{observables}} \quad \longrightarrow \quad \underbrace{m_i(\mu), g(\mu), \ldots}_{\text{parameters}}  \quad \longrightarrow \quad \underbrace{ O_{N_{\LL+1}},\ldots}_{\text{predictions}} \,.
\end{align}
The Lagrangian plays the role of an intermediary, allowing one to relate observables to each other. The $S$-matrix program of the 1960's avoided any use of the Lagrangian, and related observables directly to each other using analyticity and unitarity.

Given a QFT Lagrangian $\LL$, including a renormalization procedure, you can calculate $S$-matrix elements. No additional outside input is needed, and the calculation is often automated. For example, in QED, it is not necessary to know that the theory is the low-energy limit of the Standard Model (SM), or to consult an oracle to obtain the value of certain loop graphs. All predictions of the theory are encoded in the Lagrangian. A renormalizable theory has only a finite number of terms in the Lagrangian, and hence only a finite number of parameters. One can compute observables to arbitrary accuracy, at least in principle, and obtain parameter-free predictions.

The above discussion applies to EFTs as well, including the last bit about a finite number of parameters, provided that one works to a \emph{finite} accuracy $\delta^n$ in the power counting parameter. As an example, consider the Fermi theory of weak interactions, which we discuss in more detail in Sec.~\ref{sec:fermi}. The EFT Lagrangian in the lepton sector is
\begin{align}
\LL &= \LL_\text{QED} - \frac{4 G_F}{\sqrt 2} (\overline e \gamma^\mu P_L  \nu_e)(  \overline \nu_\mu \gamma_\mu P_L  \mu) + \ldots\,,
\label{10a}
\end{align}
where $P_L=(1-\gamma_5)/2$, and $G_F=1.166 \times 10^{-5}\,\text{GeV}^{-2}$ has dimensions of inverse mass-squared. As in QCD or QED, one can calculate  $\mu$-decay directly using eqn~(\ref{10a}) without using any external input, such as knowing eqn~(\ref{10a}) was obtained from the low-energy limit of the SM. The theory is renormalized as in eqn~(\ref{5a},\ref{5},\ref{6}). The main difference is that the Lagrangian eqn~(\ref{10a}) has an infinite series of operators (only one is shown explicitly), with coefficients which absorb the divergences of loop graphs. The expansion parameter of the theory is $\delta=G_F  p^2$. 
To a fixed order in $\delta$, the theory is just like a regular QFT. However, if one wants to work to higher accuracy, more operators must be included in $\LL$, so that there are more parameters. If one insists on infinitely precise results, then there are an infinite number of terms and an infinite number of parameters. Thus an EFT is just like a regular QFT, supplemented by a power counting argument that tells you what terms to retain to a given order in $\delta$. The number of experimental inputs used to fix the Lagrangian parameters increases with the order in $\delta$.
In the $\mu$-decay example, $G_F$ can be fixed by the muon lifetime. The Fermi theory then gives a parameter-free prediction for the decay distributions, such as the electron energy spectrum, electron polarization, etc.

The parameters of the EFT Lagrangian eqn~(\ref{10a}) can be obtained from low-energy data. The divergence structure of the EFT is \emph{different} from that of the full theory, of which the EFT is a low-energy limit. This is not a minor technicality, but a fundamental difference. It is crucial in many practical applications, where IR logs can be summed by transitioning to an EFT.

In cases where the EFT is the low-energy limit of a weakly interacting full theory, e.g. the Fermi theory as the low-energy limit of the SM, one constructs the EFT Lagrangian to reproduce the same $S$-matrix as the original theory, a procedure known as matching.  The full and effective theory are equivalent; they are different ways of computing the same observables. The change in renormalization properties means that fields in the EFT are not the same as fields in the full theory, even though they are often denoted by the same symbol. Thus the electron field $e$ in eqn~(\ref{10a}) is not the same as the field $e$ in the SM Lagrangian. The two agree at tree-level, but at higher orders, one has to explicitly compute the relation between the two. A given high-energy theory can lead to multiple EFTs, depending on the physical setting. For example, $\chi$PT, HQET, NRQCD and SCET are all EFTs based on QCD.

\section{Inputs}\label{sec:inputs}

I said at the start of the lectures that it was ``obvious'' that low-energy dynamics was insensitive to the short-distance properties of the theory. This is true provided the input parameters are obtained from low-energy processes computed using the EFT. QED plus QCD with five flavors of quarks is the low-energy theory of the SM below the electroweak scale. The input couplings can be determined from measurements below 100\,GeV.

Now suppose, instead, that the input couplings are fixed at high-energies, and their low-energy values are determined by computation. Given the QED coupling $\alpha(\mu_H)$ at a scale $\mu_H > m_t$ above the top-quark mass, for example, we can determine the low-energy value $\alpha(\mu_L)$ for $\mu_L$ smaller than $m_t$. In this case, $\alpha(\mu_L)$ is sensitive to high energy parameters, such as heavy masses including the top-quark mass. For example, if we vary the top-quark mass, then
\begin{align}
\label{3.14}
m_{t}\, \frac{{\rm d}}{ {\rm d} m_{t}} \left[ \frac{1 }{ {\alpha}(\mu_L)}\right] = - \frac{1 }{ 3 \pi}\,,
\end{align}
where $\mu_L < m_t$, and we have kept $\alpha(\mu_H)$ for $\mu_H > m_t$ fixed. Similarly, if we keep the strong coupling $\alpha_s(\mu_H)$ fixed for $\mu_H > m_t$, then the proton mass is sensitive to $m_t$,
\begin{align}
\label{3.15}
m_p \propto m_t^{2/27}.
\end{align}

The bridge-builder mentioned in the introduction would have a hard time designing a bridge if the density of steel depended on the top-quark mass via eqn~(\ref{3.15}). Luckily, knowing about the existence of top-quarks is not necessary. The density of steel is an experimental observable, and its measured value is used in the design. The density is measured in lab experiments at low-energies, on length scales of order meters, not in LHC collisions. How the density depends on $m_t$ or possible BSM physics is irrelevant. There is no sensitivity to high-scale physics if the inputs to low-energy calculations are from low-energy measurements. The short distance UV parameters are not ``more fundamental'' than the long-distance ones. They are just parameters. For example, in QED, is $\alpha(\mu > m_t)$ more fundamental than $\alpha_{\text{QED}}=1/(137.036)$ given by measuring the Coulomb potential as $r \to \infty$? It is $\alpha_{\text{QED}}$, for example, which is measured in quantum Hall effect experiments.

Combining low-energy EFTs with high-energy inputs mixes different scales, and leads to problems. The natural parameters of the EFT are those measured at low energies. Using high-energy inputs forces the EFT to use inputs that do not fit naturally into the framework of the theory. We will return to this point in Sec.~\ref{sec:naturalness}.

Symmetry restrictions from the high-energy theory feed down to the low-energy theory. QCD (with $\theta=0$) preserves $C$, $P$ and $CP$, and hence so does $\chi$PT. Causality in QFT leads to the spin-statistics theorem. This is a restriction which is imposed in quantum mechanics, and follows because the quantum theory is the non-relativistic limit of a QFT.

\begin{exercise}
Verify eqn~(\ref{3.14}) and eqn~(\ref{3.15}).
\end{exercise}

\section{Integrating Out Degrees of Freedom}

The old-fashioned view is that EFTs are given by integrating out high momentum modes of the original theory, and thinning out degrees of freedom as one evolves from the UV to the IR~\cite{Kadanoff:1966wm,Wilson:1971dc,Wilson:1973jj}. That is not what happens in the EFTs discussed in this school, which are used to describe experimentally observable phenomena, and it is not the correct interpretation of renormalization-group evolution in these theories.

In SCET, there are different collinear sectors of the theory labelled by null vectors $n_i=(1,\mathbf{n}_i)$,  $\mathbf{n}_i^2=1$. Each collinear sector of SCET is the same as the full QCD Lagrangian, so SCET has multiple copies of the original QCD theory, as well as ultrasoft modes that couple the various collinear sectors. The number of degrees of freedom in SCET is much larger than in the original QCD theory.
In $\chi$PT, the EFT is written in terms of meson and baryon fields, whereas QCD is given in terms of quarks and gluons. Mesons and baryons are created by composite operators of quarks and gluons, but there is no sense in which the EFT is given by integrating out short-distance quarks and gluons.

The renormalization group equations are a consequence of the $\mu$ independence of the theory. Thus varying $\mu$ changes nothing measurable;  $S$-matrix elements are $\mu$ independent. Nothing is being integrated out as $\mu$ is varied, and the theory at different values of $\mu$ is the same. The degrees of freedom do not change with $\mu$. The main purpose of the renormalization group equations is to sum logs of ratios of scales, as we will see in Sec.~\ref{sec:rge}.

It is much better to think of EFTs in terms of the physical problem you are trying to solve, rather than as the limit of some other theory. The EFT is then constructed out of the dynamical degrees of freedom (fields) that are relevant for the problem. The focus should be on what you want, not on what you don't want.

\chapter{Power Counting}

The EFT functional integral is
\begin{align}
\int \mathcal{D}\phi \ e^{i S}\,,
\end{align}
so that the action $S$ is dimensionless. The EFT action is the integral of a local Lagrangian density
\begin{align}
S &= \int \rd^\d x\ \LL(x)\,,
\end{align}
(neglecting topological terms), so that in $\d$ spacetime dimensions, the Lagrangian density has mass dimension
$\d$,
\begin{align}
\left[ \LL(x) \right] &= \d\,,
\end{align}
and is the sum
\begin{align}
\LL(x) &= \sum_i c_i \, O_i(x)\,,
\end{align}
of local, gauge invariant, and Lorentz invariant operators $O_i$ with coefficients $c_i$. The operator dimension will be denoted by $\opdim$, and its coefficient has dimension $\d-\opdim$.

The fermion and scalar kinetic terms are
\begin{align}
S &= \int \rd^\d x\ \bar \psi \ i \slashed{\partial}\ \psi, & S &= \int \rd^\d x\ \frac 12 \partial_\mu \phi\, \partial^\mu \phi,
\end{align}
so that dimensions of fermion and scalar fields are
\begin{align}
\left[ \psi \right] &= \frac12 (\d-1), &
\left[ \phi \right] &= \frac12 (\d-2).
\end{align}
The two terms in the covariant derivative $D_\mu=\partial_\mu + i g A_\mu$ have the same dimension, so
\begin{align}
\label{2.7}
\left[D_\mu \right] &=1, &
\left[ gA_\mu \right] &= 1 \,.
\end{align}
The gauge field strength $X_{\mu\nu}=\partial_\mu A_\nu - \partial_\nu A_\mu + \ldots$ has a single derivative of $A_\mu$, so $A_\mu$ has the same dimension as a scalar field. This determines, using eqn~(\ref{2.7}), the dimension of the gauge coupling $g$,
\begin{align}
\left[A_\mu \right] &= \frac12 (\d-2), &
\left[g\right]=\frac12(4-\d)\,.
\end{align}
In $\d=4$ spacetime dimensions,
\begin{align}
\left[ \phi \right] &= 1, &
\left[ \psi \right] &= 3/2, &
\left[ A_\mu \right] &= 1, &
\left[D \right] &=1, & 
[g] &=0 \,.
\end{align}
In $\d=4-2\epsilon$ dimensions, $\left[g\right]=\epsilon$, so in dimensional regularization, one usually uses a dimensionless coupling $g$ and writes the coupling in the Lagrangian as $g \mu^{\epsilon}$, as in eqn~(\ref{6}).

The only gauge and Lorentz invariant operators with dimension $\opdim \le \d=4$ that can occur in the Lagrangian are
\begin{align}
\label{2.10all}
\opdim=0:\quad& 1 \nn
\opdim=1:\quad & \phi \nn
\opdim=2:\quad& \phi^2 \nn
\opdim=3:\quad & \phi^3, \bar \psi \psi \nn
\opdim=4:\quad & \phi^4,\ \phi\, \bar \psi \psi,\ D_\mu \phi\, D^\mu \phi,\ \bar \psi \ i \slashed{D}\ \psi,\ X_{\mu \nu}^2\,.
\end{align}
Other operators, such as $D^2\phi$ vanish upon integration over $\rd^\d x$, or are related to operators already included eqn~(\ref{2.10all}) by integration by parts.  In $\d=4$ spacetime dimensions, fermion fields can be split into left-chiral and right-chiral fields which transform as irreducible representations of the Lorentz group. The projection operators are $P_L=(1-\gamma_5)/2$ and $P_R=(1+\gamma_5)/2$. Left-chiral fermions will be denoted $\psi_L = P_L \psi$, etc.

Renormalizable interactions have coefficients with mass dimension $\ge 0$, and eqn~(\ref{2.10all}) lists the allowed renormalizable interactions in four spacetime dimensions. The distinction between renormalizable and non-renormalizable operators should be clear after Sec.~\ref{sec:4.2}.

In $\d=2$ spacetime dimensions
\begin{align}
\left[ \phi \right] &= 0, &
\left[ \psi \right] &= 1/2, &
\left[ A_\mu \right] &= 0, & \left[D \right] &=1, & [g]&=1,
\end{align}
so an arbitrary potential $V(\phi)$ is renormalizable, as is the  $\left( \bar \psi \psi \right)^2$ interaction, so that the sine-Gordon and Thirring models are renormalizable. In $\d=6$ spacetime dimensions,
\begin{align}
\left[ \phi \right] &= 2, &
\left[ \psi \right] &= 5/2, &
\left[ A_\mu \right] &= 2, & \left[D \right] &=1, & [g]&=-1.
\end{align}
The only allowed renormalizable interaction in six dimensions is $\phi^3$. There are no renormalizable interactions above six dimensions.\footnote{There are exceptions to this in strongly coupled theories where operators can develop large anomalous dimensions.}

\begin{exercisebn}

In $\d=4$ spacetime dimensions, work out the field content of Lorentz-invariant operators with dimension $\opdim$ for $\opdim=1,\ldots,6$. At this point, do not try and work out which operators are independent, just the possible structure of allowed operators. Use the notation $\phi$ for a scalar, $\psi$ for a fermion, $X_{\mu\nu}$ for a field strength, and $D$ for a derivative. For example, an operator of type $\phi^2 D$ such as $\phi D_\mu \phi$ is not allowed because it is not Lorentz-invariant. An operator of type $\phi^2 D^2$ could be either $D_\mu \phi D^\mu \phi$ or $\phi D^2 \phi$, so a $\phi^2 D^2$ operator is allowed, and we will worry later about how many independent $\phi^2 D^2$ operators can be constructed.

\end{exercisebn}

\begin{exercisenb}

For $\d=2,3,4,5,6$ dimensions, work out the field content of operators with dimension $\opdim \le \d$, i.e. the ``renormalizable'' operators.

\end{exercisenb}

\section{EFT Expansion}

The EFT Lagrangian follows the same rules as the previous section, and has an expansion in powers of the operator dimension
\begin{align}
\label{2.17}
\LL_\text{EFT} = \sum_{\opdim \ge 0 ,i} \frac{ c_i^{(\opdim)} O_i^{(\opdim)} }{ \Lambda^{\opdim-d}}  =
\sum_{\opdim \ge 0} \frac{ \LL_\opdim}{ \Lambda^{\opdim-d}}
\end{align}
where $O_i^{(\opdim)}$ are the allowed operators of dimension $\opdim$. All operators of dimension $\opdim$ are combined into the dimension $\opdim$ Lagrangian $\LL_\opdim$. The main difference from the previous discussion is that one does not stop at $\opdim=\d$, but includes operators of arbitrarily high dimension. A scale $\Lambda$ has been introduced so that the coefficients $c_i^{(\opdim)} $ are dimensionless. $\Lambda$ is the short-distance scale at which new physics occurs, analogous to $1/a$ in the multipole expansion example in Sec.~\ref{sec:mult}. As in the multipole example, what is relevant for theoretical calculations and experimental measurements is the product $c_\opdim \Lambda^{\d-\opdim}$, not $c_\opdim$ and $\Lambda^{\d-\opdim}$ separately. $\Lambda$ is a convenient device that makes it clear how to organize the EFT expansion.

In $\d=4$,
\begin{align}
\LL_\text{EFT}  = \LL_{\opdim \le 4} + \frac{\LL_5 }{ \Lambda} + \frac {\LL_6 }{ \Lambda^2} + \ldots
\end{align}
$\LL_\text{EFT}$ is given by an infinite series of terms of increasing operator dimension. An important point is that the $\LL_\text{EFT}$ has to be treated as an expansion in powers of $1/\Lambda$. If you try and sum terms to all orders, you violate the EFT power counting rules, and the EFT breaks down.

\section{Power Counting  and Renormalizability}\label{sec:4.2}

Consider a scattering amplitude $\mathscr{A}$ in $\d$ dimensions, normalized to have mass dimension zero. If one works at some typical momentum scale $p$, then a single insertion of an operator of dimension $\opdim$ in the scattering graph gives a contribution to the amplitude of order
\begin{align}
\mathscr{A} \sim \left(  \frac{p}{\Lambda} \right)^{\opdim-\d}
\end{align}
by dimensional analysis. The operator has a coefficient of mass dimension $1/\Lambda^{\opdim-\d}$ from eqn~(\ref{2.17}), and the remaining dimensions are produced by kinematic factors such as external momenta to make the overall amplitude dimensionless. An insertion of a set of higher dimension operators in a tree graph leads to an amplitude
\begin{align}
\label{2.20a}
\mathscr{A} &\sim \left(  \frac{p}{\Lambda} \right)^n
\end{align}
with
\begin{align}
\label{2.20}
n&=\sum_i (\opdim_i-\d), &  n&=\sum_i (\opdim_i-4) \ \text{in $\d=4$ dimensions},
\end{align}
where the sum on $i$ is over all the inserted operators. This follows from  dimensional analysis, as for a single insertion. Equation~(\ref{2.20}) is known as the EFT power counting formula. It gives the $(p/\Lambda)$ suppression of a given graph.

The key to understanding EFTs is to understand why eqn~(\ref{2.20}) holds for \emph{any} graph, not just tree graphs. The technical difficulty for loop graphs is that the loop momentum $k$ is integrated over all values of $k$, $-\infty \le k \le \infty$, where the EFT expansion in powers of $k/\Lambda$ breaks down.  Nevertheless, eqn~(\ref{2.20}) still holds. The validity of eqn~(\ref{2.20}) for any graph is explained in Sec.~\ref{sec:5.3}.

The first example of a power counting formula in an EFT was Weinberg's power counting formula for $\chi$PT. This is covered in Pich's lectures, and is closely related to eqn~(\ref{2.20}). Weinberg counted powers of $p$ in the numerator, whereas we have counted powers of $\Lambda$ in the denominator. The two are obviously related.

The power counting formula eqn~(\ref{2.20}) tells us how to organize the calculation. If we want to compute $\mathscr{A}$ to leading order, we only use $\LL_{\opdim \le \d}$, i.e.\ the renormalizable Lagrangian. In $\d=4$ dimensions, $p/\Lambda$ corrections are given by graphs with a single insertion of $\LL_5$; $(p/\Lambda)^2$ corrections are given by graphs with a single insertion of $\LL_6$, or two insertions of $\LL_5$, and so on. As mentioned earlier, we do not need to assign a numerical value to $\Lambda$ to do a systematic calculation. All we are using is eqn~(\ref{2.20}) for a fixed power $n$.

We can now understand the difference between renormalizable theories and EFTs. In an EFT, there are higher dimension operators with dimension $\opdim > \d$. Suppose we have a single dimension five operator (using the $\d=4$ example). Graphs with two insertions of this operator produce the same amplitude as a dimension six operator. In general, loop graphs with two insertions of $\LL_5$ are divergent, and we need a counterterm which is an $\LL_6$ operator. Even if we set the coefficients of $\LL_6$ to zero in the renormalized Lagrangian, we still have to add a $\LL_6$ counterterm with a $1/\epsilon$ coefficient. Thus the Lagrangian still has a coefficient $c_6(\mu)$. $c_6(\mu)$ might vanish at one special value of $\mu$, but in general, it evolves with $\mu$ by the renormalization group equations, and so it will be non-zero at a different value of $\mu$. There is nothing special about $c_6=0$ if this condition does not follow from a symmetry. Continuing in this way, we generate the infinite series of terms in eqn~(\ref{2.17}). We can generate operators of arbitrarily high dimension by multiple insertions of operators with $\opdim-\d>0$.

On the other hand, if we start only with  operators in $\LL_{\opdim \le \d}$, we do not generate any new operators, only the ones we have already included in $\LL_{\opdim \le \d}$. The reason is that $\opdim - \d \le 0$ in eqn~(\ref{2.20}) so we only generate operators with $\opdim \le \d$. Divergences in a QFT are absorbed by local operators, which have $\opdim \ge 0$. Thus new operators generated by loops have $0 \le \opdim \le \d$, and have already been included in $\LL$. We do not need to add counterterms with negative dimension operators, such as $1/\phi^2(x)$, since there are no divergences of this type. In general, renormalizable terms are those with $0\le \opdim \le d$, i.e.\ the contribution to $n$ in eqn~(\ref{2.20}) is non-positive.

Renormalizable theories are a special case of EFTs, where we formally take the limit $\Lambda \to \infty$. Then all terms in $\LL$ have dimension $\opdim \le \d$. Scattering amplitudes can be computed to arbitrary accuracy, as there are no $p/\Lambda$ corrections. Theories with operators of dimensions $\opdim > \d$ are referred to as non-renormalizable theories, because an infinite number of higher dimension operators are needed to renormalize the theory. We have seen, however, that as long one is interested in corrections with some maximum value of $n$ in eqn~(\ref{2.20}), there are only a finite number of operators that contribute, and non-renormalizable theories (i.e. EFTs) are just as good as renormalizable ones.

\section{Photon-Photon Scattering}
\begin{figure}
\begin{center}
\includegraphics[width=3cm]{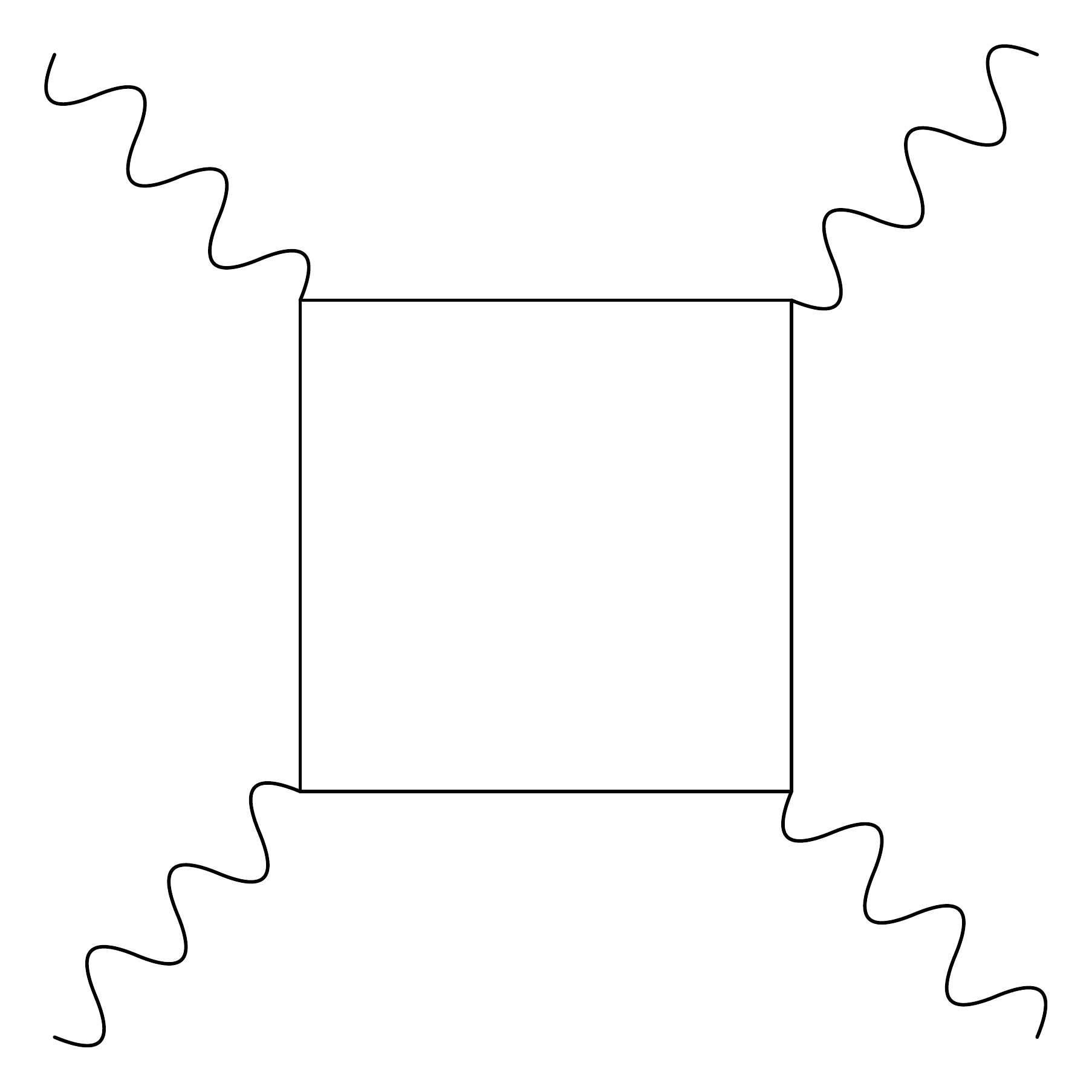}\hspace{2cm}
\raise0.5cm\hbox{\includegraphics[width=2cm]{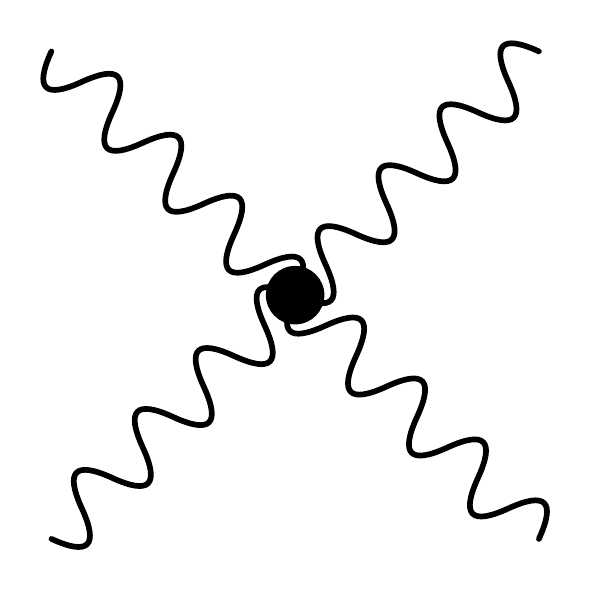}}
\caption{\label{fig:gamgam} The left figure is the QED contribution to the $\gamma\gamma$ scattering amplitude from an electron loop. The right figure is the low-energy limit of the QED amplitude treated as a local $F_{\mu \nu}^4$ operator in the Euler-Heisenberg Lagrangian.}
\end{center}
\end{figure}
We now illustrate the use of the EFT power counting formula eqn~(\ref{2.20}) with some simple examples, which show the power of eqn~(\ref{2.20}) when combined with constraints from gauge invariance and Lorentz invariance.

Consider $\gamma\gamma$ scattering at energies much lower than the electron mass, $E \ll m_e$. At these low energies, the only dynamical degrees of freedom in the EFT are photons. Classical electromagnetism without charged particles is a free theory, but in QED, photons can interact via electron loops, as shown in Fig.~\ref{fig:gamgam}. In the EFT, there are no dynamical electrons, so the $4\gamma$ interaction due to electron loops is given by a series of higher dimension operators involving photon fields. The lowest dimension interactions that preserve charge conjugation are given by dimension eight operators, so the EFT Lagrangian has the expansion
\begin{align}
\label{2.21}
\LL &= -\frac{1}{4} F_{\mu \nu}F^{\mu \nu} + \frac {\alpha^2}{  m_e^4} \left[ c_1
\left(F_{\mu \nu}F^{\mu \nu}\right)^2 +  c_2
\left(F_{\mu \nu} \tilde F^{\mu \nu}\right)^2 \right] + \ldots\,.
\end{align}
This is the Euler-Heisenberg Lagrangian~\cite{Heisenberg:1935qt}.
We can compare eqn~(\ref{2.21}) with the general form eqn~(\ref{2.17}). We have used $m_e$ for the scale $\Lambda$, since we know that the higher dimension operators are generated by the electron loop graph in QED shown in Fig.~\ref{fig:gamgam}. Since QED is perturbative, we have included a factor of  $e^4$ from the vertices, and  $1/16\pi^2$ from the loop, so that $c_{1,2}$ are pure numbers.

The scattering amplitude computed from eqn~(\ref{2.21}) in the center-of-mass frame is
\begin{align}
\mathscr{A}  \sim  \frac{ \alpha^2 \omega^4 }{ m_e^4}\,,
\end{align}
where $\omega$ is the photon energy. The $\alpha^2/m_e^4$ factor is from the Lagrangian, and the $\omega^4$ factor is because each field-strength tensor is the gradient of $A_\mu$, and produces a factor of $\omega$. The scattering cross section $\sigma$ is proportional to $\abs{\mathscr{A}}^2$, and has mass dimension $-2$. The phase space integral is thus $\propto 1/\omega^2$ to get the correct dimensions,  since $\omega$ is the only dimensionful parameter in the low-energy problem. The cross section is then
\begin{align}
\label{2.23}
\sigma   \sim \left( \frac{\alpha^2 \omega^4 }{ m_e^4 } \right)^2
\frac{1 }{ \omega^2} \frac {1 }{ 16 \pi} \sim \frac{\alpha^4 \omega^6 }{ 16 \pi m_e^8}\,.
\end{align}
The $1/(16\pi)$ will be explained in Sec.~\ref{sec:nda}. The $\omega^6$ dependence of the cross section follows from the lowest operator being of dimension eight, so that $\mathscr{A} \propto 1/m_e^4$, and $\sigma \propto 1/m_e^8$,
\begin{align}
A \propto \frac{1}{m_e^4} \Rightarrow \sigma \propto \omega^6\,.
\end{align}
If we had assumed (incorrectly) that gauge invariance was not important and written the interaction operator generated by Fig.~\ref{fig:gamgam} as the dimension four operator
\begin{align}
\LL &=  c\, \alpha^2  (A_\mu A^\mu)^2
\end{align}
the cross section would be $\sigma \sim \alpha^4/(16 \pi \omega^2)$ instead. The ratio of the two estimates is $(\omega/m_e)^8$. For $\omega \sim 1$\,eV, the ratio is $10^{48}$! 

An explicit computation~~\cite{Euler:1935zz,Euler:1936aa,Heisenberg:1935qt} gives
\begin{align}
c_1 = \frac{1 }{ 90}, \qquad c_2=\frac {7 }{ 360},
\end{align}
and~~\cite{Liang:2011sj}
\begin{align}
\sigma  = \frac{ \alpha^4 \omega^6 }{ 16 \pi m_e^8} \frac {15568}{ 10125}\,.
\end{align}
Our estimate eqn~(\ref{2.23}) is quite good (about 50\% off), and was obtained with very little work.

For scalar field scattering, the interaction operator would be $\phi^4$, so that $\sigma \sim 1/(16\pi \omega^2)$, whereas Goldstone bosons such as pions have interactions $ \Pi^2 (\partial \Pi)^2/f^2$, so that $\sigma \sim \omega^4/(16 \pi f^4)$. Cross sections can vary by many orders of magnitude ($10^{48}$ between scalars and gauge bosons), so dimensional estimates such as this are very useful to decide whether a cross section is experimentally relevant before starting on a detailed calculation.

\section{Proton Decay}

Grand unified theories violate baryon and lepton number. The lowest dimension operators constructed from SM fields which violate baryon number are dimension six operators, 
\begin{align}
\label{2.28}
\LL \sim  \frac{qqql}{M_G^2}.
\end{align}
These operators violate baryon number $B$ and lepton number $L$, but conserve $B-L$. The operator eqn~(\ref{2.28}) leads to the proton decay amplitude $p \to e^+ \pi^0$
\begin{align}\label{4.26}
\mathscr{A} \sim \frac{1}{M_G^2}\,,
\end{align}
and the proton decay rate
\begin{align}
\label{2.30}
\Gamma \sim \frac{m_p^5}{16 \pi M_G^4}\,.
\end{align}
In eqn~(\ref{2.30}), we have obtained a decay rate of the correct dimensions using the only scale in the decay rate calculation, the proton mass $m_p$, and the rule of $1/(16\pi)$ for the final state phase space discussed in Sec.~\ref{sec:nda}. The proton lifetime is
\begin{align}
\tau =\frac{1}{\Gamma} \sim  \left(\frac{M_G}{10^{15} \, \hbox{GeV}}\right)^4 \times 10^{30} \ \hbox{years}
\end{align}
EFT power counting provides a natural explanation for baryon number conservation. In the SM, baryon number is first violated at dimension six, leading to a long proton lifetime.

If baryon number were violated at dimension five (as happens in some supersymmetric models), eqn~(\ref{4.26}) would be replaced by
$\mathscr{A} \sim 1/M_G$, and the proton decay rate is
\begin{align}
\Gamma \sim \frac{m_p^3}{16 \pi M_G^2}\,.
\end{align}
The proton lifetime is very short,
\begin{align}
\tau =\frac{1}{\Gamma} \sim  \left(\frac{M_G}{10^{15} \, \hbox{GeV}}\right)^2 \times 1 \ \hbox{years},
\end{align}
and is ruled out experimentally.

\section{$n-\overline n$ Oscillations}

In some theories, baryon number is violated but lepton number is not. Then proton decay is forbidden. The proton is a fermion, and so its decay products must contain  a lighter fermion. But the only fermions lighter than the proton carry lepton number, so proton decay is forbidden. These theories do allow for a new experimental phenomenon, namely $n - \overline n$ oscillations, which violates only baryon number.

The lowest dimension operator  that leads to $n -\overline n$ oscillations, is the $\Delta B=2$ six-quark operator
\begin{align}
\LL \sim  \frac{q^6}{M_G^5},
\end{align}
which is dimension nine, and suppressed by five powers of the scale $M_G$ at which the operator is generated. This leads to an oscillation amplitude
\begin{align}
\mathscr{A} \sim \left(\frac{m_n}{M_G}\right)^5\,,
\end{align}
which is strongly suppressed.

\section{Neutrino Masses}

The lowest dimension operator in the SM which gives a neutrino mass is the $\Delta L=2$ operator of dimension five (see Sec.~\ref{sec:dim5}),
\begin{align}
\label{4.33}
\LL \sim  \frac{ (H^\dagger \ell)(H^\dagger \ell)}{M_S},
\end{align}
generated at a high scale $M_S$ usually referred to as the seesaw scale. eqn~(\ref{4.33}) gives a Majorana neutrino mass of order
\begin{align}
m_\nu  \sim \frac{v^2}{M_S}
\end{align}
when $SU(2) \times U(1)$ symmetry is spontaneously broken by $v \sim 246$\,GeV. Using $m_\nu \sim 10^{-2}$ eV leads to a seesaw scale $M_S \sim 6\times 10^{15}$~GeV. Neutrinos are light if the lepton number violating scale $M_S$ is large.

\section{Rayleigh Scattering}

The scattering of photons off atoms at low energies also can be analyzed using our power counting results. Here low energies means energies small enough that one does not excite the internal states of the atom, which have excitation energies of order electron-Volts.

The atom can be treated as a neutral particle of mass $M$, interacting with the electromagnetic field. Let $\psi(x)$ denote a field operator that creates an atom at the point $x$. Then the effective Lagrangian for the atom is
\begin{align}
\label{2.36}
\LL = \psi^\dagger\left(i \partial_t - \frac{\partial^2 }{ 2M} \right) \psi + \LL _{\rm
int},
\end{align}
where $\LL _{\rm int}$ is the interaction term. From eqn~(\ref{2.36}), we see that $\left[\psi\right]=3/2$. Since the atom is neutral, covariant derivatives acting on the atom are ordinary derivatives, and do not
contain gauge fields. The gauge field interaction term is a function of the electromagnetic field strength $F_{\mu\nu}=({\bf E},{\bf B})$. Gauge invariance forbids terms which depend only on the vector potential $A_\mu$. At low energies, the dominant interaction is one which involves the lowest dimension operators,
\begin{align}
\label{2.37}
\LL _{\rm int} = a_0^3 \ \psi^\dagger \psi \left(c_E \mathbf{E}^2 +c_B \mathbf{B}^2\right)\,.
\end{align}
An analogous $\mathbf{E\cdot B}$ term is forbidden by parity conservation. The  operators in eqn~(\ref{2.37})  have  $\opdim=7$, so we have written their coefficients as dimensionless constants times $a_0^3$. $a_0$ is the size of the atom, which controls the interaction of photons with the atom, and $[a_0]=-1$. The photon only interacts with the atom when it can resolve its charged constituents, the electron and nucleus, which are separated by $a_0$, so $a_0$ plays the role of $1/\Lambda$ in eqn~(\ref{2.37}).

The interaction eqn~(\ref{2.37}) gives the scattering amplitude
\begin{align}
\mathscr{A} \sim a_0^3 \omega^2\,,
\end{align}
since the electric and magnetic fields are gradients of the vector potential, so each factor of $\bf E$ or $\bf B$ produces a factor of $\omega$. The scattering cross-section is proportional to $\abs{\mathscr{A}}^2$. This has the correct dimensions to be a cross-section, so the phase-space is dimensionless, and
\begin{align}\label{4.38}
\sigma \propto a_0^6\ \omega^4.
\end{align}
Equation~(\ref{4.38}) is the famous $\omega^4$ dependence of the Rayleigh scattering cross-section, which explains why the sky is blue---blue light is scattered 16 times more strongly than red light, since it has twice the frequency.

The argument above also applies to the interaction of low-energy gluons with $Q \bar Q$ bound states such as the $J/\psi$ or $\Upsilon$. The Lagrangian is eqn~(\ref{2.37}) where $\mathbf{E}^2$ and $\mathbf{B}^2$ are replaced by their QCD analogs, $\mathbf{E}^A \cdot \mathbf{E}^A$ and $\mathbf{B}^A \cdot \mathbf{B}^A$. The scale $a_0$ is now the radius of the QCD bound state. The Lagrangian can be used to find the interaction energy of the $Q\bar Q$ state in nuclear matter. The $\psi$ field is a color singlet, so the only interaction with nuclear matter is via the the gluon fields. The forward scattering amplitude off a nucleon state is
\begin{align}\label{4.39}
\mathscr{A} &= a_0^3\braket{ N | c_E \mathbf{E}^A \cdot \mathbf{E}^A  +c_B \mathbf{B}^A \cdot \mathbf{B}^A | N}
\end{align}
Equation~(\ref{4.39}) is a non-perturbative matrix element of order $\lqcd^2$. It turns out that it can evaluated in terms of the nucleon mass and the quark momentum fraction measured in DIS~\cite{Luke:1992tm}. The binding energy $U$ of the $Q\bar Q$ state is related to $\mathscr{A}$ by
\begin{align}
U &= \frac{n \mathscr{A}}{2 M_N} \,,
\end{align}
where $n$ is the number of nucleons per unit volume in nuclear matter. The $1/(2M_N)$ prefactor is because nucleon states in field theory are normalized to $2M_N$ rather than to $1$, as in quantum mechanics. Just using dimensional analysis, with $n \sim \lqcd^3$, $\mathscr{A} \sim a_0^3 \lqcd^2$, and neglecting factors of two,
\begin{align}
U &= \frac{a_0^3 \lqcd^5}{M_N}\,.
\end{align}
With $a_0 \sim 0.2 \times 10^{-15}$\,m for the $J/\psi$, and $\lqcd \sim 350$\,MeV, the binding energy is $U \sim 5$\,MeV.

\section{Low energy weak interactions}\label{sec:fermi}

The classic example of an EFT is the Fermi theory of low-energy weak interactions. The full (UV) theory is the SM, and we can match onto the EFT by transitioning to a theory valid at momenta small compared to $M_{W,Z}$. Since the weak interactions are perturbative, the matching can be done order by order in perturbation theory.

The $W$ boson interacts with quarks and leptons via the weak current:
\begin{align}
j^\mu_W &= V_{ij}\ (\bar u_i\, \gamma^\mu\, P_L\, d_j)  +  (\bar \nu_\ell\, \gamma^\mu\, P_L\, \ell) ,
\end{align}
where $u_i=u,c,t$ are up-type quarks, $d_j=d,s,b$ are down-type quarks, and $V_{ij}$ is the CKM mixing matrix. There is no mixing matrix in the lepton sector because we are using neutrino flavor eigenstates, and neglecting neutrino masses.
\begin{figure}
\begin{center}
\includegraphics[width=3cm]{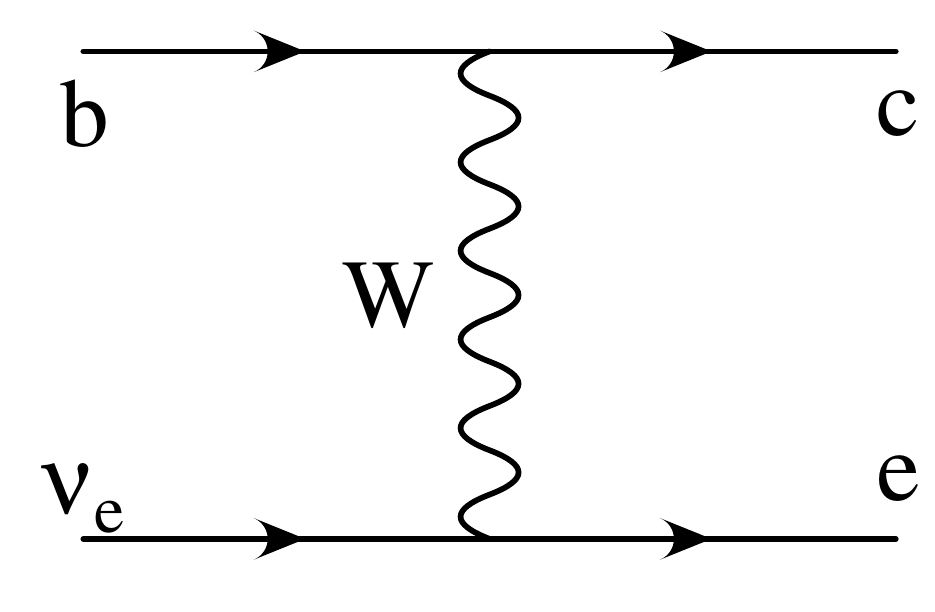} 
\caption{\label{fig:tree} Tree-level diagram for semileptonic $b \to c$ decay.}
\end{center}
\end{figure}

The tree-level amplitude for semileptonic $b \to c$ decay from Fig.~\ref{fig:tree} is
\begin{align}
\mathscr{A} &= \left(\frac{-ig}{\sqrt2}\right)^2 V_{cb} 
\left(\bar c\, \gamma^\mu\, P_L\, b\right)
\left(\bar \ell\, \gamma^\nu\, P_L\, \nu_{\ell}\right)
\left(\frac{-ig_{\mu\nu}}{p^2-M_W^2}\right),
\end{align}
where $g/\sqrt{2}$ is the $W$ coupling constant.
For low momentum transfers, $p \ll M_W$, we can expand the $W$ propagator,
\begin{align}
\label{2.45}
\frac{1}{p^2-M_W^2} = -\frac{1}{M_W^2}\left(1+\frac{p^2}{M_W^2} + \frac{p^4}{M_W^4}
+ \ldots\right),
\end{align}
giving different orders in the EFT expansion parameter $p/M_W$. Retaining only the first term gives
\begin{figure}
\begin{center}
\includegraphics[width=3cm]{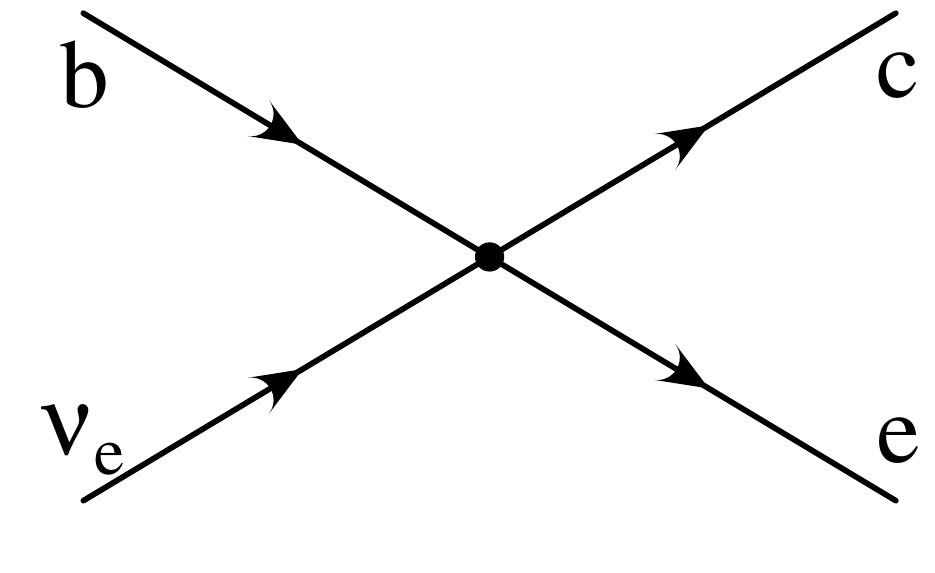} 
\end{center}
\caption{\label{fig:tree2} $b \to c$ vertex in the Fermi theory.}
\end{figure}
\begin{align}
\mathscr{A} &= \frac i{M_W^2}\left(\frac{-ig}{\sqrt2}\right)^2 V_{cb} \
\left(\bar c\, \gamma^\mu\, P_L\, b\right)\left(\bar \ell\, \gamma_\mu\, P_L\, \nu_\ell
\right)+\mathcal{O}\left(\frac{1}{M_W^4}\right)\,,
\end{align}
which is the same amplitude as that produced by the local Lagrangian
\begin{align}
\label{2.47}
\LL &= -\frac{g^2}{2 M_W^2}V_{cb} \
\left(\bar c\, \gamma^\mu\, P_L\, b\right)\left(\bar \ell\, \gamma_\mu\, P_L\, \nu_\ell
\right)+\mathcal{O}\left(\frac{1}{M_W^4}\right).
\end{align}
eqn~(\ref{2.47}) is the lowest order Lagrangian for semileptonic $b \to c$ decay in the EFT, and is represented  by the vertex in Fig.~\ref{fig:tree2}. It is usually written, for historical reasons,  in terms of $G_F$
\begin{align}
\label{2.48}
\frac{G_F}{\sqrt2} \equiv \frac{g^2}{8 M_W^2}=\frac{1}{2v^2},
\end{align}
where $v \sim 246$\,GeV is the scale of electroweak symmetry breaking,
\begin{align}
\label{2.49}
\LL & = -\frac{4 G_F}{\sqrt2}\, V_{cb}\,
\left(\bar c\, \gamma^\mu\, P_L\, b\right) \left(\bar \ell\, \gamma^\mu\, P_L\, \nu_\ell\right)\,.
\end{align}
Similarly, the $\mu$ decay Lagrangian is
\begin{align}
\label{2.50}
\LL & = -\frac{4 G_F}{\sqrt2}\, 
\left(\bar \nu_\mu\, \gamma^\mu\, P_L\, \mu\right) \left(\bar e\, \gamma^\mu\, P_L\, \nu_e\right)\,.
\end{align}
The EFT Lagrangian eqn~(\ref{2.49},\ref{2.50}) is the low-energy limit of the SM. The EFT no longer has dynamical $W$ bosons, and the effect of $W$ exchange in the SM has been included via dimension-six four-fermion operators. The procedure used here is referred to as ``integrating out'' a heavy particle, the $W$ boson.

The Lagrangian eqn~(\ref{2.49},\ref{2.50}) has been obtained by expanding in $p/M_W$, i.e.\ by treating $M_W$ as large compared with the other scales in the problem. Weak decays computed using eqn~(\ref{2.49},\ref{2.50}) still retain the complete dependence on low energy scales such as $m_b$, $m_c$ and $m_\ell$. Using eqn~(\ref{2.49}) gives the $b$ lifetime,
\begin{align}
\label{4.50}
\Gamma(b \to c \ell \overline \nu_\ell) = \frac{\abs{V_{cb}}^2 G_F^2 m_b^5}{192\pi^3}\ f\left( \frac{m_c^2}{m_b^2}\right),
\end{align}
where we have neglected $m_\ell$, and
\begin{align}
f\left(\rho \right) = 1 -8 \rho+8 \rho^3-\rho^4-12 \rho^2 \ln \rho,\qquad \rho=\frac{m_c^2}{m_b^2}.
\end{align}
Equation~(\ref{4.50}) gives the full $m_c/m_b$ dependence of the decay rate, but drops terms of order $m_b/M_W$ and $m_c/M_W$. The full $m_\ell/m_b$ dependence can also be included by retaining $m_\ell$ in the decay rate calculation. The use of the EFT Lagrangian eqn~(\ref{2.49}) simplifies the calculation. We could have achieved the same simplification by computing Fig.~\ref{fig:tree} in the SM, and expanding the amplitude using eqn~(\ref{2.45}).  The true advantages of EFT show up in higher order calculations including radiative corrections from loop graphs, which cannot be computed by simply expanding the SM amplitude.

The Fermi Lagrangian can be used to compute electroweak scattering cross sections such as the neutrino cross section. Here we give a simple dimensional estimate of the cross section,
\begin{align}
\sigma &\sim \frac{1}{16 \pi} \left( \frac{4 G_F}{\sqrt 2} \right)^2 E^2_\text{CM} \sim \frac{1}{2\pi} G_F^2 E^2_\text{CM} \,,
\end{align}
where the $G_F$ factor is from the weak interaction Lagrangian, $1/(16 \pi)$ is two-body phase space, and $E_\text{CM}$ gives $\sigma$ the dimensions of a cross section. For neutrino scattering off a fixed target, $E_{CM}^2 = 2 E_\nu M_T$, so neutrino cross sections grow linearly with the neutrino energy. Neutrino cross sections are weak as long as $E_\nu$ is much smaller the electroweak scale.

\begin{exercise}

Compute the decay rate $\Gamma( b \to c e^- \overline \nu_e)$ with the interaction Lagrangian
\begin{align*}
L &= -\frac{4 G_F}{\sqrt 2}V_{cb} ( \overline c \gamma^\mu P_L b)(\overline \nu_e \gamma_\mu P_L e)
\end{align*}
with $m_e \to 0$, $m_\nu \to 0$, but retaining the dependence on $\rho = m_c^2/m_b^2$. It is convenient to write the three-body phase space in terms of the variables $x_1=2E_e/m_b$ and $x_2=2 E_\nu/m_b$.

\end{exercise}

\section{$M_W$ vs $G_F$}

The weak interactions have two parameters, $g$ and $M_W$, and the Fermi Lagrangian in eqn~(\ref{2.49}) depends only on the combination $G_F$ in eqn~(\ref{2.48}). Higher order terms in the expansion eqn~(\ref{2.45}) are of the form
\begin{align}
\label{2.52}
-\frac{4G_F}{\sqrt 2} \left[ 1 + \frac{p^2}{M_W^2} + \ldots \right]
=-\frac{2}{v^2} \left[ 1 + \frac{p^2}{M_W^2} + \ldots \right]
\end{align}
so that the EFT momentum expansion is in powers of $\delta=p/M_W$, even though the first term in Eq~(\ref{2.52}) is $\propto 1/v^2$. The expansion breaks down for $p \sim M_W = g v/2$, which is smaller than $v \sim 246$\,GeV.

Despite the theory having multiple scales $M_W$ and $v$,  we can still use our EFT power counting rules of Sec.~\ref{sec:4.2}. From the $\mu$ decay rate computed using eqn~(\ref{2.50})
\begin{align}
\label{2.53}
\Gamma(\mu \to e  \nu_\mu \overline \nu_e) = \frac{G_F^2 m_\mu^5}{192\pi^3}\,,
\end{align}
and the experimental value of the $\mu$ lifetime $2.197 \times 10^{-6}$\,s, we obtain $G_F \sim 1.16 \times 10^{-5}\, \text{GeV}^{-2}$. Using $G_F \sim 1/\Lambda^2$  gives $\Lambda \sim 300$\,GeV. This indicates that we have an EFT with a scale of order $\Lambda$. This is similar to the multipole expansion estimate for $a$.

We can then use the power counting arguments of Sec.~\ref{sec:4.2}. They show that the leading terms in the decay amplitude are single insertions of dimension-six operators, the next corrections are two insertions of dimension-six or one insertion of dimension-eight operators, etc. None of these arguments care about the precise value of $\Lambda$. They allow one to group terms in the expansion of similar size. 

Dimension-eight corrections are $p^2/\Lambda^2$ suppressed. In $\mu$-decay, this implies that dimension-eight corrections are suppressed by $m_\mu^2/\Lambda^2$. The power counting estimate using either $\Lambda \sim M_W$ or $\Lambda \sim v$ shows that they are very small corrections. We can check that these corrections are small from \emph{experiment}. The Lagrangian eqn~(\ref{2.50}) predicts observables such as the phase-space distribution of $\mu$ decay events over the entire Dalitz plot, the polarization of the final $e^-$, etc.\  Comparing these predictions, which neglect dimension-eight contributions, with experiment provides a test that eqn~(\ref{2.50}) gives the correct decay amplitude. Very accurate experiments which are sensitive to deviations from the predictions of eqn~(\ref{2.50}), i.e.\ have an accuracy $m_\mu^2/M_W^2 \sim 10^{-6}$,  can then be used to probe dimension-eight effects, and determine the scale $M_W$.

Historically, when the SM was developed, $G_F$ was fixed from $\mu$ decay, but the values of $M_W$ and $M_Z$ were not known. Their values \emph{were not needed} to apply the Fermi theory to low-energy weak interactions.  The value of  $M_Z$ was determined by studying the energy dependence of parity violation in electron scattering through $\gamma-Z$ interference effects. This fixed the size of the dimension-eight $p^2/M_Z^2$ terms in the neutral current analog of eqn~(\ref{2.52}), and determined the scale at which the EFT had to be replaced by the full SM, with dynamical gauge fields.

\chapter{Loops}

The real power of EFTs becomes apparent when computing loop corrections. There are several tricky points that must be understood before EFTs can be used at the loop level, which are explained in this section. 

\begin{figure}
\begin{center}
\includegraphics[width=2cm]{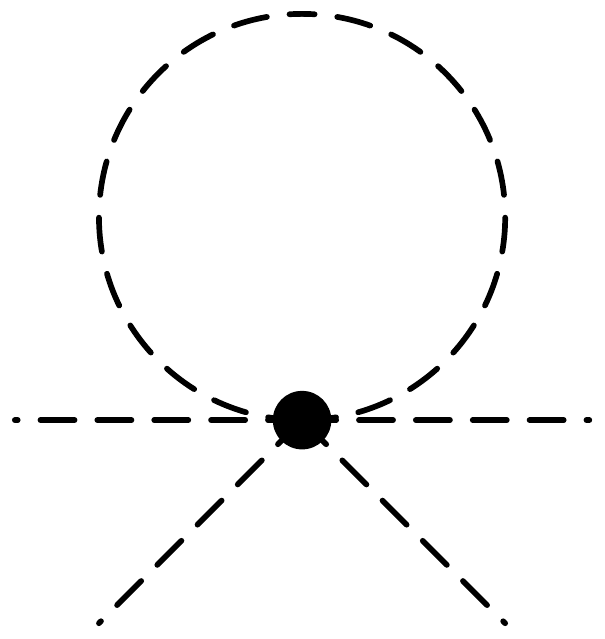}
\end{center}
\caption{\label{fig:loop1} One-loop correction to $\phi\phi$ scattering from a $\phi^6$ interaction.}
\end{figure}
For simplicity consider an EFT of a scalar field $\phi$, with a higher dimension operator
\begin{align}
\LL &= \LL_{\opdim \le 4} + \frac{c_6}{\Lambda^2} \frac{1}{6!} \phi^6 \,.
\label{3.1}
\end{align}
The dimension-six operator gives a contribution to $\phi-\phi$ scattering from the graph in Fig.~\ref{fig:loop1},
\begin{align}
\label{3.2}
\mathscr{A} &= -\frac{c_6}{2\Lambda^2} \int \frac{{ d^4 k}}{(2\pi)^4}\frac{1}{k^2-m_\phi^2}\,.
\end{align}
The EFT is valid for $k < \Lambda$, so we can use a momentum-space cutoff $\Lambda_c < \Lambda$. The scalar mass $m_\phi$ is much smaller than $\Lambda_c$, since $\phi$ is a particle in the EFT. Neglecting $m_\phi$, the integral gives
\begin{align}
\label{3.3}
\mathscr{A} &\approx -\frac{c_6}{2\Lambda^2} \frac{\Lambda_c^2}{16\pi^2}.
\end{align}
The integral eqn~(\ref{3.2}) is quadratically divergent, which gives the quadratic cutoff dependence in eqn~(\ref{3.3}). Similarly, a dimension eight operator $\phi^4 (\partial_\mu \phi)^2$ with coefficient $c_8/\Lambda^4$ has an interaction vertex $k^2/\Lambda^4$, and gives a contribution
\begin{align}
\label{3.4}
\mathscr{A}  &=-\frac{c_8}{\Lambda^4} \int \frac{{ d^4 k}}{(2\pi)^4}\frac{k^2}{k^2-m_\phi^2}\approx -\frac{c_8}{\Lambda^4} \frac{\Lambda_c^4}{16\pi^2}\,,
\end{align}
since the integral is quartically divergent.

The results eqn~(\ref{3.3},\ref{3.4}) lead to a violation of the power counting formula eqn~(\ref{2.20}), and the EFT expansion in powers of $1/\Lambda$ breaks down, since $\Lambda_c$ is the same order as $\Lambda$. Loops with insertions of higher dimension operators give contributions of leading order in the $1/\Lambda$ expansion, which need to be resummed.  One could try using $\Lambda_c \ll \Lambda$, but this turns out not to work. Firstly, $\Lambda_c$ is an artificial scale that has been introduced, with no connection to any physical scale. In the end, all $\Lambda_c$ dependence must cancel. For example, the weak interactions would require introducing a cutoff scale $m_b \ll \Lambda_c \ll M_W$ to keep the power divergences in eqn~(\ref{3.3},\ref{3.4}) under control, and this would be an artificial scale that cancels in final results.  Furthermore, cutoffs do not allow one to sum large logarithms, which is one of the main reasons why EFTs are used in the first place, since we are restricted to $\Lambda_c \ll \Lambda$. A cutoff has other problems as well, it violates important symmetries such as gauge invariance and chiral symmetry. In fact, nobody has successfully performed higher order log-resummation in EFTs with non-Abelian gauge interactions using a cutoff. Wilson proposed a set of axioms~\cite{Wilson:1972cf} for good regulators which are discussed in Ref.~\cite[Chapter 4]{Collins:1984xc}.

 Often, you will see discussions of EFTs where high momentum modes with $k > \Lambda_c$ are integrated out, and the cutoff is slowly lowered to generate an infrared theory. While ideas like this were historically useful, this is not the way to think of an EFT, and it is not the way EFTs are actually used in practice. 
 
Let us go back to a loop graph such as eqn~(\ref{3.1}), and for now, retain the cutoff $\Lambda_c$. In addition to the contribution shown in eqn~(\ref{3.3}), the loop graph also contains non-analytic terms in $m_\phi$. In more complicated graphs, there would also be non-analytic terms in the external momentum $p$. Loop graphs have a complicated analytic structure in $p$ and $m_\phi$, with branch cuts, etc.\ The discontinuities across branch cuts from logs in loop graphs are related to the total cross section via the optical theorem. The non-analytic contributions are crucial to the EFT, and are needed to make sure the EFT respects unitarity. The non-analytic part of the integral can be probed by varying $m_\phi$ and $p$, and arises from $k \sim m_\phi, p$, i.e.\ loop momenta of order the physical scales in the EFT. For loop momenta of order $\Lambda_c$, $m_\phi,p \ll \Lambda_c$, one can expand in $m_\phi$ and $p$, and the integral gives analytic but $\Lambda_c$ dependent contributions such as eqn~(\ref{3.3}).

The high-momentum part of the integral is analytic in the IR variables, and has the same structure as amplitudes generated by local operators. This is the concept of locality mentioned in the introduction. Thus the integral has non-analytic pieces we want, plus local pieces that depend on $\Lambda_c$. The cutoff integral is an approximation to the actual integral in the full theory. Thus the local pieces computed as in eqn~(\ref{3.3}) are not the correct ones. In fact, in theories such as $\chi$PT where the UV theory is not related perturbatively to the EFT, the UV part of the integral is meaningless. Luckily, locality saves the day. The local pieces have the same structure as operators in the EFT Lagrangian, so they can be absorbed into the EFT Lagrangian coefficients.  The EFT coefficients are then adjusted to make sure the EFT gives the correct $S$-matrix, a procedure referred to as ``matching.'' The difference in UV structure of the full theory and the EFT is taken care of by the matching procedure. In the end, we only need the EFT to reproduce the non-analytic dependence on IR variables; the analytic dependence is absorbed into Lagrangian coefficients.
An explicit calculation is given in Sec.~\ref{sec:5.5}. 

To actually use EFTs in practice, we need a renormalization scheme that automatically implements the procedure above---i.e.\ it gives the non-analytic IR dependence without any spurious analytic contributions that depend on $\Lambda_c$. Such a scheme also maintains the EFT power counting, since no powers of a high scale $\Lambda_c$ appear in the numerator of  loop integrals, and cause the EFT expansion to break down. Dimensional regularization is a regulator that satisfies the required properties. It has the additional advantage that it maintains gauge invariance and chiral symmetry.

\section{Dimensional Regularization}\label{sec:5.1}

The basic integral we need is
\begin{align}
\label{3.5}
 \mu^{2\epsilon} \int \frac{\rd^\d k}{(2\pi)^\d} \frac{\left(k^2\right)^a}{\left(k^2-M^2\right)^b}
&= \frac{i \mu^{2\epsilon}}{\left(4\pi\right)^{\d/2}} \frac{(-1)^{a-b} \Gamma(\d/2+a) \Gamma(b-a-\d/2)}{\Gamma(\d/2) \Gamma(b)}
\left(M^2\right)^{\d/2+a-b}\,
\end{align}
where $\d=4-2\epsilon$. The $\mu^{2\epsilon}$ prefactor arises from $\mu^\epsilon$ factors in coupling constants, as in eqn~(\ref{6}). Equation~(\ref{3.5}) is obtained by analytically continuing the integral from values of $a$ and $b$ where it is convergent. Integrals with several denominators can be converted to eqn~(\ref{3.5}) by combining denominators using Feynman parameters. 

\begin{exercise}

Verify eqn~(\ref{3.5}) by first analytically continuing to Euclidean space, and then switching to spherical polar coordinates in $\d$ dimensions.

\end{exercise}

The integral eqn~(\ref{3.5}) is then expanded in powers of $\epsilon$. As an example,
\begin{align}
\label{3.6}
I&= \mu^{2\epsilon} \int \frac{\rd^\d k}{(2\pi)^\d} \frac{1}{\left(k^2-M^2\right)^2}
= \frac{i \mu^{2\epsilon}}{\left(4\pi\right)^{2-\epsilon}} \frac{\Gamma(\epsilon)}{\Gamma(2)}
\left(M^2\right)^{-\epsilon}\,, \nn
&=
\frac{i}{16 \pi^2} \left[\frac{1}{\epsilon} -\gamma+ \log \frac{4 \pi \mu^2}{M^2} + \mathcal{O}\left(\epsilon\right) \right]\,,
\end{align}
where $\gamma=0.577$ is Euler's constant. In the \MSbar\ scheme, we make the replacement
\begin{align}
\label{3.7}
\mu^2 = \bar \mu^2 \frac{e^{\gamma}}{4\pi}\,,
\end{align}
so that 
\begin{align}
\label{3.6a}
I &=
\frac{i}{16 \pi^2} \left[\frac{1}{\epsilon} + \log \frac{\bar\mu^2}{M^2} + \mathcal{O}\left(\epsilon\right) \right]\,.
\end{align}
The $1/\epsilon$ part, which diverges as $\epsilon \to 0$, is cancelled by a counterterm, leaving the renormalized integral
\begin{align}
\label{3.8}
 I + \text{c.t.}
&=
\frac{i}{16 \pi^2} \log \frac{\bar\mu^2}{M^2} \,.
\end{align}
The replacement eqn~(\ref{3.7}) removes $\log 4\pi$ and $-\gamma$ pieces in the final result.

There are several important features of dimensional regularization:
\begin{itemize}

\item $\bar \mu$ only appears as $\log \bar \mu$, and there are no powers of $ \bar \mu$. The only source of $\bar \mu$ in the calculation is from powers of $\mu^\epsilon$ in the coupling constants, and expanding in $\epsilon$ shows that only $\log \mu$ (and hence $\log \bar \mu$) terms occur.

\item Scaleless integrals vanish,
\begin{align}
\label{3.9}
 \mu^{2\epsilon} \int \frac{\rd^\d k}{(2\pi)^\d} \frac{\left(k^2\right)^a}{\left(k^2\right)^b}
&= 0\,.
\end{align}
This follows using eqn~(\ref{3.5}) and taking the limit $M \to 0$. Since integrals in dimensional regularization are defined by analytic continuation, the limit $M \to 0$ is taken assuming $\d/2+a-b >0$ so that the limit vanishes. Analytically continuing to  $\d/2+a-b \le 0$, the integral remains $0$. The vanishing of scaleless integrals plays a very important role in calculations using dimensional regularization.

\item There are no power divergences. For example, the quadratically divergent integral
\begin{align}
\label{3.10}
 \mu^{2\epsilon} \int \frac{\rd^\d k}{(2\pi)^\d} \frac{1}{\left(k^2-m^2\right)}
&= -\frac{i \mu^{2\epsilon}}{\left(4\pi\right)^{\d/2}} \Gamma(-1+\epsilon) \left(m^2\right)^{1-\epsilon} \nn
&= \frac{i}{16 \pi^2} \left[\frac{m^2}{\epsilon} + m^2 \log \frac{\bar \mu^2}{m^2} + m^2 + \mathcal{O}\left(\epsilon\right)   \right]\,,
\end{align}
depends only on powers of the IR scale $m$. There is no dependence on any UV scale (such as a cutoff), nor any power-law dependence on $\bar\mu$. Similarly, the integral
\begin{align}
\label{3.11}
 \mu^{2\epsilon} \int \frac{\rd^\d k}{(2\pi)^\d} \frac{\left(k^2\right)}{\left(k^2-m^2\right)}
&= \frac{i \mu^{2\epsilon}}{\left(4\pi\right)^{\d/2}} \frac{\Gamma(3-\epsilon) \Gamma(-2+\epsilon)}{\Gamma(2-\epsilon) \Gamma(1)}
\left(m^2\right)^{2-\epsilon} \nn
&= \frac{i}{16 \pi^2} \left[\frac{m^4}{\epsilon} + m^4 \log \frac{\bar \mu^2}{m^2} + m^4 + \mathcal{O}\left(\epsilon\right)   \right]\,,
\end{align}
so the quartic divergence of the integral turns into the IR scale $m$ to the fourth power.

\end{itemize}

The structure of the above integrals is easy to understand. Evaluating integrals using dimensional regularization is basically the same as evaluating  integrals using the method of residues. Values of $\d,a,b$ are assumed such that the integrand vanishes sufficiently fast as $k \to \infty$ that the contour at infinity can be thrown away. The integrand is then given by the sum of residues at the poles. The location of the poles is controlled by the denominators in the integrand, which only depend on the physical scales in the low-energy theory, such as particle masses and external momenta. Dimensional regularization automatically gives what we want---it keeps all the dependence on the physical parameters, and throws away all unphysical dependence on high-energy scales. It is the simplest physical regulator, and the one used in all higher order calculations.

\newpage

\section{No Quadratic Divergences}\label{sec:quad}

\begin{figure}
\begin{center}
\includegraphics[width=2cm]{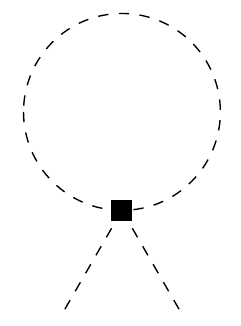}
\caption{\label{fig:quad}One loop correction to the Higgs mass from the $-\lambda (H^\dagger H)^2$ interaction.}
\end{center}
\end{figure}

Let us look at the scalar graph Fig.~\ref{fig:quad} which gives a correction to the Higgs mass in the SM,
\begin{align}
\delta m_H^2 &=-12 \lambda \mu^{2\epsilon}  \int \frac{\rd^d k}{(2\pi)^d} \frac{1}{\left(k^2-m_H^2\right)} \,,
\end{align}
where $\lambda$ is the Higgs self-coupling. You will have heard endless times that Fig.~\ref{fig:quad} gives a correction
\begin{align}
\label{3.13}
\delta m_H^2 \propto \Lambda^2\,,
\end{align}
to the Higgs mass
that depends quadratically on the cutoff. This is supposed to lead to a naturalness problem for the SM, because the Higgs is so much lighter than $\Lambda$, which is taken to be at the GUT scale or Planck Scale. The naturalness problem also goes by the names of hierarchy problem or fine-tuning problem.

The above argument for the naturalness problem is \emph{completely bogus}. The regulator used for the SM is dimensional regularization, which respects  gauge invariance. The actual value of the integral is eqn~(\ref{3.10}). Adding the renormalization counterterm cancels the $1/\epsilon$ piece, resulting in a correction to the Higgs mass
\begin{align}\label{5.15}
\delta m_H^2 &= -12 \lambda m_H^2  \left[  \log \frac{m_H^2}{\bar \mu^2} + 1 \right]\,,
\end{align}
which is proportional to the Higgs mass. There is no quadratic mass shift proportional to the cutoff; there is no cutoff. The argument eqn~(\ref{3.13}) is based on a regulator that violates gauge invariance and the Wilson axioms, and which is never used for the SM in actual calculations. Bad regulators lead to bad conclusions. 

\begin{exercise}

Compute the one-loop scalar graph Fig.~\ref{fig:quad} with a scalar of mass $m$ and interaction vertex $-\lambda \phi^4/4!$ in the \MSbar\ scheme. Verify the answer is of the form eqn~(\ref{5.15}). The overall normalization will be different, because this exercise uses a real scalar field, and $H$ in the SM is a complex scalar field. 

\end{exercise}

\section{Power Counting Formula}\label{sec:5.3}

We can now extend the power counting formula eqn~(\ref{2.20}) to include loop corrections. If we consider a loop graph with an insertion of EFT vertices with coefficients of order $1/\Lambda^a$, $1/\Lambda^b$, etc. then any amplitude (including loops) will have the $\Lambda$ dependence
\begin{align}
\frac{1}{\Lambda^a} \frac{1}{\Lambda^b} \ldots = \frac{1}{\Lambda^{a+b+ \ldots} }
\end{align}
simply from the product of the vertex factors. The discussion of Sec.~\ref{sec:5.1}  shows that the only scales which can occur in the numerator after doing the loop integrals are from  poles in  Feynman propagator denominators. These poles are at scales in the EFT, none of which is parametrically of order $\Lambda$. Thus there are no compensating factors of $\Lambda$ in the numerator, i.e.\ the power of $\Lambda$ is given by the vertex factors alone, so eqn~(\ref{2.20}),  also holds for loop graphs.

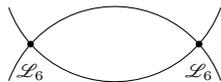
\begin{figure}
\begin{center}
\begin{tikzpicture}[scale=0.25]

\draw (0,-4)+(20:6) arc (20:160:6);
\draw (0,4)+(200:6) arc (200:340:6);

\filldraw (4.48,0) circle (0.15);
\filldraw (-4.48,0) circle (0.15);

\draw (4.48,-1.5) node [align=center] { $\scriptstyle \LL_6$};
\draw (-4.48,-1.5) node [align=center] { $\scriptstyle \LL_6$};

\end{tikzpicture}
\end{center}

\caption{\label{fig:5.3} Graph with two insertions of dimension-six operators, which requires a dimension-eight counterterm.}
\end{figure}

Loop graphs in general are infinite, and the infinities ($1/\epsilon$ poles) are cancelled by renormalization counterterms. The EFT must include all operators necessary to absorb these divergences. From $n=\sum_i (\opdim_i-4)$, we see that if there is an operator with $\opdim >4$, we will generate operators with arbitrary high dimension. Thus an EFT includes all possible higher dimension operators consistent with the symmetries of the theory. Dimension-six operators are needed to renormalize graphs with two insertions of dimension-five operators; dimension-eight operators are needed to renormalize graphs with two insertions of dimension-six operators (see Fig.~\ref{fig:5.3}), etc.
and we have to keep the entire expansion in higher dimension operators
\begin{align}
\LL_\text{EFT}  = \LL_{\opdim \le 4} + \frac{\LL_5 }{ \Lambda} + \frac{\LL_6 }{ \Lambda^2} + \ldots\,.
\end{align}
Even if we focus just on the low-dimension operators, it is understood that the higher dimension operators are still present. It also makes no sense to set their coefficients to zero. Their coefficients depend on $\bar \mu$, and on other choices such as the gauge-fixing term, etc.\ and so setting them to zero is a random unmotivated choice which will no longer hold at a different value of $\bar \mu$ unless the operator is forbidden by a symmetry.


\section{An Explicit Computation}

We now analyze a simple toy example, and explicitly compute a one-loop amplitude in the full theory, in the EFT, and discuss the matching between the two. The toy example is a two-scale integral that will be evaluated using EFT methods. The entire argument applies almost without change to a practical example, the derivation of the HQET Lagrangian to one-loop~\cite{Manohar:1997qy}.

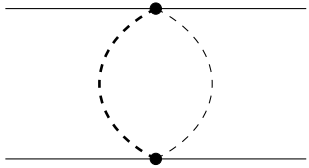
\begin{figure}
\begin{center}
\begin{tikzpicture}[scale=0.5]

\draw (-4,2) -- (4,2);
\draw (-4,-2) -- (4,-2);

\filldraw (0,2) circle (0.15);
\filldraw (0,-2) circle (0.15);

\draw[dashed] (0,2) .. controls (2,1) and (2,-1) .. (0,-2);
\draw[dashed,line width=1pt] (0,2) .. controls (-2,1) and (-2,-1) .. (0,-2);

\end{tikzpicture}
\end{center}
\caption{\label{fig:full} A graph that gives a loop integral of the form eqn~(\ref{3.17}). The solid lines are light external fields. The thin dashed line is a light particle with mass $m$. The thick dashed line is a heavy particle of mass $M$ that is not in the EFT.}
\end{figure}

Consider the integral
\begin{align}
\label{3.17}
I_F&=g^2 \mu^{2\epsilon} \int \frac{{\rm d}^\d k}{(2\pi)^\d}\ \frac{1}{(k^2-m^2)(k^2-M^2)} 
\end{align}
where we will take $m \ll M$. $M$ is the UV scale, and $m$ is the IR scale. Integrals such as eqn~(\ref{3.17}) arise in loop calculations of graphs with intermediate heavy and light particles, such as in Fig.~\ref{fig:full}. In eqn~(\ref{3.17}), we have set the external momenta to zero to get a simple integral which we can analyze to all orders in $m/M$.

The integral can be done exactly in $\d=4-2\epsilon$ dimensions
\begin{align}
\label{3.18}
I_F&=g^2\mu^{2\epsilon}\int \frac{{\rm d}^\d k}{(2\pi)^\d}\ \frac{1}{(k^2-m^2)(k^2-M^2)} \nn
&= \frac{ig^2}{16\pi^2}\left[\frac{1}{\epsilon} - \log \frac{M^2}{\bar\mu^2} 
+ \frac{m^2}{M^2-m^2}\log\frac{m^2}{M^2} +1\right] \,,
\end{align}
where we have switched to the \MSbar\ scheme using eqn~(\ref{3.7}).
$I_F$ is a relatively simple integral because there are only two mass scales in the denominator. An integral with three denominators with
unequal masses gives rise to dilogarithms.

The heavy particle $M$ can be integrated out, as was done for the $W$ boson. The heavy particle propagator is expanded in a power series,
\begin{align}
\label{3.19}
\frac{1}{k^2-M^2} = -\frac{1}{M^2}\left(1+\frac{k^2}{M^2} + \frac{k^4}{M^4}
+ \ldots\right).
\end{align}

The loop graph in the EFT is a series of contributions, one from each term in eqn~(\ref{3.19}),
\begin{align}
\label{3.20}
I_\text{EFT}&=g^2 \mu^{2\epsilon}\int \frac{{\rm d}^\d k}{(2\pi)^\d}\ \frac{1}{ (k^2-m^2)}\left[-\frac{1}{M^2}-\frac{k^2}{M^4}-\frac{k^4}{M^6}-\ldots\right] \nn
&= \frac{ig^2}{16\pi^2 M^2 }\left[-\frac{m^2}{\epsilon}
+ m^2 \log \frac{m^2}{\bar \mu^2}-m^2\right] 
+ \frac{ig^2}{16\pi^2 M^4 }\left[-\frac{m^4}{\epsilon}
+ m^4 \log \frac{m^2}{\bar \mu^2}-m^4\right]  \nn
&+ \frac{ig^2}{16\pi^2 M^6 }\left[-\frac{m^6}{\epsilon}
+ m^6 \log \frac{m^2}{\bar\mu^2}-m^6\right] 
+\ldots\,.
\end{align}
The series in eqn~(\ref{3.20}) is sufficiently simple in this example that we can sum it up,
\begin{align}
\label{3.21}
I_\text{EFT}&= \frac{ig^2}{16\pi^2}\left[-\frac{1}{\epsilon}\ \frac{m^2}{M^2-m^2}
+  \frac{m^2}{M^2-m^2}\log \frac{m^2}{\bar\mu^2}-\frac{m^2}{M^2-m^2}\right]\,,
\end{align}
to compare with $I_F$. However, it is best to think of $I_\text{EFT}$ in the expanded form eqn~(\ref{3.20}), since the EFT is an expansion in powers of $1/M$.

There are several important points to note:
\begin{itemize}

\item The two results $I_F$ and $I_\text{EFT}$ are different. The order of integration and expansion matters.

\item The $1/\epsilon$ terms do not agree, they are cancelled by counterterms which differ in the full and EFT. The two theories have \emph{different} counterterms and hence \emph{different} anomalous dimensions. In our example, the $1/\epsilon$ terms in eqn~(\ref{3.20}) give the anomalous dimensions of the $1/M^2$, $1/M^4$, $1/M^6$, etc.\ operators. Each operator has its own anomalous dimension. 

\item The full theory and the EFT  are independent theories adjusted to give the same $S$-matrix. One can use different regulators or gauge-fixing for the two theories. 

\item The $\log m^2$ terms, which are non-analytic in the IR scale, agree in the two theories. This is the part of $I_F$ which \emph{must} be reproduced in the EFT.

\item The $\log M^2$ non-analytic terms in $M$ are not present in the EFT integral. This must be the case, because in the EFT calculation, we integrated an expression which was a power series in $1/M$, and had no non-analytic terms in $M$.

\item The difference between $I_F$ and $I_{\text{EFT}}$ is from the UV part of the integral, and  is local in the IR mass scale $m$, so that $I_F-I_{\text{EFT}}$ is local (i.e.\ analytic) in $m$. This difference is called the matching contribution to the Lagrangian, and is included in the EFT result by absorbing it into shifts of the EFT Lagrangian coefficients.

\item $I_F$ has $\log M^2/m^2$ terms, which involve the ratio of the UV and IR scales. These logs can be summed using the RGE in the EFT.

\end{itemize}

\begin{exercise}\label{ex5.2}

Compute $I_F$ and $I_{\text{EFT}}$ given in eqns~(\ref{3.18},\ref{3.20})  in dimensional regularization in $\d=4-2\epsilon$ dimensions. Both integrals have UV divergences, and the $1/\epsilon$ pieces are cancelled by  counterterms. Determine the counterterm contributions $I_{F,\text{ct}}$, $I_{\text{EFT},\text{ct}}$ to the two integrals.

\end{exercise}


\section{Matching}\label{sec:5.5}

The infinite parts of $I_F$ and $I_\text{EFT}$ are cancelled by counterterms in the full theory and the EFT, respectively. The difference of the  two renormalized integrals is the matching contribution
\begin{align}
\label{3.22}
I_M &=\left[ I_F + I_{F,\text{c.t.}} \right] - \left[ I_\text{EFT} + I_{\text{EFT,c.t.}} \right] \nn
&= \frac{ig^2}{16\pi^2}\left[\left( \log \frac{\bar \mu^2}{M^2}+1\right)
+ \frac{m^2}{M^2}\left(  \log \frac{\bar\mu^2}{M^2}+1\right)+\ldots\right].
\end{align}
The terms in parentheses are matching corrections to terms of order 1, order $1/M^2$, etc.\  from integrating out the heavy particle with mass $M$. They are analytic in the IR scale $m$. In our simple toy example, the $(m^2/M^2)^r$ corrections are corrections to the coefficient of the $\chi^4$ operator, where $\chi$ is the external field in Fig.~\ref{fig:full}. If the mass $m$ is generated from a $\lambda \phi^4/4!$ interaction when a light field $\phi$ gets a vacuum expectation value $\vev{\phi}=v$, $m^2=\lambda v^2/3$, then one can treat $m^2$ as $\lambda \phi^2/3$, and the series eqn~(\ref{3.22}) is an expansion in $\chi^4 (\phi^2)^r$ operators of increasing dimension. For this reason, we refer to the $1/M$ expansion as being in operators of increasing dimension.

The logarithm of the ratio of IR and UV scales $m$ and $M$ can be written as
\begin{align}
\label{3.23}
\log \frac{m^2}{M^2} = \underbrace{ - \log \frac{M^2}{\bar \mu^2} }_{\text{matching}} + 
\underbrace{\log \frac{m^2}{\bar \mu^2}}_{\text{EFT}}\,,
\end{align}
where the scales have been separated using $\bar \mu$. The first piece is in the matching condition eqn~(\ref{3.22}), and the second in the EFT result eqn~(\ref{3.21}). We have separated a two-scale calculation into two one-scale calculations. A single scale integral is far easier to compute than a multi-scale integral, so the two-step calculation is much easier to do in practice.

\begin{exercise}\label{ex5.4}

Compute $I_M \equiv \left( I_F + I_{F,\text{ct}} \right) - \left( I_{\text{EFT}}  + I_{\text{EFT},\text{ct}} \right)$ and show that it is analytic in $m$.

\end{exercise}

\section{Summing Large Logs}

The full theory result $I_F$ has $\log M^2/m^2$ terms, which is the ratio of a UV and an IR scale. At higher orders, one gets additional powers of the log,
\begin{align}
\label{3.24}
\left[ \frac{g^2}{16\pi^2}  \log \frac{M^2}{m^2} \right]^n\,.
\end{align}
If $M \gg m$, perturbation theory can break down when $g^2/(16 \pi^2) \log M^2/m^2 \sim 1$.  QCD perturbation theory often breaks down because of such logs, and it is necessary to sum these corrections.

In the EFT approach, $I_F$ has been broken into two pieces, the matching $I_M$ and the EFT result $I_\text{EFT}$. $I_M$ only involves the high scale $M$, and logs in $I_M$ depend on the ratio $M/ \bar \mu$. These logs are not large if we choose $\bar \mu \sim M$. $I_M$ can be computed in perturbation theory with $\bar \mu \sim M$, and perturbation theory is valid as long as $g^2/(16\pi^2)$ is small, a much weaker condition than requiring $g^2/(16 \pi^2) \log M^2/m^2$ to be small. 

Similarly, $I_\text{EFT}$ only involves the scale $m$, and logs in $I_\text{EFT}$ are logs of the ratio $m/\bar  \mu$. The EFT logs are not large if we choose $\bar \mu \sim m$. Thus we can compute $I_M$ and $I_\text{EFT}$ if we use two different $\bar\mu$ values. The change in $\bar \mu$ is accomplished by using the renormalization group equations in the EFT.

\section{A Better Matching Procedure}\label{sec:5.7}

While we argued that single-scale integrals were much easier to evaluate than multi-scale ones, the way we computed $I_M$ as the difference $I_F-I_\text{EFT}$ still required first computing the multi-scale integral $I_F$. And if we know $I_F$, don't we essentially have the answer we want anyway? Why bother with constructing an EFT in the first place?

It turns out there is a much simpler way to compute the matching that does not rely on first computing $I_F$. $I_F$ and $I_\text{EFT}$ both contain terms non-analytic in the infrared scale, but the difference $I_M$ is analytic in $m$,
\begin{align}
\underbrace{ I_M(m) }_\text{analytic} &= \underbrace{ I_F(m) }_\text{non-analytic} -\underbrace{ I_\text{EFT}(m) }_\text{non-analytic} \,.
\end{align}
Therefore, we can compute $I_M$ by expanding $I_F-I_\text{EFT}$ in an expansion in the IR scale $m$. This drops the non-analytic pieces, but we know they cancel in $I_F-I_\text{EFT}$. 

The expansion of $I_F$ is
\begin{align}
\label{3.25}
I_F^{(\text{exp})} &=g^2 \mu^{2\epsilon} \int \frac{{\rm d}^\d k}{(2\pi)^\d}\ \frac{1}{k^2-M^2} \left[\frac{1}{k^2} + \frac{m^2}{k^4} + \ldots \right] .
\end{align}
The expansion of $I_\text{EFT}$ is
\begin{align}
\label{3.26}
I_\text{EFT}^{(\text{exp})} &=g^2 \mu^{2\epsilon}\int \frac{{\rm d}^\d k}{(2\pi)^\d}\  \left[\frac{1}{k^2} + \frac{m^2}{k^4} + \ldots \right]  \left[-\frac{1}{M^2}-\frac{k^2}{M^4}-\ldots\right].
\end{align}
Both $I_F^{(\text{exp})}$ and $I_\text{EFT}^{(\text{exp})}$ have to be integrated term by term. The expansions $I_F^{(\text{exp})}$ and
$I_\text{EFT}^{(\text{exp})}$ drop non-analytic terms in $m$, and do not sum to give $I_F$ and $I_\text{EFT}$. However, the non-analytic terms in $m$ cancel in the difference, so $I_F^{(\text{exp})}-I_\text{EFT}^{(\text{exp})}$ does sum to give $I_M$.

Non-analytic terms in dimensional analysis arise from contributions of the form
\begin{align}
\frac{1}{\epsilon} m^\epsilon &= \frac{1}{\epsilon} + \log m + \ldots
\end{align}
in integrals done using dimensional regularization. In eqns~(\ref{3.25},\ref{3.26}), we first expand in the IR scale $m$, and then expand in $\epsilon$.
In this case,
\begin{align}
\frac{1}{\epsilon} m^\epsilon &= \frac{1}{\epsilon} \left[ m^\epsilon\Bigr|_{m=0} + \epsilon m^{\epsilon-1}\Bigr|_{m=0}  + \ldots  \right].
\end{align}
In dimensional regularization, the $m=0$ limit of all the terms in the square brackets vanishes. Expanding in $m$ sets all non-analytic terms in $m$ to zero.

$I_\text{EFT}^{(\text{exp})}$ has to be integrated term by term. Each term is a scaleless integral, and vanishes. For example the first term in the product is
\begin{align}
\label{3.27}
&g^2 \mu^{2\epsilon}\int \frac{{\rm d}^\d k}{(2\pi)^\d}\  \left[\frac{1}{k^2}  \right]  \left[-\frac{1}{M^2}\right]
= -\frac{1}{M^2} g^2 \mu^{2\epsilon}\int \frac{{\rm d}^\d k}{(2\pi)^\d}\  \frac{1}{k^2} =0\,.
\end{align}
This is not an accident of our particular calculation, but completely general. $I_\text{EFT}$ was given by expanding the integrand of $I_F$ in inverse powers of the UV scale $M$. $I_\text{EFT}^{(\text{exp})}$ is given by taking the result and expanding the integrand in powers of the IR scale $m$. The resulting integrand has all scales expanded out, and so is scaleless and vanishes. $I_F^{(\text{exp})}$, on the other hand, now only depends on the UV scale $M$; the IR scale $m$ has been expanded out. Integrating term by term reproduces eqn~(\ref{3.22}) for the matching integral $I_M$. Thus the matching is given by evaluating $I_F$ with all IR scales expanded out. This is a much easier way to compute $I_M$ than computing $I_F$ and $I_{\text{EFT}}$ and taking the difference.

\begin{exercisebn}

Compute $I_F^{(\text{exp})}$, i.e.\ $I_F$ with the IR $m$  scale expanded out
\begin{align*}
I_F^{(\text{exp})} &= -i \mu^{2\epsilon} \int \frac{\rd^d k}{(2\pi)^d} \frac{1}{(k^2-M^2)}\left[\frac{1}{k^2} + \frac{m^2}{k^4} + \ldots  \right] \,.
\end{align*}
Note that the first term in the expansion has a $1/\epsilon$ UV divergence, and the remaining terms have $1/\epsilon$ IR divergences.

\end{exercisebn}

\begin{exercisenb}

Compute $I_F^{(\text{exp})} + I_{F,\text{ct}}$ using $I_{F,\text{ct}}$ determined in  Exercise~\ref{ex5.2}. Show that the UV divergence cancels, and the remaining $1/\epsilon$ IR divergence is the same as the UV  counterterm $I_{\text{EFT},ct}$ in the EFT.

\end{exercisenb}

Something remarkable has happened. We have taken $I_F$, and expanded term by term in inverse powers of $1/M$, i.e.\ by assuming $k \ll M$, to get $I_\text{EFT}$. Then we have taken the original $I_F$ and expanded term by term in powers of $m$, i.e.\ by assuming $k \gg m$, to get $I_F^{(\text{exp})}=I_M$. The sum of the two contributions is exactly the original integral $I_F$. Adding two different expansions of the same integrand  recovers the original result, not \emph{twice} the original result. The agreement is exact. One might worry that we have counted the region $m \ll k \ll M$ in both integrals. But this double-counting region is precisely $I_\text{EFT}^{(\text{exp})}$, and vanishes in dimensional regularization. It does not vanish with other regulators, such as a cutoff. One can understand why the EFT method works by using the analogy of dimensional regularization with integration using the method of residues. The $I_F$ integrand has UV poles at $M$ and IR poles at $m$. Expanding out in $1/M$ to get $I_\text{EFT}$ leaves only the IR poles. Expanding out in $m$ leaves only the UV poles in $I_M$. The sum of the two has all poles, and gives the full result.

Dimensional regularized integrals are evaluated with $k$ set by the physical scales in the problem. There are no artificial scales as in a cutoff regulator that lead to spurious power divergences which have to be carefully subtracted away. 

The method of regions~\cite{Beneke:1997zp} is a very clever procedure for evaluating Feynman integrals which is closely related to the above discussion. One finds momentum regions which lead to poles in the integrand, expands in a power series in each region, and integrates term-by-term using dimensional regularization. Adding up the contributions of all the momentum regions gives the original integrand. In our example, the two regions were the hard region $k \sim M$, and the soft region $k \sim m$. The method of regions provides useful information to formulate an EFT, but it is not the same as an EFT. In an EFT, one has a Lagrangian, and the EFT amplitudes are given by computing graphs using Feynman rules derived from the Lagrangian. One cannot add or subtract modes depending on which momentum region contributes to an EFT graph. For example, in HQET, graphs get contributions from momenta of order $m_b$, and of order $m_c$. Nevertheless, HQET only has a single gluon field, not separate ones for each scaling region. In the method of regions, the contribution of different regions can depend on how loop momenta are routed in a Feynman graph, though the total integral given by summing all regions remains unchanged. In an EFT, the Lagrangian and Feynman rules do not depend on the momentum routing used.

\section{UV and IR Divergences}\label{sec:5.8}

Let us look in more detail at the $1/\epsilon$ terms. The original integral $I_F$ can have both UV and IR divergences. In our example, it only has a UV divergence. The terms in $I_\text{EFT}$ are expanded in $k^2/M^2$ and become more and more UV divergent. The terms in $I_F^{(\text{exp})}$ are expanded in $m^2/k^2$ and become more and more IR divergent. Dimensional regularization regulates both the UV and IR divergences. It will be useful to separate the divergences into UV and IR, and label them by $\eUV$ or $\eIR$. In reality, there is only one $\epsilon=\eUV=\eIR$ given by $\epsilon=(4-\d)/2$. At higher loops, one has to be careful about mixed divergences which are the product of IR and UV divergences.

The log divergent (in $\d=4)$ scaleless integral vanishes
\begin{align}
\int \frac{{\rm d}^\d k}{(2\pi)^\d}\ \frac{1}{k^4} &=0.
\end{align}
It is both UV and IR divergent, and can be split into UV divergent and IR divergent integrals
\begin{align}
\label{3.29}
\int \frac{{\rm d}^\d k}{(2\pi)^\d}\ \frac{1}{k^4}=\int \frac{{\rm d}^\d k}{(2\pi)^\d}\ \left[\frac{1}{k^2(k^2-m^2)} - \frac{m^2}{k^4(k^2-m^2)}\right]\,,
\end{align}
by introducing an arbitrary mass scale $m$. The first term is UV divergent, and the second is IR divergent.
Using $\eUV,\eIR$, and evaluating the pieces, eqn~(\ref{3.29}) becomes
\begin{align}
\label{3.29b}
\int \frac{{\rm d}^\d k}{(2\pi)^\d}\ \frac{1}{k^4} &= \frac{i}{16\pi^2}\left[ \frac{1}{\eUV}  - \frac{1}{\eIR} \right]=0.
\end{align}
Log divergent scaleless integrals vanish because of the cancellation of $1/\eUV$ with $1/\eIR$. Power law divergent scaleless integrals simply vanish, and do not produce $1/\epsilon$ poles, e.g.\
\begin{align}
\label{3.29a}
\int \frac{{\rm d}^\d k}{(2\pi)^\d}\ \frac{1}{k^2} &= 0\,,&
\int \frac{{\rm d}^\d k}{(2\pi)^\d}\ 1 &= 0\,,
\end{align}
so there are no quadratic or quartic divergences in dimensional regularization.

Let us go back to our matching example. $I_F$ and $I_\text{EFT}$ have the same IR behavior, because the EFT reproduces the IR of the full theory. Now consider a particular term in $I_F^{(\text{exp})}$ with coefficient $m^r$,
\begin{align}
\label{3.31}
I_F^{(\text{exp})}(m) &=  \sum_r m^r \ I_F^{(r)} \,.
\end{align}
We have expanded out the IR scale $m$, so there can be IR divergences which would otherwise have been regulated by $m$. The integral is a single scale integral depending only on $M$, and has the form
\begin{align}
\label{3.31a}
I_F^{(r)} &=  \frac{A^{(r)}}{\eUV} + \frac{B^{(r)}}{\eIR} + C^{(r)}\,,
\end{align}
where $A^{(r)}$ is the UV divergence, $B^{(r)}$ is the IR divergence, and $C^{(r)}$ is the finite part. For example from eqn~(\ref{3.25})
\begin{align}
\label{3.32a}
I_F^{(0)} &=g^2  \mu^{2\epsilon} \int \frac{{\rm d}^\d k}{(2\pi)^\d}\ \frac{1}{k^2-M^2} \frac{1}{k^2}= \frac{ig^2}{16\pi^2} \left[ 
\frac{1}{\eUV}+ \log \frac{\bar\mu^2}{M^2}+1\right], \nn
I_F^{(2)} &=g^2 \mu^{2\epsilon} \int \frac{{\rm d}^\d k}{(2\pi)^\d}\ \frac{1}{k^2-M^2} \frac{1}{k^4}= \frac{ig^2}{16\pi^2}\frac{1}{M^2}\left[
\frac{1}{\eIR}+ \log \frac{\bar\mu^2}{M^2}+1\right],
\end{align}
so that
\begin{align}
\label{3.36}
A^{(0)} &= \frac{ig^2}{16\pi^2},  & A^{(2)} &= 0, \nn
B^{(0)} &= 0, & B^{(2)} &= \frac{ig^2}{16\pi^2}\frac{1}{M^2}, \nn
C^{(0)} &= \frac{ig^2}{16\pi^2} \left[ \log \frac{\bar\mu^2}{M^2}+1\right], &
C^{(2)} &= \frac{ig^2}{16\pi^2} \frac{1}{M^2} \left[ \log \frac{\bar\mu^2}{M^2}+1\right]\,.
\end{align}
Now look at the terms in $I_\text{EFT}^{(\text{exp})}$,
\begin{align}
\label{3.32b}
I_\text{EFT}^{(\text{exp})} (m) &=  \sum_r m^r \ I_\text{EFT}^{(r)} \,.
\end{align}
$I_\text{EFT}^{(\text{exp})}$ is a scaleless integral, and vanishes. However, we can still pick out the log divergent terms, and write $0$ in the form
eqn~(\ref{3.29b}). 
In general, we have
\begin{align}
I_\text{EFT}^{(r)} &=  -\frac{B^{(r)}}{\eUV} + \frac{B^{(r)}}{\eIR} =0\,,
\end{align}
and there is no finite piece, since the integral vanishes. $B^{(r)}$ is the \emph{same} as in eqn~(\ref{3.31a}), because the two integrals have the same IR divergence, so the $1/\eIR$ terms must agree.

In our example, from eqn~(\ref{3.26}),
\begin{align}
\label{3.32}
I_\text{EFT}^{(0)} &=0\ \hbox{since there is no $m^0/k^4$ term} ,\nn
I_\text{EFT}^{(2)} &=-g^2 \frac{1}{M^2} \mu^{2\epsilon} \int \frac{{\rm d}^d k}{(2\pi)^d}\  \frac{1}{k^4}
= - \frac{i}{16 \pi^2}\frac{1}{M^2}\left[ \frac{1}{\eUV}  - \frac{1}{\eIR} \right] ,
\end{align}
so that
\begin{align}
\label{3.40}
B^{(0)} &= 0, & B^{(2)} &= \frac{ig^2}{16\pi^2}\frac{1}{M^2},
\end{align}
which agree with $B^{(0)}$ and $B^{(2)}$ in  eqn~(\ref{3.36}), as expected. The renormalized expression for $I_F^{(r)}$ is given by adding the full theory counterterm $-A^{(r)} /\eUV$,
\begin{align}
\label{3.36a}
I_F^{(r)} + I_{F,\text{c.t.}}^{(r)} &=  \frac{B^{(r)} }{\eIR} + C^{(r)} \,,
\end{align}
and the renormalized expression for $I_\text{EFT}^{(r)}$ by adding the EFT counterterm $B^{(r)} /\eUV$,
\begin{align}\label{3.36c}
I_\text{EFT}^{(r)} + I_\text{EFT,c.t.}^{(r)}  &=  \frac{B^{(r)} }{\eIR}\,.
\end{align}
Note that one does not cancel IR divergences by counterterms. The difference of eqn~(\ref{3.36a}) and eqn~(\ref{3.36c}) is
\begin{align}
I_M^{(r)} &= \left[ I_F^{(r)} + I_{F,\text{c.t.}}^{(r)}  \right] - \left[ I_\text{EFT}^{(r)} + I_\text{EFT,c.t.}^{(r)} \right] = C^{(r)} .
\end{align}
The infrared divergences cancel between the two, leaving only the finite part $C^{(r)}$. 

\begin{exercisebn}

Compute $I_{\text{EFT}}^{(\text{exp})}$, i.e.\ $I_{\text{EFT}}$ with the IR $m$ scale expanded out. Show that it is a scaleless integral which vanishes. Using the known UV divergence from Exercise~\ref{ex5.2}, write it in the form
\begin{align*}
I_{\text{EFT}}^{(\text{exp})} &=-B \frac{1}{16\pi^2} \left[ \frac{1}{\eUV}-\frac{1}{\eIR} \right]\,,
\end{align*}
and show that the IR divergence agrees with that in $I_F^{(\text{exp})} + I_{F,ct}$.

\end{exercisebn}

\begin{exercisenb}

Compute $\left(I_F^{(\text{exp})} + I_{F,ct}\right)-\left(I_{\text{EFT}}^{(\text{exp})} + I_{\text{EFT},ct} \right)$ and show that all the $1/\epsilon$ divergences (both UV and IR) cancel, and the result is equal to $I_M$ found in Exercise~\ref{ex5.4}.

\end{exercisenb}

This gives the prescription for the matching condition: Expand $I_F$ in IR scales, and keep only the finite part. However, we have obtained some new information. The anomalous dimension in the full theory is proportional to the UV counterterm $-A$. The anomalous dimension in the EFT is proportional to the EFT counterterm $B$, which can be different from $A$. By the argument just given, $B$ is the IR divergence of the full theory. By using an EFT, we have converted IR divergences (i.e.\ the $\log m$ terms) in the full theory into UV divergences in the EFT. This converts IR logs into UV logs, which can be summed using the renormalization group. In the EFT, $\log M/m$ terms in the full theory are converted to $\log \bar \mu/m$ terms, since $M \to \infty$ in the EFT. These are summed by the EFT renormalization group equations.

\begin{exercise}

Make sure you understand why you can compute $I_M$ simply by taking $I_F^{(\text{exp})} $ and dropping all $1/\epsilon$ terms (both UV and IR).

\end{exercise}

Finally, if we do the EFT calculation without expanding out the IR scale $m$, then the EFT calculation is no longer IR divergent and can have a finite part,
\begin{align}
I_\text{EFT}^{(r)} &=  -\frac{B^{(r)}}{\eUV}  + D^{(r)}\,,
\end{align}
where the UV divergence remains the same as before. The finite part of the full amplitude $I_F$ has been split into $C^{(r)}+D^{(r)}$, with $C^{(r)}$ from the matching and $D^{(r)}$ from the EFT. In our example,
\begin{align}
D^{(0)} &= 0, \nn
D^{(2)} &= \frac{ig^2}{16\pi^2} \frac{1}{M^2}\left[ \log \frac{m^2}{\mu^2}-1 \right],
\end{align}
from eqn~(\ref{3.20}).

\section{Summary}

It has taken a while to get to the final answer, but we can now summarize our  results. The general procedure is simple to state:
\begin{itemize}
\item Compute the full theory graphs expanding in all IR scales. The integrals are single-scale integrals involving only the high scale $M$. Drop the $1/\epsilon$ terms from both UV and IR divergences. This gives $C^{(r)}(\mu)$. To avoid large logarithms, $\mu$ should be chosen to be of order the high scale $M$. The starting values of the EFT coefficient at the high scale are $C^{(r)}(\mu\sim M)$.
\item Evolve the EFT down from $\mu \sim M$ to a low scale $\mu \sim m$ using the renormalization group equations in the EFT. This sums logs of the ratios of scales, $\ln M/m$.
\item Compute in the EFT using $\mu \sim m$. There are no large logs in the EFT calculation.
\item Combine the pieces to get the final result.
\end{itemize}
One computation has been broken  up into several much simpler calculations, each of which involves a single scale.

\begin{exercisebn}

Compute the QED on-shell electron form factors $F_1(q^2)$ and $F_2(q^2)$ expanded to first order in $q^2/m^2$ using dimensional regularization to regulate the IR and UV divergences.  This gives the one-loop matching to heavy-electron EFT.   Note that it is much simpler to \emph{first} expand and then do the Feynman parameter integrals. A more difficult version of the problem is to compute the on-shell quark form factors in QCD, which gives the one-loop matching to the HQET Lagrangian. For help with the computation, see Ref.~\cite{Manohar:1997qy}. Note that in the non-Abelian case, using background field gauge is helpful because the amplitude respects gauge invariance on the external gluon fields.

\end{exercisebn}


\begin{exercisenn}

\item The SCET matching for the vector current $\overline \psi \gamma^\mu \psi$ for the Sudakov form factor is a variant of the previous problem. Compute $F_1(q^2)$ for on-shell massless quarks,  in pure dimensional regularization with $Q^2=-q^2 \not =0$. Here $Q^2$ is the big scale, whereas in the previous problem $q^2$ was the small scale. The spacelike calculation $Q^2>0$ avoids having to deal with the $+i 0^+$ terms in the Feynman propagator which lead to imaginary parts. The timelike result can then be obtained by analytic continuation.

\end{exercisenn}

\begin{exercisenb}

\item Compute the SCET matching for timelike $q^2$, by analytically continuing the previous result. Be careful about the sign of the imaginary parts.

\end{exercisenb}

\section{RG Improved Perturbation Theory}\label{sec:rge}

We have mentioned several times that renormalization group improved perturbation theory is better than fixed order perturbation theory. To understand the difference, consider an example where an operator coefficient $c(\mu)$ satisfies the one-loop renormalization group equation
\begin{align}
\label{3.45}
\mu \frac{\rd}{\rd \mu}c(\mu) &=\left[ \gamma_0  \frac{g^2(\mu)}{16\pi^2}   + \mathcal{O}\left(\frac{g^2(\mu)}{16\pi^2} \right)^2 \right] c(\mu) ,
\end{align}
where $\gamma_0$ is a constant. The evolution of $g(\mu)$ is given by the $\beta$-function equation
\begin{align}\label{3.46}
\mu \frac{\rd g(\mu)}{\rd \mu} &= -b_0 \frac{g^3(\mu)}{16\pi^2} +  \mathcal{O}\left[\frac{g^5(\mu)}{(16\pi^2)^2}  \right] .
\end{align}
As long as $g^2(\mu)/(16\pi^2)$ is small, we can integrate the ratio of eqn~(\ref{3.45}) and eqn~(\ref{3.46}) to get
\begin{align}
\label{3.47}
\frac{c(\mu_1)}{c(\mu_2)} &= \left[ \frac{\alpha_s(\mu_1)}{\alpha_s(\mu_2)} \right]^{-\gamma_0/(2b_0)}, & \alpha_s(\mu) &= \frac{g^2(\mu)}{4\pi}.
\end{align}
Integrating eqn~(\ref{3.45},\ref{3.46}) term by term, or equivalently, expanding eqn~(\ref{3.47}) gives
\begin{align}
\label{3.48}
\frac{c(\mu_1)}{c(\mu_2)} &= 1 + \gamma_0 \frac{\alpha_s(\mu_1)}{4\pi} \log \frac{\mu_1}{\mu_2} 
-\frac12 \gamma_0 (2b_0 -\gamma_0) \left[\frac{\alpha_s(\mu_1)}{4\pi} \log \frac{\mu_1}{\mu_2} \right]^2 \nn
& +\frac16 \gamma_0 (2b_0 -\gamma_0) (4b_0 -\gamma_0) \left[\frac{\alpha_s(\mu_1)}{4\pi} \log \frac{\mu_1}{\mu_2} \right]^3 + \ldots
\end{align}
The renormalization group sums the leading log (LL) series $\alpha_s^n \log^n$, as can be seen from eqn~(\ref{3.48}). One can show that the higher order corrections in eqn~(\ref{3.45},\ref{3.46}) do not contribute to the leading log series, since they are suppressed by $\alpha_s/(4\pi)$ without a log. Including the two-loop terms gives the next-to-leading-log (NLL) series $\alpha_s^n \log^{n-1}$, the three-loop terms give the NNLL series $\alpha_s^n \log^{n-2}$, etc.

The change in $g(\mu)$ and $c(\mu)$ can be very large, even if $\alpha_s(\mu)$ is small. For example, in the strong interactions, $\alpha_s(M_Z) \approx 0.118$ and $\alpha_s(m_b) \approx 0.22$, a ratio of about two. Even though both values of $\alpha_s$ are small, weak decay operator coefficients also change by about a factor of two between $M_Z$ and $m_b$, as shown below.

\subsection{Operator Mixing}

Summing logs using the renormalization group equations allows us to include operator mixing effects in a systematic way. This is best illustrated by the simple example of non-leptonic weak 
$b \to c$ decays via the effective Lagrangian
\begin{align}
L = -\frac{4 G_F}{\sqrt2}V_{cb} V_{ud}^*\
\left(c_1 O_1 + c_2 O_2 \right),
\end{align}
where the two operators and their tree-level coefficients at $\mu=M_W$ are
\begin{align}
O_1 &=\left(\bar c^\alpha\, \gamma^\mu\, P_L\, b_\alpha\right)\left(\bar d^\beta\, \gamma_\mu\, P_L\, u_\beta
\right), & c_1 &= 1 + \mathcal{O}\left(\alpha_s\right), \\
O_2 &=\left(\bar c^\alpha\, \gamma^\mu\, P_L\, b_\beta\right)\left(\bar d^\beta\, \gamma_\mu\, P_L\, u_\alpha\
\right),& c_2 &=0+ \mathcal{O}\left(\alpha_s\right),
\end{align}
where $\alpha$ and $\beta$ are color indices. Since the $W$ boson is color-singlet, only $O_1$ is produced by the tree-level graph. $O_2$ is generated by loop graphs involving gluons, which are suppressed by a power of $\alpha_s$.

\begin{figure}
\centering
\includegraphics[width=2cm]{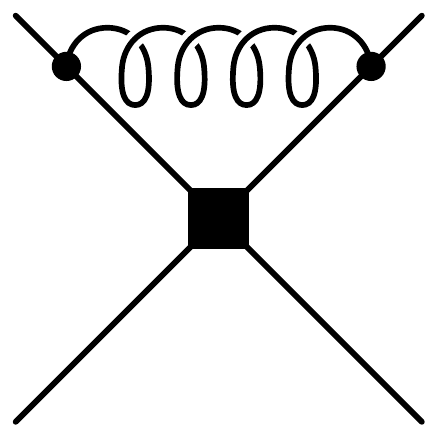}
\caption{\label{fig:anomdim}. Graph contributing to the anomalous dimension of $O_1$ and $O_2$. One has to sum over gluon exchange between all possible pairs of lines, and also include wavefunction corrections.}
\end{figure}

The renormalization group equations can be computed from the one-loop graph in Fig.~\ref{fig:anomdim}~\cite{Gilman:1979bc},
\begin{align}
\label{5.52}
\mu \frac{\rd}{\rd \mu} \left[ \begin{array}{cc} c_1 \\ c_2 \end{array} \right]
=  \frac{\alpha_s}{4\pi} \left[ \begin{array}{cc} -2 & 6 \\ 6 & -2 \end{array} \right]  \left[ \begin{array}{cc} c_1 \\ c_2 \end{array} \right].
\end{align}

\begin{exercise}

Compute the anomalous dimension mixing matrix in eqn~(\ref{5.52}).

\end{exercise}

The anomalous dimension matrix is not diagonal, which is referred to as operator mixing. In this simple example, the equations can be integrated by taking the linear combinations $c_\pm = c_1 \pm c_2$,
\begin{align}
\mu \frac{\rd}{\rd \mu} \left[ \begin{array}{cc} c_+ \\ c_- \end{array} \right]
&=  \frac{\alpha_s}{4\pi} \left[ \begin{array}{cc} 4 & 0 \\ 0 & -8 \end{array} \right]  \left[ \begin{array}{cc} c_+ \\ c_- \end{array} \right],
\end{align}
which decouples the equations. The solution is
\begin{align}
\label{3.54}
\frac{c_+(\mu_1)}{c_+(\mu_2)} &= \left[ \frac{\alpha(\mu_1)}{\alpha(\mu_2)} \right]^{-6/23}, &
\frac{c_-(\mu_1)}{c_-(\mu_2)} &= \left[ \frac{\alpha(\mu_1)}{\alpha(\mu_2)} \right]^{12/23},
\end{align}
using eqn~(\ref{3.47}), with $b_0=11-2/3 n_f=23/3$ and $n_f=5$ dynamical quark flavors. With $\alpha_s(m_b) \sim 0.22$ and $\alpha_s(M_Z) \sim 0.118$,
\begin{align}
\label{3.55}
\frac{c_+(m_b)}{c_+(M_W)} &= 0.85, &
\frac{c_-(m_b)}{c_-(M_W)} &=  1.38,
\end{align}
so that
\begin{align}
\label{3.55a}
c_1(m_b) &\approx 1.12, & c_2(m_b) &\approx -0.27\,.
\end{align}
A substantial $c_2$ coefficient is obtained at low scales, even though the starting value is $c_2(M_W)=0$. 

Equation~(\ref{3.48}) for the general matrix case is
\begin{align}
\label{3.58}
\mathbf{c}(\mu_1) &= \biggl[ 1 + \bm{\gamma}_0 \frac{\alpha_s(\mu_1)}{4\pi} \log \frac{\mu_1}{\mu_2} 
-\frac12 \bm{\gamma}_0 (2b_0 -\bm{\gamma}_0) \left[\frac{\alpha_s(\mu_1)}{4\pi} \log \frac{\mu_1}{\mu_2} \right]^2 \nn
& +\frac16 \bm{\gamma}_0 (2b_0 -\bm{\gamma}_0) (4b_0 - \bm{\gamma}_0) \left[\frac{\alpha_s(\mu_1)}{4\pi} \log \frac{\mu_1}{\mu_2} \right]^3 + \ldots \biggr] \mathbf{c}(\mu_2),
\end{align}
where $\bm{\gamma}_0$ is a matrix and $\mathbf{c}$ is a column vector. Equation~(\ref{3.58}) shows that $c_2(m_b)$ in eqn~(\ref{3.55a}) is a leading-log term, even though it starts at $c_2(M_W)=0$.  In examples with operator mixing, it is difficult to obtain the leading-log series eqn~(\ref{3.58}) by looking at graphs in the full theory. The method used in practice to sum the leading-log series is by integrating anomalous dimensions in the EFT.

The above discussion of renormalization group equations and operator mixing also holds in general EFTs. The EFT Lagrangian is an expansion in higher dimension operators,
\begin{align}
\LL &=  \LL_{\opdim \le 4} + \frac{1}{\Lambda} c^{(5)}_i O^{(5)}_i + \frac{1}{\Lambda^2} c^{(6)}_i O^{(6)}_i  + \ldots\,.
\end{align}
The running of the coupling constants in $\LL_{\opdim \le 4}$ is given by the usual $\beta$-functions of the low-energy theory, e.g.\ by the QCD and QED $\beta$-functions. The other terms in $\LL$ are higher dimension operators, and their anomalous dimensions are computed in the same way as eqn~(\ref{5.52}) for the weak interactions. The additional piece of information we have is the EFT power counting formula. This leads to RGE equations of the form
\begin{align}
\mu \frac{\rd}{\rd \mu} c^{(5)}_i  &= \gamma^{(5)}_{ij} c^{(5)}_j\,, \nn
\mu \frac{\rd}{\rd \mu} c^{(6)}_i  &= \gamma^{(6)}_{ij} c^{(6)}_j +  \gamma_{ijk}\, c^{(5)}_j c^{(5)}_k\,,
\end{align}
and in general
\begin{align}
\mu \frac{\rd}{\rd \mu} c^{(D)}_i  &= \gamma_{i j_1 j_2 \ldots j_r} c^{(D_1)}_{j_1} \ldots c^{(D_r)}_{j_r} \,,
\end{align}
with $D-4 = \sum_i (D_i-4)$, where the anomalous dimensions $\gamma$ are functions of the coupling constants in $\LL_{\opdim \le 4}$. The renormalization group equations are \emph{non-linear}. Graphs with two insertions of a dimension-five operator need a dimension-six counterterm leading to the $c^{(5)}_j c^{(5)}_k$ term in the anomalous dimension for $c^{(6)}_i$, etc.  In the presence of mass terms such as $m_H^2$, one also gets mixing to $D-4 < \sum_r (D_r-4)$ operators, e.g. 
\begin{align}
\mu \frac{\rd}{\rd \mu} c^{(4)}_i  &= m_H^2 \gamma^{(6 \to 4)}_{ij} c^{(6 )}_j + \ldots \,.
\end{align}
as in SMEFT~\cite{Jenkins:2013zja}.

\chapter{Field Redefinitions and Equations of Motion}

\section{LSZ Reduction Formula}\label{sec:LSZ}

\begin{figure}
\begin{center}
\includegraphics[width=2.5cm]{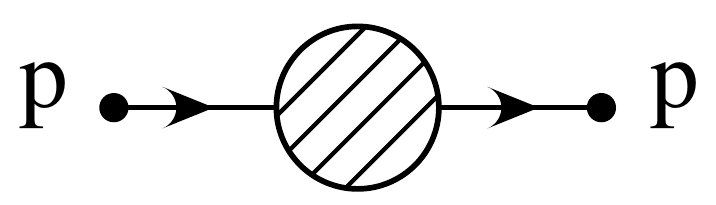}
\caption{\label{fig:prop} Two-point function $D(p)$.}
\end{center}
\end{figure}

Experimentally observable quantities in field theory are $S$-matrix elements, whereas what one computes from the functional integral are correlation functions of quantum fields. The LSZ reduction formula relates the two. For simplicity, we discuss a theory with  a scalar field $\phi(x)$. The momentum space Green's functions are defined by
\begin{figure}
\begin{center}
\includegraphics[width=4cm]{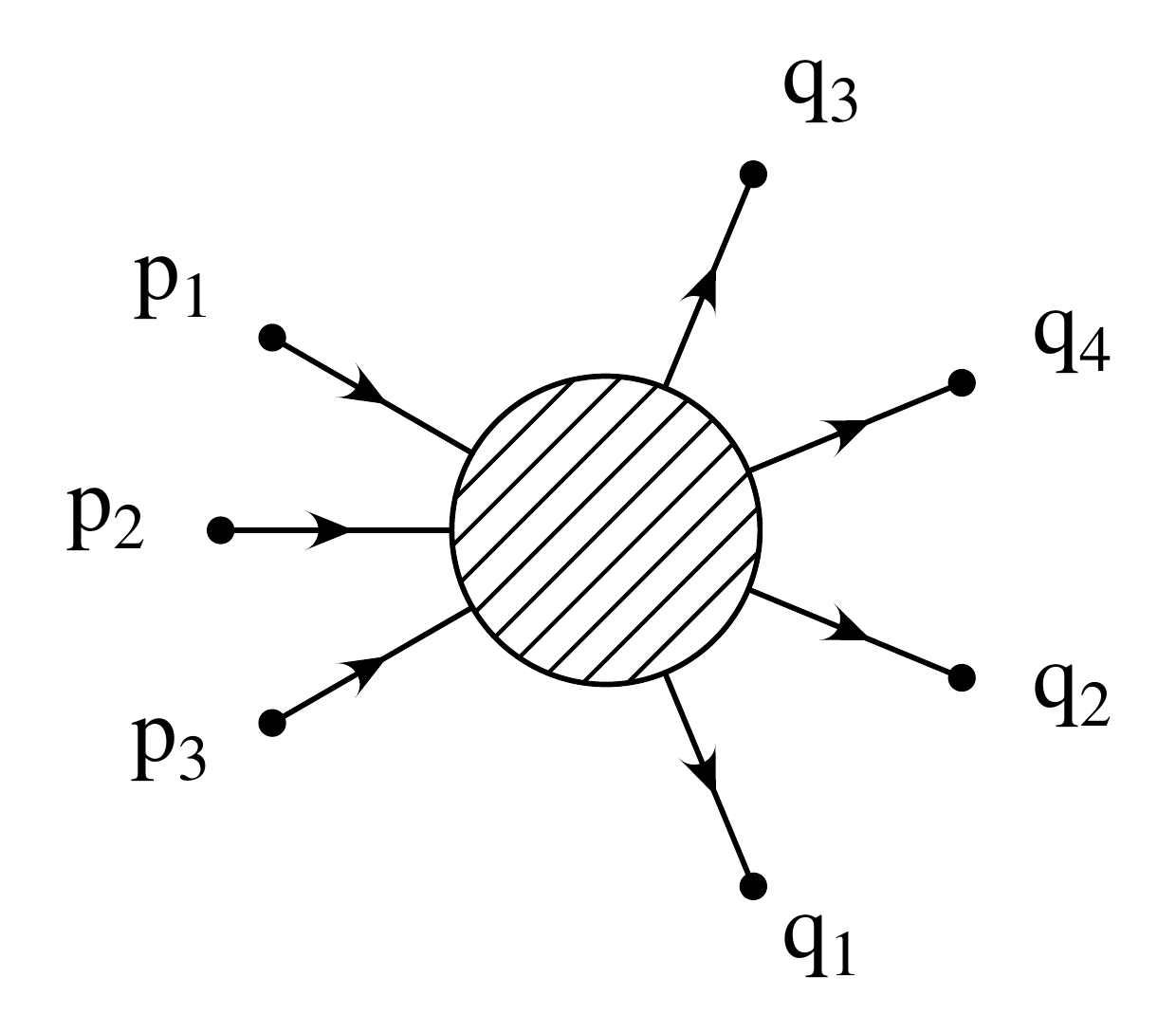}
\end{center}
\caption{\label{fig:green} Green's function with 3 incoming particles and 4 outgoing particles.}
\end{figure}
\begin{align}
\label{6.1}
&G(q_1,\ldots,q_m;p_1,\ldots,p_n) \nn
&=
\prod_{i=1}^m  \int \rd^4 y_i\ e^{i q_i \cdot y_i}\prod_{j=1}^n \int \rd^4 x_j\ e^{-i p_j \cdot x_j} \braket{0| T \left\{ \phi(y_1) \ldots \phi(y_m) \phi (x_1) \ldots \phi(x_n) \right\} |0}
\end{align}
where the momenta $p_i$ are incoming, and momenta $q_i$ are outgoing, as shown in Fig.~\ref{fig:green}. These Green's functions can be computed in perturbation theory using the usual Feynman diagram expansion. The $\phi$ propagator in Fig.~\ref{fig:prop} is a special case of eqn~(\ref{6.1}),
\begin{align}
\label{6.2}
D(p) &= \int \rd^4 x\ e^{i p \cdot x} \braket{0| T \left\{ \phi (x)  \phi (0) \right\} |0}\,.
\end{align}
If the field $\phi(x)$ can produce a single particle state $\ket{p}$ with invariant mass $m$ from the vacuum,
\begin{align}
\label{6.3}
\braket{p | \phi(x)|0} \not=0\,,
\end{align}
then the propagator $D(p)$ has a pole at $p^2=m^2$,
\begin{align}
D(p) &\sim  \frac{i \, \mathcal{R}}{p^2-m^2+i \epsilon} + \text{non-pole terms}.
\end{align}
$\phi$ is called an interpolating field for $\ket{p}$.
The wavefunction  factor $\mathcal{R}$ is defined by
\begin{align}
\lim_{\substack{p^2\to m^2 \\ p^0 > 0}} \left(p^2-m^2\right) D(p) & \equiv i \, \mathcal{R}\,.
\end{align}
$\mathcal{R}$ is finite, since $D(p)$, the renormalized propagator, is finite.

The $S$-matrix is computed from the Green's function by picking out the poles for each particle,
\begin{align}\label{6.7}
& \lim_{\substack{q_i^2\to m^2 \\ q_i^0 > 0} } \lim_{\substack{p_j^2\to m^2 \\ p_j^0 > 0}}\prod_{i=1}^m  \left(q_i^2-m^2\right)   \prod_{j=1}^n \left(p_j^2-m^2\right) G(q_1,\ldots,q_m;p_1,\ldots,p_n) \nn
&=\prod_{i=1}^m  \left(i \sqrt \mathcal{R}_i \right) \prod_{j=1}^n \left(i\, \sqrt \mathcal{R}_j \right) \ \ {}_{\text{out}}\! \braket{q_1,\ldots,q_m |p_1,\ldots, p_n }_{\text{in}},
\end{align}
i.e.\ the $n+m$ particle pole of the Green's function gives the $S$-matrix up to wavefunction normalization factors. Equation~(\ref{6.7}) is called the LSZ reduction formula~\cite{Lehmann:1954rq}. The only complication for fermions and gauge bosons is that one has to contract with spinors $u(p,s), v(p,s)$ and polarization vectors $\epsilon^{(s)}_\mu(p)$.

The important feature of eqn~(\ref{6.7}) is that the derivation only depends on eqn~(\ref{6.3}), so that any interpolation field can be used.  Particle states are given by the physical spectrum of the theory, and  Green's functions are given by correlation functions of fields. $S$-matrix elements, which are the physical observables, depend on particle states, not fields. Fields and particles are not the same. 

\section{Field Redefinitions}

It is now easy to see why field redefinitions do not change the $S$-matrix. The LSZ reduction formula does not care what field is used. To understand this in more detail, consider the functional integral
\begin{align}
\label{6.7a}
Z[J] &= \int D \phi\ e^{ i \int L[\phi]  + J \phi}.
\end{align}
The Green's functions 
\begin{align}
\braket{0| T \left\{ \phi(x_1) \ldots \phi(x_r)\right\} |0} &=\frac{ \int D \phi\ \phi(x_1) \ldots \phi(x_r) \ e^{ i S(\phi)} }{ \int D \phi\ e^{ i S(\phi)} } \,,
\end{align}
are given by
\begin{align}
\braket{0| T \left\{ \phi(x_1) \ldots \phi(x_r)\right\} |0} &= \left. \frac{1}{Z[J]}\ \frac{\delta}{i \, \delta J(x_1)}\ldots \frac{\delta}{i \, \delta J(x_r)}\ Z[J] \ \right|_{J=0} \,.
\end{align}

Consider a local field redefinition,
\begin{align}
\label{6.10}
\phi(x) &= F[ \phi^\prime(x) ]\,,
\end{align}
such as
\begin{align}
\phi(x) &= \phi^\prime(x)+c_1 \partial^2\phi^\prime(x) +c_2 \phi^\prime(x)^3\,.
\end{align}
The field redefinition $F[ \phi^\prime(x) ]$ can involve  integer powers of $\phi$ and a finite number of derivatives. 
Then $L^\prime$ defined by
\begin{align}
L[\phi(x)]=L[F[\phi^\prime(x))]=L^\prime[\phi^\prime(x)]\,,
\end{align}
is the new Lagrangian after the field redefinition eqn~(\ref{6.10}). 

The functional integral $Z^\prime$ with the new field $\phi^\prime(x)$ and
Lagrangian $L^\prime$
\begin{align}
\label{6.7aa}
Z^\prime [J] &= \int D \phi^\prime\ e^{ i \int L^\prime[\phi^\prime]  + J \phi^\prime} =  \int D \phi\ e^{ i \int L^\prime[\phi]  + J \phi}\,,
\end{align}
gives correlation functions of $\phi^\prime$ computed using $L^\prime[\phi^\prime]$, or equivalently, correlation functions of $\phi$ computed using $L^\prime[\phi]$, since $\phi^\prime$ is a dummy integration variable and can be replaced by $\phi$.
The original functional integral eqn~(\ref{6.7a}) under the change of variables eqn~(\ref{6.10}) becomes
\begin{align}\label{6.13}
Z[J] &= \int D \phi^\prime\  \abs{\frac{\delta F}{\delta \phi^\prime}} e^{ i \int L^\prime[\phi^\prime]  + J F[\phi^\prime]}\,.
\end{align}
The Jacobian $\abs{{\delta F}/{\delta \phi^\prime}}$ is unity in dimensional regularization, except for the special case of a fermionic
chiral transformation, where there is an anomaly~\cite{Fujikawa:1979ay}.  Neglecting anomalies, and dropping  primes on the dummy variable $\phi^\prime$ gives
\begin{align}\label{6.14}
Z[J] &= \int D \phi\   e^{ i \int L^\prime[\phi]  + J F[\phi]}.
\end{align}
Thus $Z[J]$, which gives the Green's functions of $\phi$ computed using Lagrangian $L[\phi]$ by eqn~(\ref{6.7a}), also gives the Green's functions of $F[\phi]$ computed using Lagrangian $L^\prime[\phi]$. In contrast, $Z^\prime[J]$ gives the correlation functions of $\phi$ computed using the new Lagrangian  $L^\prime[\phi]$. The two correlation functions are different, so Green's functions change under a field redefinition. However, the $S$-matrix remains unchanged. $Z[J]$ computes the $S$-matrix using Lagrangian $L^\prime[\phi]$ and $F[\phi]$ as the interpolating field, by eqn~(\ref{6.14}). $Z^\prime[J]$ computes
the $S$-matrix using Lagrangian $L^\prime[\phi]$ and $\phi$ as the interpolating field, by eqn~(\ref{6.7aa}). The $S$-matrix does not care about the choice of interpolating field (i.e.\ field redefinition) as long as
\begin{align}
\braket{p | F[\phi] | 0} \not= 0,
\end{align}
so a field redefinition leaves the $S$-matrix unchanged.

In field theory courses, we study renormalizable Lagrangians with terms of dimension $\leqslant 4$. The only field redefinitions allowed are linear transformations,
\begin{align}
\phi_i^\prime &= C_{ij}\ \phi_j \,.
\end{align}
These are used to put the kinetic term in canonical form,
\begin{align}
\frac12 \partial_\mu \phi_i \, \partial^\mu \phi^i.
\end{align}

In an EFT, there is much more freedom to make field redefinitions, since the Lagrangian includes higher dimensional operators. One makes field redefinitions that respect the EFT power counting, e.g.\
\begin{align}
\phi \to \phi + \frac{1}{\Lambda^2} \phi^3 + \ldots
\end{align}
and work order by order in $1/\Lambda$. Field redefinitions are often used to put EFT Lagrangians in canonical form. The EFT Lagrangian is then given by matching from the full theory, followed by a field redefinition, so fields in the EFT are not the same as in the full theory.

\section{Equations of Motion}

A special case of field redefinitions is the use of equations of motion~\cite{Georgi:1991ch,Politzer:1980me}. Let $E[\phi]$ be the \emph{classical} equation of motion
\begin{align}
E[\phi] &\equiv \frac{\delta S}{\delta \phi}.
\end{align}
For example, if
\begin{align}
\LL &= \frac 12 \partial_\mu \phi \partial^\mu \phi - \frac 12 m^2 \phi^2 - \frac{1}{4!} \lambda \phi^4,
\end{align}
$E[\phi]$ is
\begin{align}
E[\phi] &=  - \partial^2 \phi(x) - m^2 \phi(x) -  \frac{1}{3!} \lambda \phi^3(x)\,.
\end{align}
Let $\theta$ be an operator with a factor of the classical equation of motion,
\begin{align}\label{6.23}
\theta[\phi] &= F[\phi] E[\phi] = F[\phi]  \frac{\delta S}{\delta \phi},
\end{align}
and consider the functional integral
\begin{align}
\label{6.24}
Z[J,\widetilde J] &= \int D \phi\ e^{ i \int L[\phi]  + J\, \phi +\widetilde J \theta[\phi]} .
\end{align}
The correlation function
\begin{align}
\braket{0| T \left\{  \phi(x_1) \ldots \phi(x_n) \theta (x) \right\} | 0}
\end{align}
with one insertion of the equation-of-motion operator $\theta$ is given by evaluating
\begin{align}
\braket{0| T \left\{  \phi(x_1) \ldots \phi(x_n) \theta (x) \right\} | 0} &=\left. \frac{1}{Z[J,\widetilde J]}\ \frac{\delta}{i \, \delta J(x_1)}\ldots \frac{\delta}{i \, \delta J(x_r)} 
\frac{\delta}{i \, \delta \widetilde J(x)}\  Z[J,\widetilde J]\ \right|_{J=\widetilde J=0}\,.
\end{align}
Make the change of variables
\begin{align}
\phi &= \phi^\prime - \widetilde J F[\phi^\prime] 
\end{align}
in the functional integral eqn~(\ref{6.24}),
\begin{align}
Z[J,\widetilde J] &=  
 \int D \phi^\prime \abs{\frac{\delta \phi}{\delta \phi^\prime}}  \ e^{ i  \int L[\phi^\prime]-\left.\frac{\delta S}{\delta \phi} \right|_{\phi^\prime}\widetilde J F[\phi^\prime]   + J \phi^\prime - J \widetilde J F[\phi^\prime] + \widetilde J \theta[\phi^\prime] + 
 \mathcal{O}(\widetilde J)^2} , \nn
&=  
 \int D \phi^\prime \abs{\frac{\delta \phi}{\delta \phi^\prime}}  \ e^{ i \int L[\phi^\prime] + J \phi^\prime - J \widetilde J F[\phi^\prime] + 
 \mathcal{O}(\widetilde J)^2} ,
\end{align}
by eqn~(\ref{6.23}).
The Jacobian
\begin{align}
\abs{\frac{\delta \phi(x)}{\delta \phi^\prime(y)}}   &= \det\left[ \delta(x-y) - \widetilde J\ 
\frac{\delta F[\phi^\prime(x)]}{\delta \phi^\prime(y)}\right]\,,
\end{align}
is unity in dimensional regularization. Relabeling the dummy integration variable as $\phi$ gives
\begin{align}
\label{6.30}
Z[J,\widetilde J] &=  
 \int D \phi  \ e^{ i \int L[\phi]+ J \phi - J \widetilde J F[\phi]  + 
 \mathcal{O}(\widetilde J)^2} .
\end{align}
Taking the $\widetilde J$ derivative and setting $\widetilde J=0$ gives, by using the equality of eqn~(\ref{6.24}) and eqn~(\ref{6.30}),
\begin{align}\label{6.31}
 \int D \phi\ \theta(x)\ e^{ i \int L[\phi]+ J \phi } &= -\int D \phi\ J(x) F[\phi(x)]\ 
 e^{ i  \int L[\phi]+ J \phi } \,.
\end{align}
Differentiating multiple times w.r.t.\ $J$ gives the equation-of-motion Ward identity
\begin{align}\label{6.32}
& \braket{0 | T \left\{ \phi(x_1) \ldots \phi (x_n) \theta(x) \right\} | 0}  \nn
&= i \sum_r  \delta(x-x_r)
\braket{0|T \left\{ \phi(x_1) \ldots \cancel{\phi(x_r)} \ldots \phi (x_n) F[\phi(x_r)] \right\} | 0} .
\end{align}
The $S$ matrix element with an insertion of $\theta$ vanishes,
\begin{align}\label{6.33}
{}_{\text{out}}\! \braket{q_1,\ldots,q_m |\theta|p_1,\ldots, p_n }_{\text{in}} = 0\,,
\end{align}
because it is given by picking out the term with $m+n$ poles on the l.h.s. of eqn~(\ref{6.32}). But the r.h.s.\ shows that the matrix element of the $r{}^{\text{th}}$ term has no pole in $p_r$, because of the $\delta$ function. Each term in the sum vanishes, leading to eqn~(\ref{6.33}). As a result, equation-of-motion operators can be dropped because they do not contribute to the $S$-matrix. 

Note that eqn~(\ref{6.33}) implies that the \emph{classical} equations of motion can be dropped. The equations of motion have quantum corrections, but the Ward identity eqn~(\ref{6.33}) is for the classical equations of motion without the quantum corrections. The Ward identity holds even for insertions of the equation-of-motion operator in loop graphs, where the particles are off-shell, and do not satisfy the classical equations of motion.

Using the equations of motion is a special case of a field redefinition. Consider the field redefinition (with $\epsilon \ll 1$):
\begin{align}\label{6.34}
\phi(x) &= \phi^\prime(x)+\epsilon\, F[\phi^\prime(x)]\,.
\end{align}
The change in the Lagrangian due to eqn~(\ref{6.34}) is
\begin{align}\label{6.35}
L[\phi] &= L[\phi^\prime] +\epsilon \,F[\phi^\prime] \frac{\delta S[\phi^\prime]}{\delta \phi^\prime}+ \mathcal{O}\left(\epsilon^2\right)= L[\phi^\prime] +\epsilon \, \theta[\phi^\prime] + + \mathcal{O}\left(\epsilon^2\right)\,.
\end{align}
We have already seen that a field redefinition leaves the $S$-matrix invariant. Thus the $S$-matrix computed with the new Lagrangian $L^\prime [\phi] = L[\phi] + \epsilon \theta [\phi]$ is the same as that computed with $L[\phi]$.\footnote{Remember $\phi$ is a dummy variable, so we can use $L^\prime[\phi]$ instead of $L^\prime[\phi^\prime]$.} Thus we can shift the Lagrangian by equation-of-motion terms. The way equations-of-motion are  used in practice is to eliminate operators with derivatives in the EFT Lagrangian.

\begin{exercise}

The classical equation of motion for $\lambda \phi^4$ theory,
\begin{align*} 
L &= \frac12 (\partial_\mu \phi)^2 - \frac 12 m^2 \phi^2 - \frac{\lambda}{4!} \phi^4\,,
\end{align*}
is
\begin{align*} 
E[\phi] &= (-\partial^2-m^2)\phi - \frac{\lambda}{3!} \phi^3\,.
\end{align*}
The EOM Ward identity for $\theta = F[\phi] E$ is eqn~(\ref{6.32}). Integrate both sides
with
\begin{align*}
\int \rd x\ e^{-i q \cdot x} \prod_i \int \rd x_i\ e^{-i p_i \cdot x_i}
\end{align*}
to get the momentum space version of the Ward identity
\begin{align*}
\braket{0 | T \left\{ \widetilde \phi(p_1) \ldots \widetilde \phi(p_n)  \widetilde \theta(q) \right \} | 0} &= i \sum_{r=1}^n 
\braket{0 | T \left\{ \widetilde \phi(p_1) \ldots \cancel{\widetilde \phi(p_r)} \ldots \widetilde \phi(p_n) \widetilde  F(q+p_r)  \right\} | 0}\,.
\end{align*}
(a) Consider the equation of motion operator
\begin{align*}
\theta_1 &= \phi \, E[\phi] = \phi (-\partial^2-m^2)\phi - \frac{\lambda}{3!} \phi^4\,,
\end{align*}
and verify the Ward identity by explicit calculation at order $\lambda$ (i.e.\ tree level) for $\phi \phi$ scattering, i.e. for $\phi \phi \to \phi \phi$. \\
(b) Take the on-shell limit $p_r^2 \to m^2$ at fixed $q \not = 0$ of
\begin{align*}
\prod_r (-i) (p_r^2-m^2) \times \text{Ward Identity}\,,
\end{align*}
and verify that both sides of the Ward identity vanish. Note that both sides do not vanish if one first takes $q=0$ and then takes the on-shell limit. \\
(c) Repeat the above calculation to order $\lambda^2$, i.e.\ one loop. \\
(d) Repeat to one loop for the equation of motion operator 
\begin{align*}
\theta_2 &= \phi^3 \, E[\phi] = \phi^3 (-\partial^2-m^2)\phi - \frac{\lambda}{3!} \phi^6\,.
\end{align*}

\end{exercise}

As an example of the use of the equations-of-motion, suppose we have an EFT Lagrangian
\begin{align}\label{6.36}
\LL &= \frac 12 \partial_\mu \phi \partial^\mu \phi - \frac 12 m^2 \phi^2 - \frac{1}{4!} \lambda \phi^4 + \frac{c_1}{\Lambda^2} \phi^3 \partial^2 \phi
+  \frac{c_6}{\Lambda^2} \phi^6 + \ldots\,.
\end{align}
Then making the field redefinition
\begin{align}\label{6.37}
\phi \to \phi +  \frac{c_1}{\Lambda^2} \phi^3\,,
\end{align}
gives the new Lagrangian
\begin{align}\label{6.38}
\LL &= \frac 12 \partial_\mu \phi \partial^\mu \phi - \frac 12 m^2 \phi^2 - \frac{1}{4!} \lambda \phi^4 + \frac{c_1}{\Lambda^2} \phi^3 \partial^2 \phi
+  \frac{c_6}{\Lambda^2} \phi^6 \nn
&+ \frac{c_1}{\Lambda^2}\phi^3\left[ -\partial^2\phi - m^2 \phi - \frac{\lambda}{3!} \phi^3 \right]  + \ldots \nn
 &= \frac 12 \partial_\mu \phi \partial^\mu \phi - \frac 12 m^2 \phi^2 - \left[ \frac{1}{4!} \lambda  + \frac{c_1}{\Lambda^2} m^2 \right]\phi^4 
+  \left[ \frac{c_6}{\Lambda^2}  - \frac{c_1}{\Lambda^2} \frac{\lambda}{3!} \right] \phi^6 + \ldots\,.
\end{align}
The two Lagrangians eqn~(\ref{6.36}) and eqn~(\ref{6.38}) give the same $S$-matrix. In eqn~(\ref{6.38}), we have eliminated the $\phi^3\partial^2 \phi$ operator at the cost of redefining the coefficients of the $\phi^4$ and $\phi^6$ operators. The EFT power counting has been maintained in going from eqn~(\ref{6.36}) to eqn~(\ref{6.38}).  It is easier to do computations with eqn~(\ref{6.38}) rather than eqn~(\ref{6.36}), because eqn~(\ref{6.38}) has fewer independent operators. In EFTs, one usually applies the equations of motion to eliminate as many operators with derivatives as possible. 

The calculation above only retained terms up to dimension six. If one works to dimension eight, one has to retain the terms quadratic in $c_1/\Lambda^2$ in the transformed Lagrangian. These terms are second order in the equation of motion. Working to second order in the  equations of motion is tricky~\cite{Jenkins:2017dyc,Manohar:1997qy,Manohar:1993qn}, and it is best to systematically use field redefinitions to eliminate operators to avoid making mistakes.

Using field redefinitions rather than the equations of motion also clears up some subtleties. For example, the fermion kinetic term is
\begin{align}
\overline \psi\, i \slashed{D}\, \psi .
\end{align}
This operator vanishes using the fermion equation of motion $i \slashed{D}\, \psi  = 0$. However, it is not possible to eliminate this term by a field redefinition, so one cannot eliminate the fermion kinetic energy using the equations of motion. One can eliminate higher order terms such as $\phi^2 \overline \psi\, i \slashed{D}\, \psi $. Another interesting example is given in Ref.~\cite{Jenkins:2017dyc}.

\begin{exercisebn}

Write down all possible $C$-even dimension six terms in eqn~(\ref{2.21}), and show how they can be eliminated by field redefinitions.

\end{exercisebn}

\begin{exercisenb}

Take the heavy quark Lagrangian
\begin{align*}
{\mathcal L}_v &= \bar Q_v \left\{ i v \cdot D + i {{\,\raise.15ex\hbox{/}\mkern-13.5mu D}}_\perp \frac{1}{ 2 m + i
v \cdot D} i {{\,\raise.15ex\hbox{/}\mkern-13.5mu D}}_\perp \right\} Q_v \nonumber \\
&= \bar Q_v \left\{ i v \cdot D - \frac{1}{2 m} {{\,\raise.15ex\hbox{/}\mkern-13.5mu D}}_\perp 
{{\,\raise.15ex\hbox{/}\mkern-13.5mu D}}_\perp +\frac {1}{4 m^2} {{\,\raise.15ex\hbox{/}\mkern-13.5mu D}}_\perp \left(iv\cdot D\right)
{{\,\raise.15ex\hbox{/}\mkern-13.5mu D}}_\perp + \ldots \right\} Q_v \nonumber
\end{align*}
and use a sequence of field redefinitions to eliminate the  $1/m^2$ suppressed $v \cdot D$ term.  The  equation of motion for the heavy quark field is $(i v \cdot D)Q_v=0$, so this example shows how to eliminate equation-of-motion operators in HQET. Here $v^\mu$ is the velocity vector of the heavy quark with $v \cdot v=1$, and
\begin{align*}
D_\perp^\mu \equiv D^\mu - (v \cdot D) v^\mu\,.
\end{align*}
If you prefer, you can work in the rest frame of the heavy quark, where $v^\mu=(1,0,0,0)$, $v \cdot D=D^0$ and $ D_\perp^\mu = (0,\mathbf{D})$. See Ref.~\cite{Manohar:1997qy} for help.

\end{exercisenb}

In general, there are are many equation-of-motion operators $E_i$. Under renormalization, these operators mix among themselves,
\begin{align}
\mu\frac{\rd}{\rd \mu} E_i &= \gamma_{ij} E_j\,,
\end{align}
where $\gamma_{ij}$ can be gauge dependent. The reason is that the l.h.s. vanishes when inserted in an $S$-matrix element, and this needs to hold for all values of $\mu$. $E_i$ are not observable quantities, and their anomalous dimensions can depend on choice of gauge. For non-equation-of-motion operators $O_i$, the anomalous dimensions take the form
\begin{align}
\mu\frac{\rd}{\rd \mu} O_i &= \gamma_{ij} O_j + \Gamma_{ik} E_k.
\end{align}
An operator $O_i$ is not an equation-of-motion operator if $O_i$ contributes to $S$-matrix elements. Under $\mu$ evolution, these operators can mix with $\{E_i\}$, since $\{E_i\}$ have zero contributions to $S$-matrix elements. Since $O_i$ are observable, $\gamma_{ij}$ is gauge independent, but $\Gamma_{ik}$ can be gauge dependent.

\begin{figure}
\centering
\includegraphics[width=2cm]{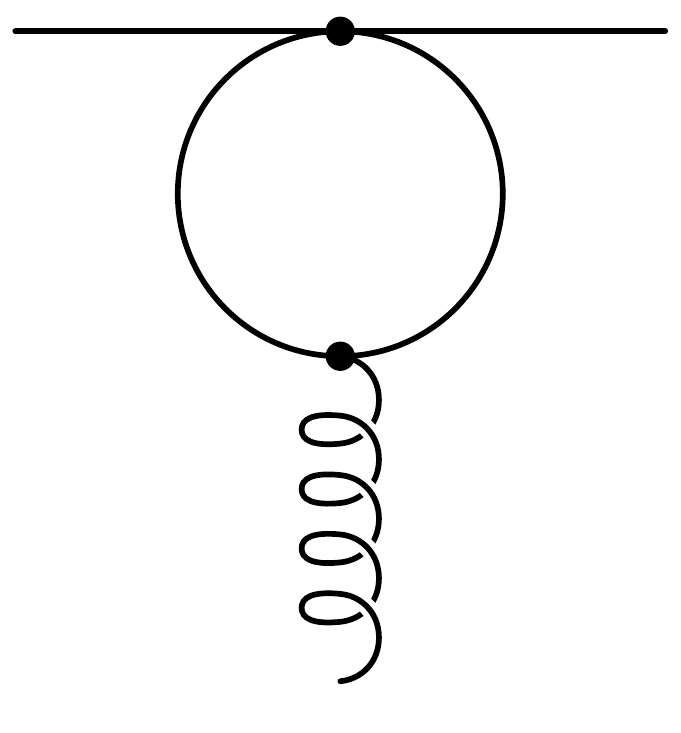}
\caption{\label{fig:penguin} Penguin graph in the weak interactions.}
\end{figure}

\begin{figure}
\centering
\includegraphics[width=2cm]{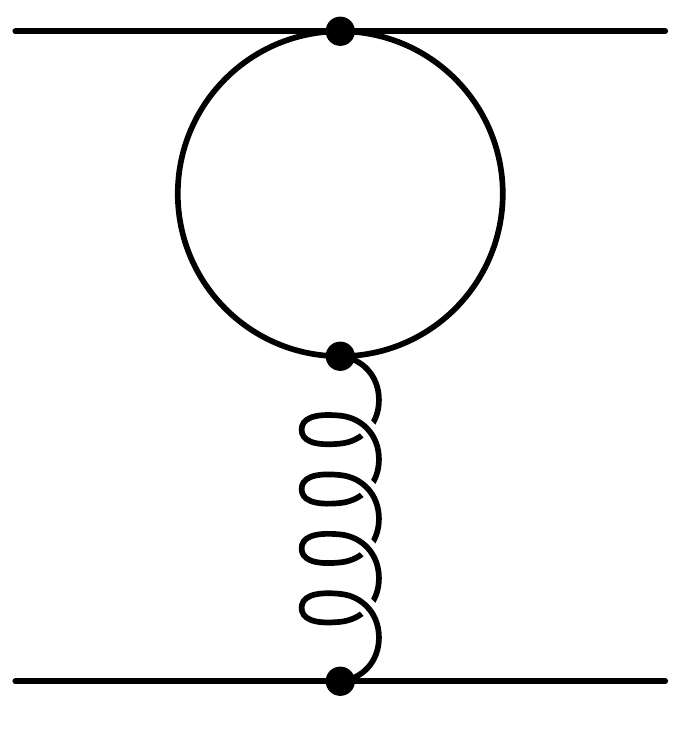}\hspace{2cm}
\raise0.375cm\hbox{\includegraphics[width=1.5cm]{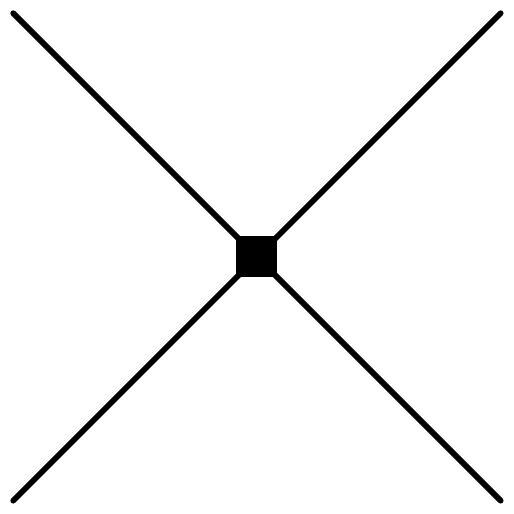}}
\caption{\label{fig:penguin2} Penguin and four-quark contribution to $qq \to qq$.}
\end{figure}

A well-known example of the use of equations-of-motion is for penguin graphs in the weak interactions~\cite{Gilman:1979bc}, shown in Fig.~\ref{fig:penguin}.
The penguin graph is divergent, and requires the counterterm
\begin{align}\label{6.42}
\LL&= \frac{4G_F}{\sqrt 2} \frac{c_P}{\epsilon} g(\overline \psi \gamma^\mu T^A \psi) \left( D^\nu F_{\mu \nu}\right)^A\,.
\end{align}
The penguin counterterm is eliminated from the Lagrangian by making a field redefinition,
\begin{align}\label{6.44}
\LL&= \frac{4G_F}{\sqrt 2}\frac{c_P}{\epsilon} g ( \overline \psi \gamma^\mu T^A \psi) \left( D^\nu F_{\mu \nu}\right)^A
\to  \frac{4G_F}{\sqrt 2} \frac{c_P}{\epsilon} g (\overline \psi \gamma^\mu T^A \psi)  g (\overline \psi \gamma_\mu T^A \psi)\,,
\end{align}
and replacing it by a four-quark operator.
The field redefinition needed for eqn~(\ref{6.44}) is
\begin{align}
A^A_\mu \to A^A_\mu -  \frac{4G_F}{\sqrt 2}\frac{c_P}{\epsilon} g \overline \psi \gamma^\mu T^A \psi\,,
\end{align}
which is a field redefinition with an infinite coefficient. Green's functions using the redefined Lagrangian eqn~(\ref{6.44}) are infinite, but the $S$-matrix is finite. There is no counterterm to cancel the penguin graph divergence, but the on-shell four-quark amplitude gets both the penguin and counterterm contributions (Fig.~\ref{fig:penguin2}) and is finite.
%


\chapter{Decoupling of Heavy Particles}\label{sec:decoupling}

Heavy particles do not decouple in a mass-independent subtraction scheme such as \MSbar. For example, the one-loop QCD $\beta$-function coefficient is $b_0=11-2/3 n_f$, where $n_f$ is the number of quark flavors. Thus $b_0$ has the same value for all $\mu$, independent of the quark masses. One expects that the top quark only contributes to the $\beta$-function for $\mu \gg m_t$, and no longer contributes when $\mu \ll m_t$, i.e.\ heavy particles decouple at low energy.

\begin{figure}
\begin{center}
\includegraphics[width=4cm]{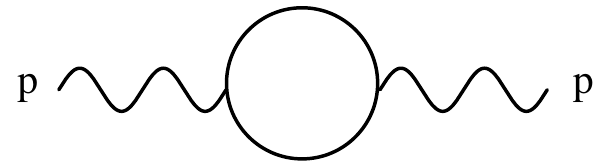}
\end{center}
\caption{\label{fig:qed} One loop contribution to the QED $\beta$-function
from a fermion of mass $m$}
\end{figure}

To understand the decoupling of heavy particles, consider the contribution of a charged lepton of mass $m$ to the one-loop $\beta$ function in QED. The diagram Fig.~\ref{fig:qed} in dimensional regularization gives
\begin{align}\label{7.1}
& i \frac{e^2}{2\pi^2}\left(p_\mu p_\nu -p^2 g_{\mu\nu}\right)
\left[ \frac{1}{6\epsilon}  - \int_0^1 dx\ x(1-x)\
\log\frac{m^2-p^2 x(1-x)}{\overline\mu^2}\right]  \nn
 \equiv & \, i\left(p_\mu p_\nu -p^2 g_{\mu\nu}\right) \Pi(p^2)
\end{align}
where $p$ is the external momentum. 

\section{Momentum-Subtraction Scheme}

Consider a mass-dependent scheme, the momentum space subtraction scheme, where one subtracts the value of the graph at a Euclidean momentum point $p^2=-\mu_M^2$, to get the renormalized vacuum polarization function,
\begin{align}
\Pi_{\text{mom}}(p^2,m^2,\mu_M^2) &= - \frac{e^2}{2\pi^2} 
\left[\int_0^1 dx\ x(1-x)\ \log\frac{m^2-p^2 x(1-x)}{m^2+\mu_M^2 x(1-x)} \right].
\end{align}

\begin{figure}
\begin{center}
\includegraphics[width=6cm]{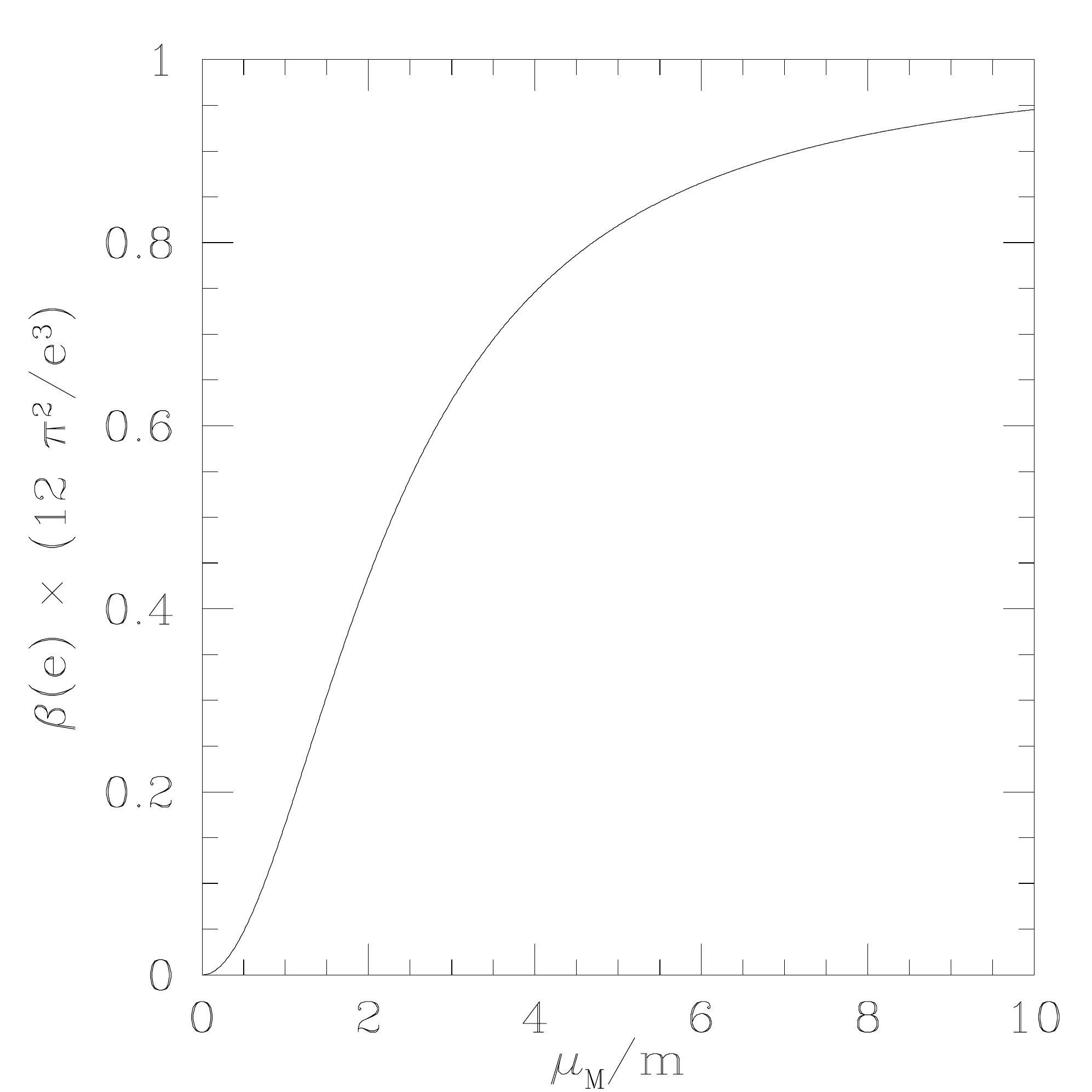}
\end{center}
\caption{\label{fig:82} Contribution of a fermion of mass $m$ to the QED $\beta$-function. The result is given for the momentum-space subtraction scheme, with renormalization scale $\mu_M$. The $\beta$ function does not attain its limiting value of $e^3/12\pi^2$ until $\mu_M \gtrsim 10\, m$. The fermion decouples for $\mu_M\ll m$.}
\end{figure}

The fermion contribution to the QED $\beta$-function is obtained by acting on $\Pi$ with $(e/2)\mu_M\, \rd/\rd \mu_M$,
\begin{align}
\beta_{\text{mom}}\left(e\right)&=-\frac{e}{ 2} \mu_M \frac{\rd }{\rd \mu_M} \frac{e^2}{ 2\pi^2}
\left[\int_0^1 dx\ x(1-x)\ \log \frac{m^2-p^2 x(1-x)}{ m^2+\mu_M^2 x(1-x)}\right]\cr
&= \frac{e^3}{ 2\pi^2}\int_0^1 dx\ x(1-x)\ \frac{\mu_M^2 x (1-x)}{ m^2+\mu_M^2 x(1-x)}.
\end{align}
The fermion contribution to the $\beta$-function is plotted in Fig.~\ref{fig:82}. When the fermion mass $m$ is small compared with the renormalization point $\mu_M$, $m\ll \mu_M$, the $\beta$-function contribution is
\begin{align}
\beta\left(e\right) \approx \frac{e^3}{ 2\pi^2}\int_0^1 dx\ x(1-x) = \frac
{e^3}
{12\pi^2}. 
\end{align}
As the renormalization point passes through $m$, the fermion decouples, and for $\mu_M\ll m$, its contribution to $\beta$ vanishes as
\begin{align}
\beta\left(e\right) \approx \frac{e^3}{2\pi^2}
\int_0^1 dx\ x(1-x) \frac{\mu_M^2 x (1-x)}{ m^2}
=\frac {e^3}{60 \pi^2} \frac{\mu_M^2}{m^2} \to 0
\end{align}
Thus in the momentum space scheme, we see the expected behavior that heavy particles decouple, which is an example of the Appelquist-Carazzone decoupling theorem~\cite{Appelquist:1974tg}.

\section{The \MSbar\ Scheme}

In the \MSbar\ scheme, one subtracts only the $1/\epsilon$ pole of eqn~(\ref{7.1}), so
\begin{align}\label{8.7}
\Pi_{\MSbar}(p^2,m^2,\overline\mu^2) &= - \frac{e^2}{2\pi^2} 
\left[\int_0^1 dx\ x(1-x)\ \log\frac{m^2-p^2 x(1-x)}{\overline \mu^2} \right].
\end{align}
The fermion contribution to the QED $\beta$-function is obtained by acting with $(e/2) \bar\mu\, \rd/\rd\bar\mu$ on $\Pi$,
\begin{align}
\beta_{\MSbar}\left(e\right)&=- \frac{e}{2}\bar \mu \frac{\rd}{\rd \bar \mu} \frac{e^2}{ 2\pi^2}
\left[\int_0^1 dx\ x(1-x) \log \frac{m^2-p^2 x(1-x)}{ \overline\mu^2}\right]\nn
&= \frac{e^3}{ 2\pi^2}\int_0^1 dx\ x(1-x) = \frac {e^3}{12\pi^2},
\end{align}
which is independent of the fermion mass and $\bar\mu$.

The fermion contribution to the $\beta$-function in the \MSbar\ scheme does not vanish as $m\gg \bar\mu$, so the fermion does not decouple as it should. There is another problem: from eqn~(\ref{8.7}), the finite part of the Feynman graph in the \MSbar\ scheme at low momentum is
\begin{align}
\Pi_{\MSbar}(0,m^2,\overline\mu^2)  &= -\frac{e^2 }{ 2\pi^2}
\left[\int_0^1 dx\ x(1-x) \log \frac{m^2}{ \bar\mu^2}\right].
\end{align}
For $\bar\mu\ll m$ the logarithm becomes large, and perturbation theory breaks down. These two problems are related.  The large finite part corrects for the fact that the value of the running coupling used at low energies is ``incorrect,'' because it was obtained using the ``wrong'' $\beta$-function.  

The two problems can be solved at the same time by integrating out heavy particles. One uses a theory including the heavy fermion as a dynamical field when $m<\bar\mu$, and a theory without the fermion field when $m>\bar\mu$. Effects of the heavy particle in the low energy theory are included via higher dimension operators, which are suppressed by inverse powers of the heavy particle mass.  The matching condition of the two theories is that $S$-matrix elements for light particle scattering in the low-energy theory must equal the $S$-matrix elements for light particle scattering in the high-energy theory. Schematically, one matches
\begin{align}
\LL^{(n_l+1)} & \to \LL^{(n_l)}\,,
\end{align}
from a theory with $n_l$ light particles and one heavy particle to a theory with $n_l$ light particles. The effects of the heavy particles are absorbed into changes in the coefficients of $\LL$. These are referred to as threshold corrections. Thus at the matching scale, $\LL$ changes, both in terms of the field content and the values of the Lagrangian coefficients. However, nothing discontinuous is going on, and the physics (i.e. $S$-matrix elements) are continuous across the threshold. The description changes, but the resulting $S$-matrix elements remain the same.

In our example, we can integrate out the heavy lepton at the matching scale $\bar\mu$. The effect of the one-loop heavy lepton graph Fig.~\ref{fig:qed} can be expanded for $p^2 \ll m^2$ as
\begin{align}\label{8.7exp}
\Pi_{\MSbar}(p^2,m^2,\overline\mu^2) &= - \frac{e^2}{2\pi^2} 
\int_0^1 dx\ x(1-x)\ \left\{  \log\frac{m^2}{\overline \mu^2}+ \log \left[ 1 - \frac{p^2}{m^2} x(1-x) \right] 
\right\} \nn
&= - \frac{e^2}{2\pi^2} 
\int_0^1 dx\ x(1-x)\ \left\{  \log\frac{m^2}{ \overline\mu^2} - \frac{p^2}{m^2} x(1-x)  + \ldots   \right\} \nn
&=-\frac16 \log \frac{m^2}{\overline\mu^2} + \frac{p^2}{30 m^2} + \mathcal{O}\left( \frac{p^4}{m^4} \right)\,.
\end{align}
The first term is included in $\LL^{(n_l)}$ by a shift in the gauge kinetic term. Rescaling the gauge field to restore the kinetic term to its canonical normalization $-F_{\mu\nu}^2/4$ gives a shift in the gauge coupling constant,
\begin{align}\label{7.11}
\frac{1}{e_L^2(\overline\mu)} &= \frac{1}{e_H^2(\overline\mu)} - \frac{1}{12\pi^2} \log \frac{m^2}{\overline\mu^2} .
\end{align}
where $e_{L}$ is the gauge coupling in the low-energy theory, and $e_H$ is the gauge coupling in the high-energy theory. The $\overline \mu$ dependence of the threshold correction is related to the difference in $\beta$-functions of the two theories.

The second term in eqn~(\ref{8.7exp}) gives a dimension six operator in the low-energy theory,
\begin{align}
\LL &=   \frac{e^2}{240 \pi^2 m^2}  \partial_\alpha F_{\mu \nu} \partial^\alpha F^{\mu \nu},
\end{align}
and so on. While the Lagrangian has changed at $\mu_M$, the $S$-matrix has not. The change in the Lagrangian is exactly the same as the contribution from Fig.~\ref{fig:qed}, which is present in the high energy theory but not in the low-energy theory.

\begin{exercisebn}

Verify that the first term in eqn~(\ref{8.7exp}) leads to the threshold correction in the gauge coupling given in eqn~(\ref{7.11}). If one matches at $\bar \mu=m$, then $e_L(\bar \mu)=e_H(\bar \mu)$, and the gauge coupling is continuous at the threshold. Continuity does not hold at higher loops, or when a heavy scalar is integrated out.

\end{exercisebn}

\begin{exercisenb}

Assume the threshold correction is of the form
\begin{align*}
\frac{1}{e_L^2(\overline\mu)} &= \frac{1}{e_H^2(\overline\mu)} + c  \log \frac{m^2}{\overline\mu^2} \,.
\end{align*}
Find the relation between $c$ and the difference $\beta_H-\beta_L$ of the $\beta$-functions in the two theories, and check that this agrees with eqn~(\ref{7.11}).

\end{exercisenb}

\chapter{Naive Dimensional Analysis}\label{sec:nda}


There is a slightly more sophisticated version of the EFT power counting formula which is referred to as naive dimensional analysis (NDA)~\cite{Manohar:1983md}. It is a power counting formula that keeps track of  the $4\pi$ factors from loop graphs. If $\phi$, $\psi$ and $X_{\mu \nu}$, $g$, $y$, $\lambda$ denote generic scalar fields, fermion fields, gauge field-strength tensors, gauge couplings, Yukawa couplings and $\phi^4$ couplings, then the NDA formula says that an operator in the EFT should be normalized as
\begin{align}
\widehat O &= f^{2} \Lambda^{2} \left[\frac{\partial}{\Lambda}\right]^{N_p} \left[\frac{\phi}{ f } \right]^{N_\phi} \left[\frac{A}{ f } \right]^{N_A} \left[\frac{\psi}{ f \sqrt{\Lambda}} \right]^{N_\psi}  \left[\frac{g}{ 4\pi }  \right]^{N_g}
 \left[\frac{y}{ 4\pi } \right]^{N_y} \left[\frac{ \lambda}{ 16\pi^2}\right]^{N_\lambda}.
\label{fnda4}
\end{align}
where $\Lambda$ and $f$ are related by
\begin{align}\label{10.2}
\Lambda = 4\pi f\,,
\end{align}
and $\Lambda$ is the scale of the EFT derivative expansion.  With this normalization,  EFT coefficients are expected to be of order unity,
\begin{align}
\LL &= \sum \widehat C_i \widehat O_i\,,
\end{align}
with $\widehat C_i \sim 1$.
A generalization of NDA to $\d$ dimensions can be found in Ref.~\cite{Gavela:2016bzc}.  From eqn~(\ref{fnda4}),
\begin{align}
\frac{D}{\Lambda} &= \frac{\partial + i g A}{\Lambda}=  \frac{\partial}{\Lambda} +i\left[ \frac{g}{4 \pi }  \right] \left[ \frac{A}{f}\right]
\label{16}
\end{align}
so that both  parts of a covariant derivative have the same power counting. 

Loop graphs in the EFT maintain the NDA form, i.e.\ an arbitrary graph with insertions of operators of the form eqn~(\ref{fnda4}) generates an operator of the same form. The proof, which relies on counting $1/(16\pi^2)$ factors from each loop and the topological identity for a connected graph $V-I+L=1$, where $V$ is the number of vertices, $I$ the number of internal lines, and $L$ the number of loops,  is left as an exercise.

\begin{exercise}\label{ex:nda}

\item Show that the power counting formula eqn~(\ref{fnda4}) for an EFT Lagrangian is self-consistent, i.e.\ an arbitrary graph with insertions of vertices of this form generates an interaction which maintains
the same form. (See \cite{Gavela:2016bzc} and \cite{Manohar:1983md}). Show that eqn~(\ref{fnda4}) is equivalent to
\begin{align*}
\widehat O &\sim \frac{\Lambda^4}{16 \pi^2 } \left[\frac{\partial}{\Lambda}\right]^{N_p}  \left[\frac{ 4 \pi\,  \phi}{ \Lambda} \right]^{N_\phi}
 \left[\frac{ 4 \pi\,  A}{ \Lambda } \right]^{N_A}  \left[\frac{ 4 \pi \,  \psi}{\Lambda^{3/2}}\right]^{N_\psi} \left[ \frac{g}{4 \pi }  \right]^{N_g}
\left[\frac{y}{4 \pi } \right]^{N_y} \left[\frac{\lambda}{16 \pi^2 }\right]^{N_\lambda}  .
\end{align*}

\end{exercise}

Using the more sophisticated power counting of eqn~(\ref{fnda4}) instead of only counting factors of $\Lambda$ makes a big difference in estimating the coefficients of higher dimension terms in the Lagrangian. For example, the four-quark dimension six operator is normalized to
\begin{align*}
\widehat O &= f^2 \Lambda^2 \frac{\left( \overline \psi \gamma^\mu \psi\right)^2}{(f \sqrt{\Lambda})^4} =
 \frac{1}{f^2} \left( \overline \psi \gamma^\mu \psi\right)^2 = \frac{16\pi^2}{\Lambda^2} \left( \overline \psi \gamma^\mu \psi\right)^2 \,.
\end{align*}
The extra $16\pi^2$ makes a difference of $\sim 150$ in the normalization of the operator.

In $\chi$PT, the Lagrangian is written in terms of
\begin{align}
U(x) &= e^{2 i \Pi(x)/f}\,,
\end{align}
where $\Pi(x)$ is a matrix of pion fields. $U(x)$ satisfies eqn~(\ref{fnda4}), since every $\Pi$ comes with a factor $1/f$. The normalization of the two-derivative term in the chiral Lagrangian is
\begin{align}
\widehat O &= \Lambda^2 f^2\ \frac{\partial U}{\Lambda} \frac{\partial U^\dagger}{\Lambda} = f^2\, \partial_\mu U \partial^\mu U^\dagger
\end{align}
which is the usual normalization of the kinetic term.
The four-derivative term is normalized to
\begin{align}\label{10.7}
\widehat O &=\Lambda^2 f^2\ \frac{\partial U}{\Lambda} \frac{\partial U^\dagger}{\Lambda} \frac{\partial U}{\Lambda} \frac{\partial U^\dagger}{\Lambda} = 
\frac{1}{16 \pi^2}\ \partial_\mu U \partial^\mu U^\dagger \partial_\mu U \partial^\mu U^\dagger\,.
\end{align}
The four-derivative coefficients in the chiral Lagrangian are usually denoted by $L_i$, and eqn~(\ref{10.7}) shows that one expects
$L_i \sim 1/(16\pi^2) \sim 4 \times 10^{-3}$, which is true experimentally (see~\cite{Pich:cs}).

The difference between $f = 93$\,MeV and $\Lambda=4\pi f = 1.2$\,GeV is very important for $\chi$PT. The value of $f$ is fixed from the experimental value of the $\pi \to \mu \overline \nu_\mu$ decay rate. If we did not keep track of the $4\pi$ factors, this would imply that the scale $\Lambda$ of $\chi$PT is $\Lambda \sim f$, and $\chi$PT  breaks down for momenta of order $f$. If this is the case, $\chi$PT  is not very useful, since the pion mass is around $140$\,MeV, so $\chi$PT breaks down for on-shell pions. Luckily, eqn~(\ref{10.2}) says that $\Lambda_\chi$, the scale of the $\chi$PT derivative expansion is $4\pi f$~\cite{Manohar:1983md} which is much larger than $f$, so that $\chi$PT is valid for $\pi-\pi$ scattering at low momentum. Loop corrections in pion $\chi$PT are of order $[m_\pi/(4 \pi f)]^2 \sim 0.014$, and are a few percent. $\chi$PT for kaons has corrections of order
$[m_K/(4 \pi f)]^2 \sim 0.2$.

The NDA formula eqn~(\ref{fnda4}) implies that if all operators in the Lagrangian are normalized using NDA, then an arbitrary loop graph  gives
\begin{align}
\delta\widehat C_i & \sim \prod_{k} \widehat C_{k}\,,
\label{46}
\end{align}
where the graph has insertions of Lagrangian terms $\widehat C_k \widehat O_k$, and produces an amplitude of the form $\widehat C_i \widehat O_i$. All the $4\pi$ factors have disappeared, and one obtains a very simple form for the amplitudes.  The results are equally valid for strongly and weakly coupled theories. 

The NDA formula eqn~(\ref{46}) also shows that in strongly coupled theories $\widehat C \lesssim 1$~\cite{Manohar:1983md}. The reason is that if $\widehat C \gg 1$, then the hierarchy of equations eqn~(\ref{46}) is unstable, because higher order contributions to $\widehat C_i$ are much larger than $\widehat C_i$. On the other hand, there is no inconsistency if $\widehat C_i \ll 1$, since all this implies is that higher order corrections are small, a sign of a weakly coupled theory. eqn~(\ref{46}) shows that an interaction becomes strongly coupled  when $\widehat C \sim 1$.  For the dimension-four interactions,  strong coupling is when gauge couplings are $g \sim 4\pi$, Yukawa couplings are $y \sim 4\pi$ and scalar self-couplings are $\lambda \sim (4\pi)^2$.

One can use NDA for cross sections as well as amplitudes. A cross section is the imaginary part of the forward scattering amplitude, so one can estimate cross sections by using NDA for the forward amplitude, and then multiplying by $\pi$, since the imaginary part comes from $\log(-1)=i \pi$. Since two-body final states give a one-loop forward scattering diagram, and $n$-body final states give a $n-1$ loop diagram, the $4\pi$ counting rules for phase space are: $1/(16 \pi)$ for the first two particles, and $1/(16\pi^2)$ for each additional particle. We  used this $4\pi$ counting rule earlier in these lectures in our estimates of cross sections.

\chapter{Invariants}

EFT Lagrangians are constructed using gauge and Lorentz invariant operators which are polynomials in the basic fields. Classifying these operators is a fun topic which is extensively studied in the mathematics invariant theory literature. I discuss invariant theory briefly in this section.
For an elementary summary, see Refs.\cite{Hanany:2010vu,Jenkins:2009dy}.
 
Start with the simple example of a theory with $N_f$ fermions with mass term
\begin{align}
\LL &= - \overline \psi_L M \psi_R + \text{h.c.}\,,
\end{align}
where $M$ is an $N_f \times N_f$ matrix.
We can make a field redefinition (ignoring anomalies),
\begin{align}\label{10.2a}
\psi_L &\to L \psi_L, &
\psi_R &\to R \psi_R,
\end{align}
under which
\begin{align}
M &\to L M R^\dagger\,.
\end{align}
Under $CP$, $M \to M^*$. The $S$-matrix is invariant under the field redefinition eqn~(\ref{10.2a}), and depends only on invariants constructed from $M$. To eliminate $R$, define
\begin{align}
X &\equiv M M^\dagger, & X &\to L X L^\dagger,
\end{align}
which transforms only under $L$. Then the invariants are
\begin{align}
I_{2n} &=\vev{X^n} \,,
\end{align}
where $2n$ is the degree of the invariant in the basic object $M$, and $\vev{\,\cdot\,}$ denotes a trace. Suppose $N_f=1$. Then $X$ is a $1\times1$ matrix, and
\begin{align}
\vev{X^2} &=I_4 = I_2^2=\vev{X}^2, & \vev{X^3}=I_6 &= I_2^3=\vev{X}^3,
\end{align}
and there is one independent invariant of every even degree, $I_{2n}=I_2^n=\vev{X}^n$. 

The Hilbert series is defined as
\begin{align}
H(q) &= \sum_{n=0}^\infty N_n q^n\,
\end{align}
where $N_n$ is the number of invariants of degree $n$, and $N_0=1$ by convention. In the $1\times 1$ matrix example,
\begin{align}\label{9.8}
H(q)&=1+q^2+q^4 +\ldots = \frac{1}{1-q^2}\,.
\end{align}
The denominator of $H(q)$ in eqn~(\ref{9.8}) tells us that there is one generator of degree two, which is $\vev{X}$, and that all invariants are given by  powers of this generator. Given $I_2$, we can determine the fermion mass, $m = \sqrt{I_2}$, as a real, non-negative number. The invariant is $CP$ even, since under $CP$, $X \to X^*$, and $\vev{X} \to \vev{X^*}=\vev{X^\dagger}=\vev{X}$ since $X$ is Hermitian, and the trace is invariant under transposition of the matrix.

The next case is $N_f=2$, with invariants
\begin{align}
\vev{X},\ \vev{X^2},\ \vev{X^3}, \ldots \,.
\end{align}
These are not all independent, because the Cayley-Hamilton theorem implies
\begin{align}
\vev{X^3} &= \frac32 \vev{X}\vev{X^2}-\frac12 \vev{X}^3\,,
\end{align}
for any $2 \times 2$ matrix.
This identity eliminates all traces of $X^n$ for $n\ge 3$. There is one invariant of degree $2$, $\vev{X}$, two of degree four $\vev{X}^2$ and $\vev{X^2}$, etc.\ The Hilbert series is
\begin{align}
H(q)&=1+q^2+2q^4 +\ldots = \frac{1}{(1-q^2)(1-q^4)}\,.
\end{align}
The denominator factors imply that all invariants are generated by products of $\vev{X}$ and $\vev{X^2}$. Given $\vev{X}$ and $\vev{X^2}$, we can find the two masses by solving
\begin{align}\label{10.12}
\vev{X} &= m_1^2+m_2^2, &
\vev{X^2} &= m_1^4+m_2^4\,.
\end{align}

For $N_f=3$, the generators are $\vev{X}$, $\vev{X^2}$, $\vev{X^3}$. Higher powers are eliminated by the Cayley-Hamilton theorem,
\begin{align}
\vev{ X^4} &= \frac1{6}\vev{X}^4 -  \vev {X}^2 \vev{X^2}
+ \frac 4 3 \vev {X^3} \vev X  + \frac 1 2 \vev{X^2}^2\,,
\end{align}
and the Hilbert series is
\begin{align}
H(q)&=1+q^2+2q^4 +\ldots = \frac{1}{(1-q^2)(1-q^4)(1-q^6)}\,.
\end{align}

\begin{exercise}

By explicit calculation, show that
\begin{align*}
\left[\frac 12 \vev{A}^2- \frac12 \vev{A^2}\right] \mathbf{1} - \vev{A}A +A^2 &= 0\,,\nn
\frac 16 \vev{A}^3-\frac12\vev{A}\vev{A^2}+\frac13\vev{A^3} &=0 \,,
\end{align*}
for a general $2 \times 2$ matrix $A$ and that
\begin{align*}
\vev{A}\vev{B}\vev{C}-\vev{A}\vev{BC}-\vev{B}\vev{AC}-\vev{C}\vev{AB}+\vev{ABC}+\vev{ACB}
&=0\,.
\end{align*}
for general $2 \times 2$ matrices $A,B,C$.
Identities analogous to this for $3 \times 3$ matrices are used in $\chi$PT to remove $L_0$ and to replace it by $L_{1,2,3}$, as discussed by Pich in his lectures~\cite{Pich:cs}.

\end{exercise}

Now consider the case of two quark types,  $u$ and $d$,  in the SM. There are two mass matrices $M_u$ and $M_d$ which transform as
\begin{align}\label{9.15}
M_u &\to L M_u R_u^\dagger\,, &
M_d &\to L M_d R_d^\dagger\,.
\end{align}
Equation~(\ref{9.15}) results because the right handed quarks $u_R$ and $d_R$ are independent fields with independent transformations $R_u$ and $R_d$ in the SM, whereas the left-handed quarks are part of a weak doublet,
\begin{align}
q_L &= \left[ \begin{array}{c} u_L \\ d_L \end{array} \right]\,,
\end{align}
so $L_u=L_d=L$. To construct invariants, we can eliminate $R_{u,d}$ by constructing
\begin{align}
X_u &= M_u M_u^\dagger, &
X_d &= M_d M_d^\dagger,
\end{align}
which transform as
\begin{align}
X_u &\to L X_u L^\dagger ,&
X_d &\to L X_d L^\dagger \,.
\end{align}
For $N_f=1$, $X_u$ and $X_d$ are numbers, and the only independent invariants are $\vev{X_u}$ and $\vev{X_d}$, and the Hilbert series is
\begin{align}
H(q) &= \frac{1}{(1-q^2)^2}\,.
\end{align}
For $N_f=2$, the independent generators are $\vev{X_u}$, $\vev{X_d}$, $\vev{X_u^2}$, $\vev{X_d^2}$ and $\vev{X_u X_d}$, and
\begin{align}\label{10.20}
H(q) &= \frac{1}{(1-q^2)^2(1-q^4)^3}\,.
\end{align}

\begin{exercise}

Show that for $N_f=2$, all invariants are generated by the independent invariants $\vev{X_u}$, $\vev{X_d}$, $\vev{X_u^2}$, $\vev{X_d^2}$ and $\vev{X_u X_d}$.

\end{exercise}

$\vev{X_u}$ and $\vev{X_u^2}$ determine the two $u$-quark masses $m_u$ and $m_c$ as in eqn~(\ref{10.12}). $\vev{X_d}$ and $\vev{X_d^2}$ determine the two $d$-quark masses $m_d$ and $m_s$. $\vev{X_u X_d}$ determines the Cabibbo angle,
\begin{align}
\vev{X_u X_d} &= (m_u^2 m_d^2+ m_c^2 m_s^2) - (m_c^2-m_u^2)(m_s^2-m_d^2)\sin^2 \theta\,.
\end{align}
If $m_u=m_c$ or if $m_d=m_s$, $\theta$ is not defined (or can be rotated away). 

All the invariants are $CP$ even, so there is no $CP$ violation in the quark sector for two quark flavors. For example, under $CP$,
\begin{align}
\vev{X_u X_d} &\to \vev{X_u^* X_d^*} = \vev{(X_u^* X_d^*)^T} = \vev{ X_d^\dagger X_u^\dagger} = \vev{X_d X_u} = \vev{X_u X_d}
\end{align}
since $X_u$ and $X_d$ are Hermitian, and the trace is invariant under transposition and cyclic permutation.

The first non-trivial example is $N_f=3$. The $CP$ even  generators are
\begin{align}
\label{9.22}
\vev{X_u},\ \vev{X_u^2},\ \vev{X_u^3},\ \vev{X_d},\ \vev{X_d^2},\ \vev{X_d^3},\ \vev{X_u X_d}, \vev{X_u^2 X_d}, \vev{X_u X_d^2},\
\vev{X_u^2 X_d^2}.
\end{align}
They determine the quark masses $m_{u,c,t}$, $m_{d,s,b}$, and the three CKM angles $\theta_{12},\theta_{13},\theta_{23}$. However, the terms in eqn~(\ref{9.22}) do not generate all the invariants. We also have the $CP$ odd invariant
\begin{align}
I_- &= \vev{X_u^2 X_d^2 X_u X_d} - \vev{X_d^2 X_u^2 X_d X_u} = \frac13 \vev{\left[X_u,X_d\right]^3}\,.
\end{align}
and the $CP$ even invariant 
\begin{align}
I_+=\vev{X_u^2 X_d^2 X_u X_d} + \vev{X_d^2 X_u^2 X_d X_u}\,.
\end{align}
$I_+$ is not independent; it  can be written as a linear combination of the lower order invariants in eqn~(\ref{9.22}). 

While $I_-$ is not a linear combination of the invariants in eqn~(\ref{9.22}), it turns out that $I_-^2$ \emph{is} a linear combination. This is an example of a relation among the invariants. There also can be relations among relations, which are known as syzygies. Thus the independent invariants are arbitrary products of powers of eqn~(\ref{9.22}) plus $I_-$ to at most the first power. This gives the Hilbert series for $N_f=3$
\begin{align}
\label{9.24}
H(q) &= \frac{1+q^{12}}{(1-q^2)^2(1-q^4)^3(1-q^6)^4(1-q^8)}\,,
\end{align}
where the $+q^{12}$ in the numerator is the contribution from $I_-$. $I_-$ is related to the  Jarlskog invariant $J$,
\begin{align}\label{9.27}
I_- &= 2 i (m_c^2-m_u^2)(m_t^2-m_c^2)(m_t^2-m_u^2)(m_s^2-m_d^2)(m_b^2-m_s^2)(m_b^2-m_d^2) J,
\end{align}
where
\begin{align}\label{9.28}
J &= \text{Im}\, \left[ V_{11} V_{12}^* V_{22} V_{21}^* \right] = c_{12} s_{12} c_{13} s_{13}^2 c_{23} s_{23} s_\delta,
\end{align}
using the CKM matrix convention of the PDG~\cite{Patrignani:2016xqp}.

The $CP$-even invariants in eqn~(\ref{9.22}) determine $J^2$, and hence $ J$ but an overall sign. The invariant $I_-$ fixes the sign. This analysis should be familiar from the study of $CP$ violation in the SM. By measuring $CP$ conserving decay rates, one can determine the lengths of the three sides of the unitarity triangle. This determines the triangle (including the area, which is a measure of $CP$ violation) up to an overall reflection, which is fixed by the sign of $J$. Thus, one can determine if $CP$ is violated only from $CP$ conserving measurements. 

\newpage

\begin{exercisebn}

 Show that the invariant 
\begin{align*}
I_- &= \vev{X_u^2 X_d^2 X_u X_d} - \vev{X_d^2 X_u^2 X_d X_u}\,,
\end{align*}
is the lowest order $CP$-odd invariant made of the quark mass matrices. Show that $I_-$ also can  be written in the form
\begin{align*}
I_- &= \frac13 \vev{\left[X_u,X_d\right]^3}\,,
\end{align*}
and explicitly work out $I_-$ in the SM using the CKM matrix convention of the PDG~\cite{Patrignani:2016xqp}. Verify eqns~(\ref{9.27},\ref{9.28}).

\end{exercisebn}

\begin{exercisenb}

Compute the Hilbert series for the ring of invariants generated by
 $x,y,z$ (each of dimension 1), and invariant under the transformation $(x,y,z) \to (-x,-y,-z)$.

\end{exercisenb}

The general structure of $H(q)$ is the ratio of a numerator $N(q)$ and a denominator $D(q)$,
\begin{align}
H(q) &= \frac{N(q)}{D(q)},
\end{align}
where the denominator $D(q)$ is a product of the form
\begin{align}
D(q) &= (1- q^{n_1})^{r_1} (1- q^{n_2})^{r_2} \ldots
\end{align}
and the numerator $N(q)$ is a polynomial with non-negative coefficients of degree $d_N$ which is palindromic, i.e.
\begin{align}
q^{d_N} N(1/q) &= N(q)\,.
\end{align}
The number of denominator factors $\sum r_i$ is the number of parameters~\cite{knop}. In eqn~(\ref{9.24}) the number of parameters is 10, which are the six masses, 3 angles and one phase.

As a non-trivial example, the lepton sector of the seesaw theory for $n_g=2$ generations has invariants generated by the mass matrices for the charged leptons $m_E$, neutrinos $m_\nu$ and the singlet Majorana mass matrix $M$. The Hilbert series is~\cite{Jenkins:2009dy}
\begin{align}
H(q) &= \frac{1+q^6+3q^8+2q^{10}+3q^{12}+q^{14}+q^{20}}{(1-q^2)^3(1-q^4)^5(1-q^6)(1-q^{10})}\,,
\end{align}
which has a palindromic numerator. The numerator is of degree twenty, and the coefficients are $1,0,0,0,0,0,1,0,3,0,2,0,3,0,1,0,0,0,0,0,1$, which is the same string read in either direction.

To construct an EFT, we have basic fields $\psi(x)$, $\phi(x)$, etc.\ which transform under various symmetries, and we want to construct invariant Lagrangians which are polynomials in the basic fields. This is a problem in invariant theory, with a few additional requirements.
\begin{itemize}
\item We can act with covariant derivatives on fields, $D_\mu \phi(x)$, to get an object that transforms the same way as $\phi(x)$ under gauge and flavor symmetries, but adds an extra Lorentz index.
\item We can drop total derivatives since they vanish when integrated to get the action. Equivalently, we are allowed to integrate by parts.
\item We can make field redefinitions or equivalently use the equations of motion to eliminate operators. 
\end{itemize}
Counting invariants including these constraints seems simple, but there is a subtlety. Terms such as
\begin{align}
\partial_\mu (\phi^\dagger \partial^\mu \phi - \partial^\mu \phi^\dagger \phi )
\end{align}
vanish because they are a total derivative, and also by using the equations of motion. We have to make sure we do not double count the terms eliminated by these two conditions. This is a non-trivial problem that was recently solved in Ref.~\cite{Henning:2015alf} using representations of the conformal group. The HQET/NRQCD dimension-eight operators were recently classified with the help of invariants~\cite{Kobach:2017xkw}.

\chapter{SMEFT}\label{sec:smeft}

\begin{table}
\renewcommand{\arraystretch}{1.5}
\setlength{\arraycolsep}{0.125cm}
\begin{align}
\begin{array}{c|c|ccc}
& \text{Lorentz} & SU(3) & SU(2) & U(1) \\
\hline
G_{\mu \nu} & (1,0)+(0,1) & 8 & 1 & 0 \\
W_{\mu \nu} & (1,0)+(0,1) & 1 & 3 & 0 \\
B_{\mu \nu} & (1,0)+(0,1) & 1 & 1 & 0 \\
H & (0,0) & 1 & 2 & \frac12 \\
q & (1/2,0) & 3 & 2 & \frac16 \\
l & (1/2,0) & 1 & 2 & -\frac 12 \\
u & (0,1/2) & 3 & 1 & \frac23 \\
d & (0,1/2) & 3 & 1 & -\frac13 \\
e & (0,1/2) & 1 & 1 & -1 \\
\end{array}
\end{align}
\caption{\label{tab:SM} Fields of the Standard Model. The Lorentz group is $SU(2)_L \times SU(2)_R$. The fermions have a generation index $n_g=1,2,3$.}
\end{table}

The SMEFT is an EFT constructed using the basic fields of the SM given in Table~\ref{tab:SM}.  For an extensive recent review, see Ref.~\cite{Brivio:2017vri}. The dimension-four terms give the usual SM Lagrangian. There is only a single $U(1)$ gauge field in the SM. In theories with multiple Abelian gauge fields,  the general kinetic energy for the $U(1)$ gauge fields has the form
\begin{align}
\LL &= -\frac14 C_{ij} F_{\mu \nu}^{(i)} F_{\mu \nu}^{(j)}\,,
\end{align}
where $C$ is a real symmetric matrix with positive eigenvalues, which is referred to as kinetic mixing~\cite{Galison:1983pa,Holdom:1985ag},

Constructing the higher dimension operators in SMEFT is not easy. It is useful to note that Lorentz invariance requires that fermion fields come in pairs. The allowed fermion bilinears written in terms of chiral fields are
\begin{align}
&\overline \psi_L \gamma^\mu \psi_L, &
&\overline \psi_R \gamma^\mu \psi_R, &
&\overline \psi_L \psi_R, &
& \overline \psi_L \sigma^{\mu \nu} \psi_R, &
&\overline \psi_R \psi_L, &
& \overline \psi_R \sigma^{\mu \nu} \psi_L .
\end{align}
One can always replace a right-handed field $\psi_R$ by its charge-conjugate left-handed field $\psi^c_L$,
\begin{align}\label{11.4}
\psi_R &= C \psi_L^{c*},
\end{align}
where $C=i \gamma^2$. Thus we can use either a right-handed $e^-_R$ field, or a left-handed $e^+_L$ field. The SMEFT is usually written using left-handed $SU(2)$ doublet fields, and right-handed $SU(2)$ singlet fields, as shown in Table~\ref{tab:SM}. 

Mass terms and dipole interactions are written in terms of left-handed field bilinears
\begin{align}
\overline \psi_R \psi_L &= \psi_L^{cT} C \psi_L, &
\overline \psi_R \sigma^{\mu \nu} \psi_L &= \psi_L^{cT} C \sigma^{\mu \nu} \psi_L.
\end{align}
In general, if there are multiple left-handed fields,  the mass and dipole operators are
\begin{align}\label{11.6}
&\psi_{Lr}^{T} C \psi_{Ls}, &
&\psi_{Lr}^{T} C \sigma^{\mu \nu} \psi_{Ls},
\end{align}
where $r,s$ are flavor indices. The mass term is symmetric in $rs$, and the dipole term is antisymmetric in $rs$. One still has to ensure that the terms in eqn~(\ref{11.6}) respect gauge invariance, so that a mass term $e^{+T}_L C e^-_L$ is allowed, but not $e^{-T}_L C e^-_L$.

Left-handed fields transform as $(1/2,0)$ under the Lorentz group, so that the fermion bilinear $ \chi_L^T C \Gamma\psi_L$ transforms as $(1/2,0) \otimes (1/2,0) = (0,0) \oplus (1,0)$. The $(0,0)$ representation is $\chi^T_L C \psi_L$ and the $(1,0)$ representation is $\chi^T_L C  \sigma^{\mu \nu}\psi_L$. The $(1,0)$ representation is self-dual because of the self-duality condition on $\sigma^{\mu \nu}P_L$,
\begin{align}\label{11.7}
\frac{i}{2} \epsilon^{\alpha \beta \mu \nu} \sigma_{\mu \nu} P_L &= \sigma^{\alpha \beta} P_L \,.
\end{align}
Similarly, the right-handed matrix satisfies the anti-self-duality condition
\begin{align}\label{11.8}
\frac{i}{2} \epsilon^{\alpha \beta \mu \nu} \sigma_{\mu \nu} P_R &= -\sigma^{\alpha \beta} P_R\,.
\end{align}

\begin{exercisebn}\label{ex:11.1}

Show that $(\psi_{Lr}^T C \psi_{Ls})$ is symmetric in $rs$ and $(\psi_{Lr}^T C \sigma^{\mu \nu} \psi_{Ls})$ is antisymmetric in $rs$. 
\end{exercisebn}

\begin{exercisenb}

Prove the duality relations eqns~(\ref{11.7},\ref{11.8}). The sign convention is $\gamma_5 = i \gamma^0 \gamma^1 \gamma^2 \gamma^3$ and $\epsilon_{0123}=+1$.

\end{exercisenb}

The lowest dimension term in the SMEFT with $\opdim >4$ is the dimension-five term
\begin{align}
	\label{eq:smeft5}
	\LL^{(5)} &= C_{\substack{5 \\ rs}}\epsilon^{ij} \epsilon^{kl}  (l_{ i r}^T\, C\, l_{k s}) H_j H_l  + \text{h.c.}\,.
\end{align}
Here $r,s$ are flavor indices, and $i,j,k,l$ are $SU(2)$ gauge indices. The coefficient $C_{\substack{5 \\ rs}}$ is symmetric in $rs$, by Exercise~\ref{ex:11.1}. $\LL^{(5)}$ is a $\Delta L=2$ interaction, and gives a Majorana mass term to the neutrinos when $H$ gets a vacuum expectation value.

It can be shown~\cite{Kobach:2016ami} that invariant operators constructed from SM fields satisfy
\begin{align}\label{kobach}
\frac12(\Delta B-\Delta L) \equiv \opdim \quad \mod 2\,.
\end{align}
Thus a $\opdim=5$ operator cannot conserve both baryon and lepton number.

\begin{exercisebn}

\item Show that eqn~(\ref{eq:smeft5}) is the unique dimension-five term in the SMEFT Lagrangian.

\end{exercisebn}

\begin{exercisenb}

Show that eqn~(\ref{eq:smeft5})  generates a Majorana neutrino mass when $H$ gets a vacuum expectation value, and find the neutrino mass matrix $M_\nu$ in terms of $C_5$ and $v$.

\end{exercisenb}

At dimension-six there are eight different operator classes, $X^3$, $H^6$, $H^4D^2$, $X^2 H^2$, $\psi^2 H^3$, $\psi^2 XH$, $\psi^2 H^2 D$ and $\psi^4$, in terms of their field content. Determining the independent operators is a non-trivial task~\cite{Buchmuller:1985jz,Grzadkowski:2010es}. Here I discuss a few aspects of the analysis.

The four-quark operators $\psi^4$ can be simplified using Fierz identities. Consider invariants made from two $\overline l \,\Gamma\, l$ bilinears. Since $l$ is a left-handed field, the only gamma-matrix allowed is $\Gamma=\gamma^\mu$. Bilinears constructed from $l$ can be either $SU(2)$ singlets or $SU(2)$ triplets, so the $l^4$ invariants are
\begin{align}
Q_{\substack{ll \\ prst}} &= (\overline l_{ip} \gamma^\mu l^i{}_r )(\overline l_{js} \gamma_\mu l^j{}_t ), \nn
Q^{(3)}_{\substack{ll \\ prst}} &= (\overline l_{ip} \gamma^\mu [\tau^a]^i{}_j l^j{}_r )(\overline l_{ks} \gamma_\mu [\tau^a]^k{}_m l^m{}_t ),
\end{align}
where $p,r,s,t$ are generation (flavor) indices and $i,j,k,m$ are weak $SU(2)$ indices.  Using the $SU(2)$ Fierz identity (Exercise~\ref{ex:nfierz})
\begin{align}
\label{9.39su2}
[\tau^a]^i{}_j  [\tau^a]^k{}_m &= 2 \delta^i_m \delta^k_j - \delta^i_j \delta^k_m,
\end{align}
the second bilinear can be written as
\begin{align}\label{12.41}
Q^{(3)}_{\substack{ll \\ prst}} &= 2 (\overline l_{ip} \gamma^\mu l^j{}_r )(\overline l_{js} \gamma_\mu  l^i{}_t )- (\overline l_{ip} \gamma^\mu l^i{}_r )(\overline l_{js} \gamma_\mu l^j{}_t ).
\end{align}
Applying the spinor Fierz identity (Exercise~\ref{ex:spinfierz})
\begin{align}
\label{9.40}
(\overline \psi_1 \gamma^\mu P_L \psi_2)
(\overline \psi_3 \gamma_\mu P_L \psi_4)
&= (\overline \psi_1 \gamma^\mu P_L \psi_4)
(\overline \psi_3 \gamma_\mu P_L \psi_2)
\end{align}
on the first term of eqn~(\ref{12.41}) gives
\begin{align}
\label{9.41}
Q^{(3)}_{\substack{ll \\ prst}} &= 2 (\overline l_{ip} \gamma^\mu  l^i{}_t )(\overline l_{js} \gamma_\mu l^j{}_r )- (\overline l_{ip} \gamma^\mu l^i{}_r )(\overline l_{js} \gamma_\mu l^j{}_t )
= 2 Q_{\substack{ll \\ ptsr}} - Q_{\substack{ll \\ prst}}\,.
\end{align}
Equation~(\ref{9.41}) implies that we do not need to include $Q^{(3)}_{\substack{ll \\ prst}}$ operators, as they are linear combinations of $Q_{ll}$ operators, so the independent $l^4$ operators are $Q_{ll}$.

For $l q$ operators,
\begin{align}
\label{9.42}
Q^{(1)}_{\substack{lq \\ prst}} &= (\overline l_{ip} \gamma^\mu l^i{}_r )(\overline q_{\alpha js} \gamma_\mu q^{\alpha j}{}_t ), \nn 
Q^{(3)}_{\substack{lq \\ prst}} &= (\overline l_{ip} \gamma^\mu [\tau^a]^i{}_j l^j{}_r )(\overline q_{\alpha ks} \gamma_\mu [\tau^a]^k{}_m q^{\alpha m}{}_t ),
\end{align}
the  identity eqn~(\ref{9.40}) cannot be used since it would produce $(\overline l q)$ bilinears. Thus both $lq$ operators in eqn~(\ref{9.42}) are independent. 

For four-quark operators
$(\overline q \gamma^\mu q)(\overline q \gamma_\mu q)$, there are four possible gauge invariants, written schematically as
\begin{align}
\label{9.43}
1 \otimes 1,\quad \tau^a \otimes \tau^a,\quad T^A \otimes T^A,\quad \tau^a T^A \otimes \tau^a T^A,
\end{align}
depending on what gauge generators are inserted in each bilinear. The $SU(N)$ version of eqn~(\ref{9.39su2}) from Exercise~\ref{ex:nfierz}
\begin{align}
\label{9.39}
[T^A]^\alpha{}_\beta [T^A]^\lambda{}_\sigma &=
\frac12 \delta^\alpha_\sigma \delta^\lambda_\beta - \frac{1}{2N} \delta^\alpha_\beta
\delta^\lambda_\sigma\,,
\end{align}
can be used for the color generators with $N=3$. One can view the index contractions for the $SU(2)$ and $SU(3)$ generators as either direct or swapped, i.e.\ in $(\overline q_1 \gamma^\mu q_2)(\overline q_3  \gamma_\mu q_4)$ contracted between $q_1,q_2$ and $q_3,q_4$, or between $q_1,q_4$ and $q_2,q_3$. Then the four possible terms in eqn~(\ref{9.43}) are
\begin{align}
\label{9.40a}
\text{direct},\quad SU(2)\ \text{swapped},\quad SU(3)\ \text{swapped},\quad \text{both swapped}.
\end{align}
The spinor Fierz identity eqn~(\ref{9.40}) exchanges the $q$ fields, so it swaps both the $SU(2)$ and $SU(3)$ indices, and hence converts
\begin{align}
\label{9.40b}
\text{direct} \leftrightarrow \text{both swapped} \qquad SU(2)\ \text{swapped} \leftrightarrow SU(3)\ \text{swapped}.
\end{align}
Thus there are only two independent invariants out of the four in eqn~(\ref{9.43}), which are chosen to be $1 \otimes 1$ and $\tau^a \otimes \tau^a$. 

For $\psi^4$ operators involving $\sigma^{\mu\nu}$, the duality relations eqns~(\ref{11.7},\ref{11.8}) can be used to eliminate $\epsilon_{\mu \nu \alpha \beta}$ contracted with  $\sigma$ matrices. One also has the relation
\begin{align}\label{10.21}
(\overline A \sigma^{\mu \nu} P_L  B)(\overline C \sigma_{\mu \nu} P_R D) &=0
\end{align}
The left-hand side is a Lorentz singlet in the tensor product $(1,0) \otimes (0,1)= (1,1)$, and so must vanish.

Using the above results, one can determine the independent  $\psi^4$ operators.

\begin{exercise}
Prove eqn~(\ref{10.21}).
\end{exercise}

\section{SMEFT Operators}

Since the SMEFT is playing an increasingly important role in current research, I will summarize the operators in SMEFT up to dimension six. The number of operators of each type is listed, and their $CP$ property is given as a subscript. For non-Hermitian operators $\O$, $\O+\O^\dagger$ is $CP$ even, and $\O-\O^\dagger$ is $CP$-odd. The flavor indices have not been included for notational simplicity. For example, including flavor indices, $Q_{eW}$ is $Q_{\substack{eW \\ p r}}$ and $Q_{ll}$ is $Q_{\substack{ll \\ prst}}$, etc.

Table~\ref{tab:oplist} gives a summary of the SMEFT operators up to dimension six. For $n_g=3$, there are 6 $\Delta L=2$ operators plus their Hermitian conjugates, 273 $\Delta B=\Delta L=1$ operators plus their Hermitian conjugates, and 2499 Hermitian $\Delta B=\Delta L=0$ operators~\cite{Alonso:2013hga}. For $n_g=1$, there are 76 Hermitian $\Delta B=\Delta L=0$ operators. In the literature, you will often see that there are 59 $\Delta B=\Delta L=0$ operators. This counts the number of operator types listed in the tables below. Some of the operators, such as $(H^\dagger H)^3$ are Hermitian, whereas others, such as
$(H^\dag H)(\bar l e H)$ are not, and count as two Hermitian operators. Hermitian operators have a real coefficient in the Lagrangian, whereas non-Hermitian operators have a complex coefficient. Counting Hermitian operators is equivalent to counting real Lagrangian parameters.

\begin{table}
\begin{align*}
\begin{array}{c|c||c|c|c||c|c|c}
\text{dim} & & \multicolumn{3}{c||}{n_g = 1} & \multicolumn{3}{c}{n_g=3} \\
\hline
& & CP\text{-even}& CP\text{-odd} & \text{Total} &  CP\text{-even}& CP\text{-odd} & \text{Total}  \\
\hline
5 & \Delta L = 2 & & & 1 & & & 6 \\ 
5 & \Delta L = -2 & & & 1 & & & 6 \\ 
\hline
6 & \Delta B = \Delta L = 1 & & & 4 & & & 273  \\
6 & \Delta B = \Delta L = -1 & & & 4 & & & 273  \\
\hline\hline
6 & X^3 & 2 & 2 & 4 & 2 & 2 & 4 \\
6 & H^6 &  1 & 0 & 1 & 1& 0 & 1\\
6 & H^4 D^2 & 2 & 0 & 2 & 2 & 0 & 2 \\
6 & X^2 H^2 &  4 & 4 & 8 & 4 & 4 & 8 \\
6 & \psi^2 H^3 & 3 & 3 & 6 & 27 & 27 & 54 \\
6 & \psi^2 X H & 8 & 8 & 16 & 72 & 72 & 144  \\
6 & \psi^2 H^2 D & 8 & 1 & 9 & 51 & 30 & 81 \\
6 & (\bar L L )(\bar LL ) & 5 & 0 & 5 & 171 & 126 & 297 \\
6 & (\bar RR )(\bar RR ) & 7 & 0 & 7 & 255 & 195 & 450 \\
6 & (\bar L L )(\bar RR ) & 8 & 0 & 8 & 360 & 288 & 648 \\
6 & (\bar L R )(\bar RL )+\text{h.c.} & 1 & 1 & 2 & 81 & 81 & 162 \\
6 & (\bar L R )(\bar LR) +\text{h.c.} & 4 & 4& 8 & 324 & 324 & 648\\
\hline
&\text{Total}\ \Delta B=\Delta L =0& 53 & 23 & 76 & 1350 & 1149 & 2499
\end{array}
\end{align*}
\caption{\label{tab:oplist} Number of operators of each type in the SMEFT up to dimension six.}
\end{table}

\subsection{Dimension $5$}\label{sec:dim5}

The dimension five operators $Q_5$ are $\Delta L=2$ operators.
%
%
\begin{align*}
\renewcommand{\arraystretch}{1.5}
\begin{array}[t]{c|c|c}
\multicolumn{3}{c}{\boldsymbol{(LL)HH+\text{h.c.}}} \\
\hline
Q_{5}  &\frac12 n_g(n_g+1)    & \epsilon^{ij} \epsilon^{k\ell} (l_{ip}^T C l_{kr} ) H_j H_\ell    \\
\hline
\text{Total} & \frac12 n_g(n_g+1) + \text{h.c.}
\end{array}
\end{align*}
%
There are $n_g(n_g+1)/2$ $\Delta L=2$ operators, and $n_g(n_g+1)/2$ $\Delta L=-2$ Hermitian conjugate operators. $CP$ exchanges the $\Delta L= \pm 2$ operators. The $\Delta L=\pm 2$ operators give a Majorana neutrino mass when the weak interactions are spontaneously broken. Since neutrino masses are very small, the $\Delta L = \pm 2$ operators are assumed to be generated at a very high scale (which could be the GUT scale).

\subsection{Dimension $6,\ \Delta B=\Delta L=1$}

The dimension six operators can be divided into several groups. The first group are the $\Delta B=\Delta L=1$ operators and their Hermitian conjugates.
%
%
\begin{align*}
\renewcommand{\arraystretch}{1.5}
\begin{array}[t]{c|c|c}
\multicolumn{3}{c}{\boldsymbol{\Delta B = \Delta L = 1 +\text{h.c.}}} \\
\hline
Q_{duql}  &n_g^4    & \epsilon^{\alpha \beta \gamma} \epsilon^{ij} (d^T_{\alpha p} C u_{\beta r} ) (q^T_{\gamma i s} C l_{jt})    \\
Q_{qque}  & \frac12 n_g^3(n_g+1)    & \epsilon^{\alpha \beta \gamma} \epsilon^{ij} (q^T_{\alpha i p} C q_{\beta j r} ) (u^T_{\gamma s} C e_{t})    \\
Q_{qqql}  & \frac13 n_g^2(2n_g^2+1)   & \epsilon^{\alpha \beta \gamma} \epsilon^{i\ell} \epsilon^{jk} (q^T_{\alpha i p} C q_{\beta j r} ) (q^T_{\gamma k s} C l_{\ell t})    \\
Q_{duue} & n_g^4     & \epsilon^{\alpha \beta \gamma} (d^T_{\alpha p} C u_{\beta r} ) (u^T_{\gamma s} C e_{t})    \\
\hline
\text{Total} & \frac 16 n_g^2 (19 n_g^2+3n_g+2) + \text{h.c.}
\end{array}
\end{align*}
%
The $\Delta B=\Delta L=1$ operators violate baryon number, and lead to proton decay. They are generated in unified theories, and are suppressed by two powers of the GUT scale.

\subsection{Dimension $6,\ X^3$}

There are 2 $CP$-even and $2$ $CP$-odd operators with three field-strength tensors. In this and subsequent tables, the $CP$ property is shown as a subscript.
\begin{align*}
\renewcommand{\arraystretch}{1.5}
\begin{array}[t]{c|c|c}
\multicolumn{3}{c}{\boldsymbol{X^3}} \\
\hline
Q_G  & 1_+ & f^{ABC} G_\mu^{A\nu} G_\nu^{B\rho} G_\rho^{C\mu}  \\
Q_{\widetilde G} & 1_-  & f^{ABC} \widetilde G_\mu^{A\nu} G_\nu^{B\rho} G_\rho^{C\mu}  \\
Q_W & 1_+ & \epsilon^{IJK} W_\mu^{I\nu} W_\nu^{J\rho} W_\rho^{K\mu} \\ 
Q_{\widetilde W} & 1_-  & \epsilon^{IJK} \widetilde W_\mu^{I\nu} W_\nu^{J\rho} W_\rho^{K\mu} \\
\hline
\text{Total} & 2_+ + 2_-
\end{array}
\end{align*}

\subsection{Dimension $6,\ H^6$}

There is a single operator involving six Higgs fields. It adds a $h^6$ interaction of the physical Higgs particle to the SMEFT Lagrangian after spontaneous symmetry breaking.
%
%
\begin{align*}
\renewcommand{\arraystretch}{1.5}
\begin{array}[t]{c|c|c}
\multicolumn{3}{c}{\boldsymbol{H^6}} \\
\hline
Q_H & 1_+ & (H^\dag H)^3 \\
\hline
\text{Total} & 1_+ 
\end{array}
\end{align*}

\newpage

\subsection{Dimension $6,\ H^4 D^2$}

\begin{align*}
\renewcommand{\arraystretch}{1.5}
\begin{array}[t]{c|c|c}
\multicolumn{3}{c}{\boldsymbol{H^4 D^2}} \\
\hline
Q_{H\Box} & 1_+ & (H^\dag H)\Box(H^\dag H) \\
Q_{H D} & 1_+  & \ \left(H^\dag D_\mu H\right)^* \left(H^\dag D_\mu H\right) \\
\hline
\text{Total} & 2_+
\end{array}
\end{align*}

%
%

\subsection{Dimension $6,\ X^2 H^2$}

\begin{align*}
\renewcommand{\arraystretch}{1.5}
\begin{array}[t]{c|c|c}
\multicolumn{3}{c}{\boldsymbol{X^2 H^2}} \\
\hline
Q_{H G}  & 1_+   & H^\dag H\, G^A_{\mu\nu} G^{A\mu\nu} \\
Q_{H\widetilde G}    & 1_-     & H^\dag H\, \widetilde G^A_{\mu\nu} G^{A\mu\nu} \\
Q_{H W}   & 1_+  & H^\dag H\, W^I_{\mu\nu} W^{I\mu\nu} \\
Q_{H\widetilde W}  & 1_-       & H^\dag H\, \widetilde W^I_{\mu\nu} W^{I\mu\nu} \\
Q_{H B}  & 1_+   &  H^\dag H\, B_{\mu\nu} B^{\mu\nu} \\
Q_{H\widetilde B}   & 1_-      & H^\dag H\, \widetilde B_{\mu\nu} B^{\mu\nu} \\
Q_{H WB} & 1_+    &  H^\dag \tau^I H\, W^I_{\mu\nu} B^{\mu\nu} \\
Q_{H\widetilde W B}    & 1_-     & H^\dag \tau^I H\, \widetilde W^I_{\mu\nu} B^{\mu\nu} \\
\hline
\text{Total} & 4_+ + 4_-
\end{array}
\end{align*}
%
The $X^2H^2$ operators are very important phenomenologically. They lead to $gg \to h$ and $h \to \gamma \gamma$ vertices, and contribute to Higgs production and decay. The corresponding SM amplitudes start at one loop, so LHC experiments are sensitive to $X^2H^2$ operators via interference effects with SM amplitudes~\cite{Grojean:2013kd,Manohar:2006gz}.

\subsection{Dimension $6,\ \psi^2 H^3$}

\begin{align*}
\renewcommand{\arraystretch}{1.5}
\begin{array}[t]{c|c|c}
\multicolumn{3}{c}{\boldsymbol{(\bar L R) H^3+ \text{h.c.}}} \\
\hline
Q_{eH}  & n_g^2          & (H^\dag H)(\bar l_p e_r H) \\
Q_{uH} & n_g^2         & (H^\dag H)(\bar q_p u_r \widetilde H ) \\
Q_{dH}  & n_g^2         & (H^\dag H)(\bar q_p d_r H)\\
\hline
\text{Total} & 3 n_g^2 + \text{h.c.}
\end{array}
\end{align*}
%
These operators are $H^\dagger H$ times the SM Yukawa couplings, and violate the relation that the Higgs boson coupling to fermions is proportional to their mass.

\subsection{Dimension $6,\ \psi^2 X H $}\label{sec:dipole6}
%
\begin{align*}
\renewcommand{\arraystretch}{1.5}
\begin{array}[t]{c|c|c}
\multicolumn{3}{c}{\boldsymbol{(\bar L R) X H + \text{h.c.}}} \\
\hline
Q_{eW}  & n_g^2    & (\bar l_p \sigma^{\mu\nu} e_r) \tau^I H W_{\mu\nu}^I \\
Q_{eB} & n_g^2         & (\bar l_p \sigma^{\mu\nu} e_r) H B_{\mu\nu} \\
Q_{uG}  & n_g^2        & (\bar q_p \sigma^{\mu\nu} T^A u_r) \widetilde H \, G_{\mu\nu}^A \\
Q_{uW}  & n_g^2        & (\bar q_p \sigma^{\mu\nu} u_r) \tau^I \widetilde H \, W_{\mu\nu}^I \\
Q_{uB}  & n_g^2        & (\bar q_p \sigma^{\mu\nu} u_r) \widetilde H \, B_{\mu\nu} \\
Q_{dG}  & n_g^2        & (\bar q_p \sigma^{\mu\nu} T^A d_r) H\, G_{\mu\nu}^A \\
Q_{dW}  & n_g^2         & (\bar q_p \sigma^{\mu\nu} d_r) \tau^I H\, W_{\mu\nu}^I \\
Q_{dB}  & n_g^2        & (\bar q_p \sigma^{\mu\nu} d_r) H\, B_{\mu\nu} \\
\hline 
\text{Total} & 8 n_g^2 + \text{h.c.}
\end{array}
\end{align*}
%
%
When $H$ gets a VEV, these operators lead to dipole operators for transitions such as $\mu \to e \gamma$, $b \to s \gamma$ and $b \to s g$.

\subsection{Dimension $6,\ \psi^2 H^2 D $}

\begin{align*}
\renewcommand{\arraystretch}{1.5}
\begin{array}[t]{c|c|c}
\multicolumn{3}{c}{\boldsymbol{\psi^2  H^2 D }} \\
\hline
Q_{H l}^{(1)} & \frac12 n_g(n_g+1)_+ + \frac12 n_g (n_g-1)_-     & (H^\dag i\overleftrightarrow{D}_\mu H)(\bar l_p \gamma^\mu l_r)\\
Q_{H l}^{(3)}  & \frac12 n_g(n_g+1)_+ + \frac12 n_g (n_g-1)_-    & (H^\dag i\overleftrightarrow{D}^I_\mu H)(\bar l_p \tau^I \gamma^\mu l_r)\\
Q_{H e}   & \frac12 n_g(n_g+1)_+ + \frac12 n_g (n_g-1)_-         & (H^\dag i\overleftrightarrow{D}_\mu H)(\bar e_p \gamma^\mu e_r)\\
Q_{H q}^{(1)}  & \frac12 n_g(n_g+1)_+ + \frac12 n_g (n_g-1)_-     & (H^\dag i\overleftrightarrow{D}_\mu H)(\bar q_p \gamma^\mu q_r)\\
Q_{H q}^{(3)} & \frac12 n_g(n_g+1)_+ + \frac12 n_g (n_g-1)_-       & (H^\dag i\overleftrightarrow{D}^I_\mu H)(\bar q_p \tau^I \gamma^\mu q_r)\\
Q_{H u}   & \frac12 n_g(n_g+1)_+ + \frac12 n_g (n_g-1)_-    & (H^\dag i\overleftrightarrow{D}_\mu H)(\bar u_p \gamma^\mu u_r)\\
Q_{H d}   & \frac12 n_g(n_g+1)_+ + \frac12 n_g (n_g-1)_-       & (H^\dag i\overleftrightarrow{D}_\mu H)(\bar d_p \gamma^\mu d_r)\\
Q_{H u d} + \text{h.c.} &  n_g^2 + \text{h.c.} & i(\widetilde H ^\dag D_\mu H)(\bar u_p \gamma^\mu d_r)\\
\hline
\text{Total} &  \frac12 n_g(9n_g+7)_++\frac12 n_g(9n_g-7)_-
\end{array}
\end{align*}
%
The $\psi^2 H^2 D$ operators modify the coupling of electroweak bosons to fermions. $(Q_{Hud} \pm Q_{Hud}^\dagger)$ are $CP$-even/odd combinations, and contribute $n_g^2$ $CP$-even and $n_g^2$ $CP$-odd operators to the total.


\subsection{Dimension $6,\ (\bar LL)(\bar LL)$}

The $\psi^4$ operators can be grouped into different sets, depending on the chirality properties of the operators. We have seen earlier why the  $(\bar LL)(\bar LL)$ invariants are the ones listed in the table.
\begin{align*}
\renewcommand{\arraystretch}{1.5}
\begin{array}[t]{c|c|c}
\multicolumn{3}{c}{\boldsymbol{(\bar LL)(\bar LL) }} \\
\hline
Q_{ll} & \frac14 n_g^2(n_g^2+3)_+ +   \frac14 n_g^2(n_g^2-1)_-     & (\bar l_p \gamma_\mu l_r)(\bar l_s \gamma^\mu l_t) \\
Q_{qq}^{(1)} & \frac14 n_g^2(n_g^2+3)_+ +   \frac14 n_g^2(n_g^2-1)_-  & (\bar q_p \gamma_\mu q_r)(\bar q_s \gamma^\mu q_t) \\
Q_{qq}^{(3)} & \frac14 n_g^2(n_g^2+3)_+ +   \frac14 n_g^2(n_g^2-1)_-  & (\bar q_p \gamma_\mu \tau^I q_r)(\bar q_s \gamma^\mu \tau^I q_t) \\
Q_{lq}^{(1)}  & \frac12 n_g^2(n_g^2+1)_+ +   \frac12 n_g^2(n_g^2-1)_-        & (\bar l_p \gamma_\mu l_r)(\bar q_s \gamma^\mu q_t) \\
Q_{lq}^{(3)}  & \frac12 n_g^2(n_g^2+1)_+ +   \frac12 n_g^2(n_g^2-1)_-                & (\bar l_p \gamma_\mu \tau^I l_r)(\bar q_s \gamma^\mu \tau^I q_t)  \\
\hline
\text{Total} & \frac14 n_g^2(7 n_g^2+13)_+ + \frac74 n_g^2(n_g^2-1)_-
\end{array}
\end{align*}

%

\subsection{Dimension $6,\ (\bar RR)(\bar RR)$}

%
\begin{align*}
\renewcommand{\arraystretch}{1.5}
\begin{array}[t]{c|c|c}
\multicolumn{3}{c}{\boldsymbol{(\bar RR)(\bar RR) }} \\
\hline
Q_{ee}    & \frac18 n_g(n_g+1)(n_g^2+n_g+2)_+ + \frac18 (n_g-1)n_g(n_g+1)(n_g+2)_-           & (\bar e_p \gamma_\mu e_r)(\bar e_s \gamma^\mu e_t) \\
Q_{uu}  & \frac14 n_g^2(n_g^2+3)_+ +   \frac14 n_g^2(n_g^2-1)_-      & (\bar u_p \gamma_\mu u_r)(\bar u_s \gamma^\mu u_t) \\
Q_{dd}   &  \frac14 n_g^2(n_g^2+3)_+ +   \frac14 n_g^2(n_g^2-1)_-     & (\bar d_p \gamma_\mu d_r)(\bar d_s \gamma^\mu d_t) \\
Q_{eu}   &  \frac12 n_g^2(n_g^2+1)_+ +   \frac12 n_g^2(n_g^2-1)_-                    & (\bar e_p \gamma_\mu e_r)(\bar u_s \gamma^\mu u_t) \\
Q_{ed}    &  \frac12 n_g^2(n_g^2+1)_+ +   \frac12 n_g^2(n_g^2-1)_-                   & (\bar e_p \gamma_\mu e_r)(\bar d_s\gamma^\mu d_t) \\
Q_{ud}^{(1)}   &  \frac12 n_g^2(n_g^2+1)_+ +   \frac12 n_g^2(n_g^2-1)_-              & (\bar u_p \gamma_\mu u_r)(\bar d_s \gamma^\mu d_t) \\
Q_{ud}^{(8)}   & \frac12 n_g^2(n_g^2+1)_+ +   \frac12 n_g^2(n_g^2-1)_-              & (\bar u_p \gamma_\mu T^A u_r)(\bar d_s \gamma^\mu T^A d_t) \\
\hline
\text{Total} & \frac18 n_g (21n_g^3+2n_g^2+31n_g+2)_+ + \frac18 n_g(n_g^2-1) (21n_g+2)_-
\end{array}
\end{align*}

%
%


\subsection{Dimension $6,\ (\bar LL)(\bar RR)$}

\begin{align*}
\renewcommand{\arraystretch}{1.5}
\begin{array}[t]{c|c|c}
\multicolumn{3}{c}{\boldsymbol{(\bar LL)(\bar RR) }} \\
\hline
Q_{le} &\frac12 n_g^2(n_g^2+1)_+ +   \frac12 n_g^2(n_g^2-1)_-               & (\bar l_p \gamma_\mu l_r)(\bar e_s \gamma^\mu e_t) \\
Q_{lu} &\frac12 n_g^2(n_g^2+1)_+ +   \frac12 n_g^2(n_g^2-1)_-               & (\bar l_p \gamma_\mu l_r)(\bar u_s \gamma^\mu u_t) \\
Q_{ld}   &\frac12 n_g^2(n_g^2+1)_+ +   \frac12 n_g^2(n_g^2-1)_-             & (\bar l_p \gamma_\mu l_r)(\bar d_s \gamma^\mu d_t) \\
Q_{qe}  & \frac12 n_g^2(n_g^2+1)_+ +   \frac12 n_g^2(n_g^2-1)_-              & (\bar q_p \gamma_\mu q_r)(\bar e_s \gamma^\mu e_t) \\
Q_{qu}^{(1)} & \frac12 n_g^2(n_g^2+1)_+ +   \frac12 n_g^2(n_g^2-1)_-         & (\bar q_p \gamma_\mu q_r)(\bar u_s \gamma^\mu u_t) \\ 
Q_{qu}^{(8)}   & \frac12 n_g^2(n_g^2+1)_+ +   \frac12 n_g^2(n_g^2-1)_-       & (\bar q_p \gamma_\mu T^A q_r)(\bar u_s \gamma^\mu T^A u_t) \\ 
Q_{qd}^{(1)} & \frac12 n_g^2(n_g^2+1)_+ +   \frac12 n_g^2(n_g^2-1)_-  & (\bar q_p \gamma_\mu q_r)(\bar d_s \gamma^\mu d_t) \\
Q_{qd}^{(8)} & \frac12 n_g^2(n_g^2+1)_+ +   \frac12 n_g^2(n_g^2-1)_-  & (\bar q_p \gamma_\mu T^A q_r)(\bar d_s \gamma^\mu T^A d_t)\\
\hline
\text{Total} & 4 n_g^2(n_g^2+1)_+ +   4  n_g^2(n_g^2-1)_- 
\end{array}
\end{align*}

%

\subsection{Dimension $6,\ (\bar LR)(\bar RL)$}

\begin{align*}
\renewcommand{\arraystretch}{1.5}
\begin{array}[t]{c|c|c}
\multicolumn{3}{c}{\boldsymbol{(\bar LR)(\bar RL)+\text{h.c.} }} \\
\hline
Q_{ledq} & n_g^4 & (\bar l_p^j e_r)(\bar d_s q_{tj})  \\
\hline
\text{Total} & n_g^4 + \text{h.c.}
\end{array}
\end{align*}

%
%

\subsection{Dimension $6,\ (\bar LR)(\bar LR)$}

\begin{align*}
\renewcommand{\arraystretch}{1.5}
\begin{array}[t]{c|c|c}
\multicolumn{3}{c}{\boldsymbol{(\bar LR)(\bar LR)+\text{h.c.} }} \\
\hline
Q_{quqd}^{(1)} & n_g^4 & (\bar q_p^j u_r) \epsilon_{jk} (\bar q_s^k d_t) \\
Q_{quqd}^{(8)} & n_g^4 & (\bar q_p^j T^A u_r) \epsilon_{jk} (\bar q_s^k T^A d_t) \\
Q_{lequ}^{(1)} & n_g^4 & (\bar l_p^j e_r) \epsilon_{jk} (\bar q_s^k u_t) \\
Q_{lequ}^{(3)} & n_g^4 & (\bar l_p^j \sigma_{\mu\nu} e_r) \epsilon_{jk} (\bar q_s^k \sigma^{\mu\nu} u_t) \\
\hline
\text{Total} & 4n_g^4 + \text{h.c.}
\end{array}
\end{align*}

%

\begin{exercise}

\item In the SMEFT for $n_g$ generations, how many operators are there of the following kind (in increasing order of difficulty): (a) $Q_{He}$ (b) $Q_{ledq}$ (c) $Q_{lq}^{(1)}$ (d) $Q_{qq}^{(1)}$ (e) $Q_{ll}$ (f) $Q_{uu}$ (g) $Q_{ee}$ \\ (h) show that there are a total of 2499 Hermitian dimension-six $\Delta B= \Delta L=0$ operators.

\end{exercise}

The NDA normalization eqn~(\ref{fnda4}) for the SMEFT leads to an interesting pattern for the operators~\cite{Gavela:2016bzc,Jenkins:2013sda},
\begin{align}\label{42}
\LL &\sim  \widehat C_H \frac{(4\pi)^4}{\Lambda^2} H^6 \nn
& + \widehat C_{\psi^2 H^3} \frac{(4\pi)^3 }{\Lambda^2} \psi^2 H^3 \nn
 & + \widehat C_{H^4 D^2} \frac{(4\pi)^2 }{\Lambda^2} H^4 D^2 + \widehat C_{\psi^2H^2 D} \frac{(4\pi)^2}{\Lambda^2}  \psi^2 H^2 D  + \widehat C_{\psi^4} \frac{(4\pi)^2}{\Lambda^2}  \psi^4 \nn
& + \widehat C_{\psi^2 X H} \frac{(4\pi) }{\Lambda^2} g \psi^2 X H \nn
& + \widehat C_{X^2 H^2} \frac{1}{\Lambda^2} g^2 X^2 H^2 \nn
& + \widehat C_{X^3} \frac{1}{(4\pi)^2\Lambda^2} g^3 X^3
\end{align}
with $4\pi$ factors ranging from $(4\pi)^4$ to $1/(4\pi)^2$, a variation of $\sim 4\times 10^6$.

The complete renormalization group equations for the SMEFT up to dimension six have been worked out~\cite{Alonso:2014zka,Alonso:2013hga,Jenkins:2013zja,Jenkins:2013wua}.  A very interesting feature of these equations is that they respect holomorphy, reminiscent of what happens in a supersymmetric gauge theory~\cite{Alonso:2014rga}. The renormalization group equations take a simpler form if written using the normalization eqn~(\ref{42}).

\section{EFT below $M_W$}

Below the electroweak scale, one can write a low energy effective theory (LEFT) with quark and lepton fields, and only QCD and QED gauge fields. The operators have been classified in Ref.~\cite{Jenkins:2017dyc,Jenkins:2017jig}. Since $SU(2)$ gauge invariance is no longer a requirement, there are several new types of operators beyond those in SMEFT. 
\begin{itemize}
\item
There are dimension-three $\nu \nu$ operators which give a Majorana neutrino mass for left-handed neutrinos.
\item There are dimension-five dipole operators. These are the analog of
the $(\bar L R) XH$ operators in sec~\ref{sec:dipole6}, which turn into dimension-five operators when $H$ is replaced by its vacuum expectation value $v$. There are 70 Hermitian $\Delta B=\Delta L=0$ dipole operators for $n_g=3$.
\item There are $X^3$ and $\psi^4$ operators as in SMEFT, but operators containing $H$ are no longer present.
\item There are $\Delta L=4$ $\nu^4$ operators, and $\Delta L=2$ $(\bar \psi \psi)\nu \nu $ four-fermion operators, as well as four-fermion $\Delta B=-\Delta L$ operators.
\item There are 3631 Hermitian $\Delta B=\Delta L=0$ dimension-six operators for $n_g=3$.
\end{itemize}

The complete renormalization group equations up to dimension-six have been worked out for LEFT~\cite{Jenkins:2017dyc,Jenkins:2017jig}. Since the theory has dimension-five operators, there are non-linear terms from two insertions of dimension-five operators for the dimension-six running. Various pieces of the computation have been studied previously~\cite{Aebischer:2015fzz,Aebischer:2017gaw,Bhattacharya:2015rsa,Buchalla:1995vs,Celis:2017hod,Cirigliano:2017azj,Cirigliano:2012ab,Crivellin:2017rmk,Davidson:2016edt,Dekens:2013zca,Falkowski:2017pss,Gonzalez-Alonso:2017iyc}.

\acknowledgements

I would like to thank Sacha Davidson, Paolo Gambino, Mikko Laine and Matthias Neubert for organizing a very interesting school, and all the students for listening patiently to the lectures, asking lots of questions, and solving homework problems on weekends instead of hiking. Sacha Davidson, in particular, made sure all the students were well looked after. I would also like to thank Elizabeth Jenkins, Andrew Kobach, John McGreevy and Peter Stoffer for carefully reading  the manuscript, and Peter Stoffer for permission to use the photographs in Fig.~\ref{fig:eclipse}. This work was supported in part by DOE Grant No.~DE-SC0009919.

\appendix

\chapter{Naturalness and The Hierarchy Problem}\label{sec:naturalness}

In the SM, most Lagrangian terms have dimension four, but there is an operator of dimension two,
\begin{align}\label{4.42}
\LL &= \lambda v^2 H^\dagger H\,.
\end{align}
$m_H^2=2 \lambda v^2$ is the mass of the physical Higgs scalar $h$. If we assume the SM is an EFT with a power counting scale $\Lambda \gg v$, then blindly applying eqn~(\ref{2.20}) gives
\begin{align}\label{4.43}
\LL \sim \Lambda^2 H^\dagger H\,.
\end{align}
The quadratic $\Lambda^2$ dependence in eqn~(\ref{4.43}) is the so-called hierarchy problem: that the Higgs mass gets a correction of order $\Lambda$. Similarly, the cosmological constant $c$, the coefficient of the dimension-zero operator $\mathbf{1}$ is of order $\Lambda^4$, whereas we know experimentally that $c \sim \left( 2.8 \times 10^{-3} \ \hbox{eV} \right)^4$. 

The power counting argument of eqn~(\ref{2.20}) does \emph{not} imply that $m_H \propto \Lambda$ or $c \propto \Lambda^4$. We have seen in Sec.~\ref{sec:quad} and eqns~(\ref{3.29a}) that there are no $\Lambda^2$ and $\Lambda^4$ contributions from loops in dimensional regularization. By construction, the EFT describes the dynamics of a theory with particles with masses $m_H$ much smaller than $\Lambda$. Since $H$ is in our EFT Lagrangian, its mass $m_H$ is a light scale, $m_H \ll \Lambda$. With this starting point, corrections to $m_H$ only depend on other light scales and possible suppression factors of $1/\Lambda$ from higher-dimension terms. There are no positive powers of $\Lambda$.

Let us look at the hierarchy problem in more detail. The usual argument is that loop corrections using a cutoff $\Lambda$ give contributions to $m_H^2$ of order $\Lambda^2$, so that the bare $m_0^2 H^\dagger H$ coupling in the Lagrangian must be fine-tuned to cancel the $\Lambda^2$ contribution, leaving a small remainder of order $v^2$. Furthermore, this cancellation is unnatural, because $m_0^2$ must be fine-tuned order-by-order in perturbation theory to cancel the $\Lambda^2$ terms from higher order corrections. There are several reasons why this argument is invalid: Firstly, Nature does not use perturbation theory, so what happens order-by-order in perturbation theory is irrelevant. Secondly, in a sensible renormalization scheme that factorizes scales properly, such as dimensional regularization, there are no $\Lambda^2$ loop contributions, and no fine-tuning is needed. Explicit computation of the Higgs mass correction in eqn~(\ref{5.15}) shows that the correction is proportional to $m_H^2$, not $\Lambda^2$.

Here is an even better argument---assume there is new BSM physics with particles at a high scale $M_G$, say the GUT scale. Then loop corrections to the renormalized mass $m^2$ are proportional to $M_G^2$, and these must be cancelled order-by-order in perturbation theory to give a Higgs mass $m_H$ much smaller than $M_G$. The order-by-order problem is irrelevant, as before. However, we still have corrections $m^2 \propto M_G^2$, even if we compute exactly. These terms show the sensitivity of IR physics (the Higgs mass $m_H$) to UV parameters ($M_G$). Recall that in the introduction, it was obvious that short- and long-distance physics factorized, and our bridge-builder did not need to know about $M_G$ to design a bridge. The sensitive dependence of $m_H$ on $M_G$ follows because we are computing low-energy observables in terms of high-energy parameters. We have already seen an example of this in Sec~\ref{sec:inputs}. The solution is to use parameters defined at the scale of the measurement. Using the EFT ideas discussed so far, it should be clear that if we do this, all $M_G$ effects are either logarithmic, and can be absorbed into running coupling constants, or are
\emph{suppressed} by powers of $1/M_G$. There are no corrections with positive powers of $M_G$.

Naturalness arguments all rely on the sensitivity of low-energy observables to high-energy (short distance) Lagrangian parameters. But treating this as a fundamental problem is based on attributing an unjustified importance to Lagrangian parameters. Lagrangian parameters are a convenient way of relating physical observables to each other, as discussed in Sec.~\ref{sec:3.3}. As an example, consider the computation of hadron properties using lattice gauge theory, with Wilson fermions. The bare quark masses $m_0$ get corrections of order $\Lambda \sim 1/a$, where $a$ is the lattice spacing. $m_0$ must be adjusted so that the physical pion mass is small (remember that one cannot measure the quark mass), and this is what is done in numerical simulations. Obtaining light Wilson fermions in the continuum limit is a numerical problem, not a fundamental one. The lattice fine-tuning required does not imply that QCD has a naturalness problem. We know this, because there are other ways to calculate in which the fine-tuning is absent. Similarly, in GUTs, there are ways to calculate in which there is no fine-tuning required for the Higgs mass.

We now consider the only version of the hierarchy problem which does not depend on how experimental observables are calculated. Assume we have a theory with two scales $M_G$, and $m_H$, which are widely separated, $m_H \ll M_G$. Here $m_H$ and $M_G$ are \emph{not} Lagrangian parameters, but experimentally measured physical scales. $m_H$ can be obtained by measuring the physical Higgs mass, $m_H \sim 125$\,GeV. $M_G$ can be measured, for example, from the proton decay rate (if the proton does decay). The hierarchy problem is simply the statement that two masses, $m_H$ and $M_G$, are very different. But suppose instead that we had a situation where $m_H$ and $M_G$ were comparable. Then we would have a different naturalness problem---$m_H$ and $M_G$ can differ by many orders of magnitude, so why are they comparable? The only physics problem is to understand why experimentally measurable quantities such as $\alpha_\text{QED}$, $m_e$, $m_\mu$, $m_B$ etc.\ have the values they have. Naturalness is not such a problem. The 40+ years of failure in searches for new physics based on naturalness, as well as the non-zero value of the cosmological constant, have shown that Nature does not care about naturalness.

\begin{figure}
\centering
\includegraphics[width=9cm,angle=90]{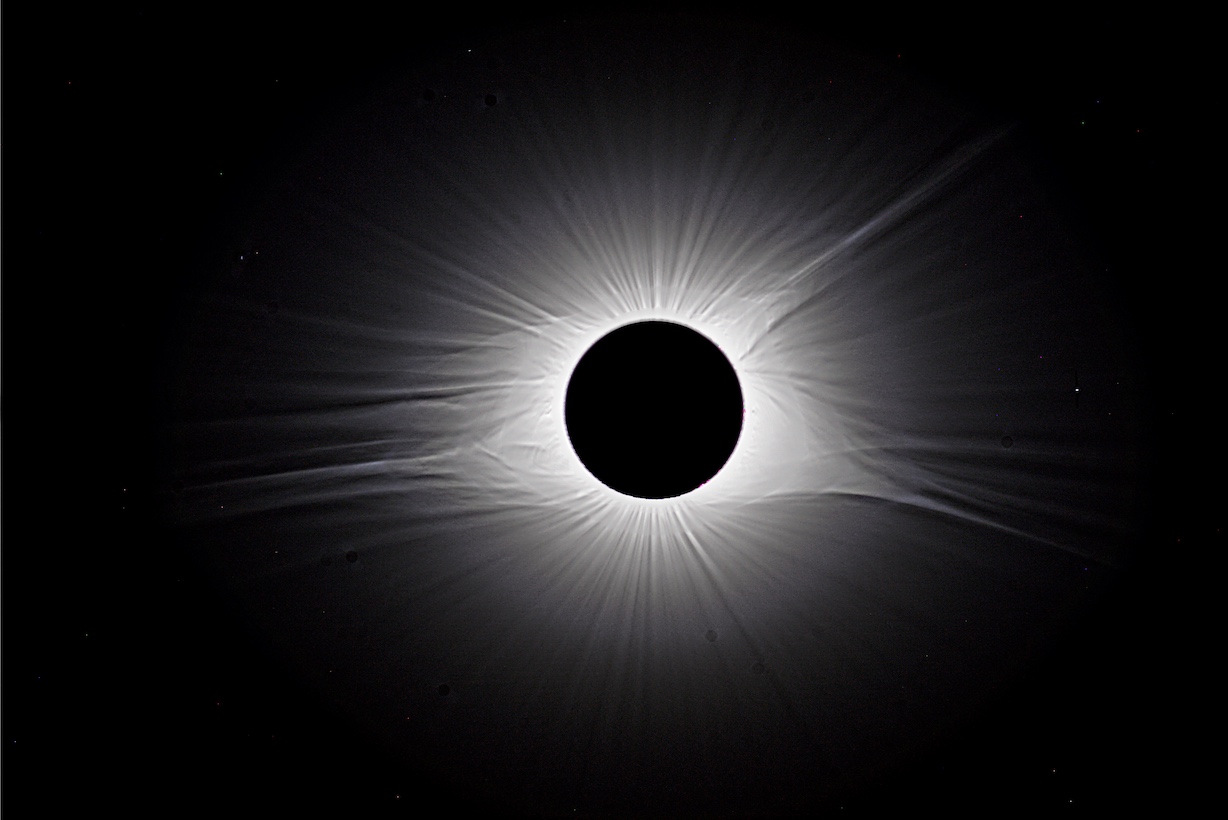}\hspace{0.5cm}
\includegraphics[width=9cm,angle=90]{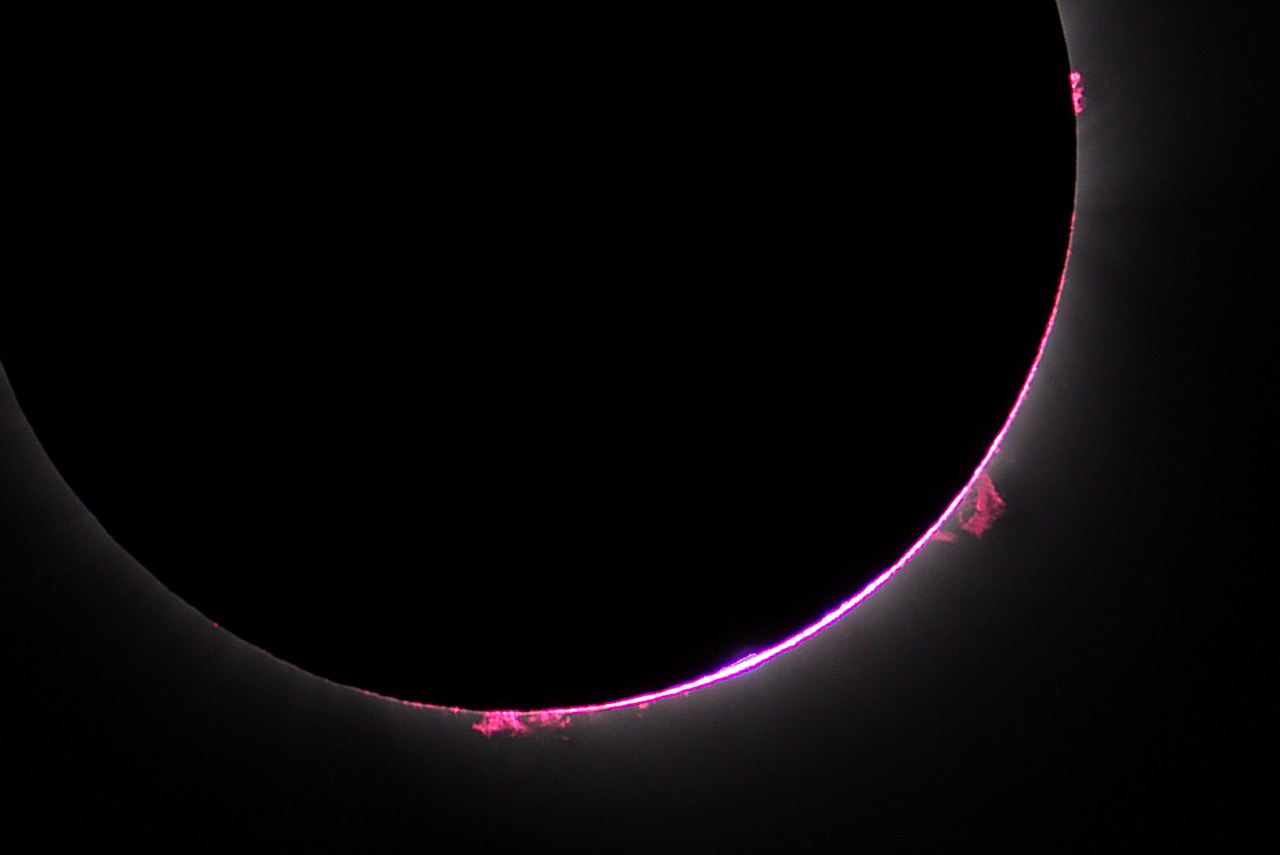}
\caption{\label{fig:eclipse} Photos of the 21 Aug 2017 solar eclipse. {\sl [Credit: P. Stoffer]}}
\end{figure}

Finally, let me comment on another fine-tuning problem that many of you are excited about. There will be a total solar eclipse on Aug 21, 2017, shortly after the  Les Houches school ends. The angular diameter of the Sun and Moon as seen from Earth are almost identical---the Moon will cover the Sun, leaving only the solar corona visible (see Fig.~\ref{fig:eclipse}). The angular diameters of the Sun and Moon are both experimentally measured (unlike in the Higgs problem where the Higgs mass parameter $m$ at the high scale $M_G$ is not measured) and the difference of angular diameters is much smaller than either.\footnote{The angular diameters are not constant, but change because of the small eccentricity of the orbit. As a result, one can have both total and annular eclipses. Furthermore, the Earth-Sun-Moon system is almost planar; otherwise there would not be an eclipse. These are two additional fine-tunings.} Do you want to spend your life solving such problems?


\bibliographystyle{OUPnum}
\bibliography{leshouches}

\end{document}